\documentclass[useAMS,usenatbib]{mn2e}

\usepackage{graphicx}\usepackage{epsfig}\usepackage{epsf}
\usepackage{amsmath}\usepackage{amssymb}\usepackage{stfloats}
\title[Eta Car light echoes]{Light-echoes from the plateau in Eta
  Carinae's Great Eruption reveal a two-stage shock-powered
  event\thanks{This paper includes data gathered with the 6.5 meter
    Magellan Telescopes located at Las Campanas Observatory, Chile.}}

\author[Smith et al.]{Nathan Smith$^{1}$\thanks{E-mail:
    nathans@as.arizona.edu}, Jennifer E.\ Andrews$^1$, Armin Rest$^2$,
  Federica B.\ Bianco$^{3,4}$, \newauthor Jose L. Prieto$^{5,6}$,Tom
  Matheson$^7$, David J. James$^8$, R.\ Chris Smith$^{9}$, \newauthor
  Giovanni Maria Strampelli$^{2,10}$, and A.\ Zenteno$^{9}$
  \\ $^{1}$Steward Observatory, University of Arizona, 933 N. Cherry
  Ave., Tucson, AZ 85721, USA \\ $^2$Space Telescope Science
  Institute, 3700 San Martin Drive, Baltimore, MD 21218, USA
  \\ $^3$Center for Urban Science and Progress, New York University, 1
  MetroTech Center, Brooklyn, NY 11201, USA \\ $^4$Center for
  Cosmology and Particle Physics, New York University, 4 Washington
  Place, New York, NY 10003, USA \\ $^5$N\'ucleo de Astronom\'ia de la
  Facultad de Ingenier\'ia, Universidad Diego Portales, Av. Ej\'ercito
  441, Santiago, Chile \\ $^6$Millennium Institute of Astrophysics,
  Santiago, Chile \\ $^7$National Optical Astronomy Observatory,
  Tucson, AZ 85719, USA \\ $^8$Event Horizon Telescope, Smithsonian
  Astrophysical Observatory MS 42, Harvard-Smithsonian Center for
  Astrophysics, 60 Garden \\ Street, Cambridge, MA 02138, USA
  \\ $^{9}$Cerro Tololo Inter-American Observatory, National Optical
  Astronomy Observatory, Colina El Pino S/N, La Serena, Chile
  \\ $^{10}$Universidad de La Laguna, Tenerife, Spain}

\begin{document}

\pagerange{\pageref{firstpage}--\pageref{lastpage}} \pubyear{2012}
\maketitle
\label{firstpage}

\begin{abstract}

  We present multi-epoch photometry and spectroscopy of a light echo
  from $\eta$ Carinae's 19th century Great Eruption.  This echo shows
  a steady decline over a decade, sampling the 1850s plateau of the
  eruption. Spectra show the bulk outflow speed increasing from
  $\sim$150 km s$^{-1}$ at early times, up to $\sim$600 km s$^{-1}$ in
  the plateau.  Later phases also develop remarkably broad emission
  wings indicating mass accelerated to more than 10,000 km s$^{-1}$.
  Together with other clues, this provides direct evidence for an
  explosive ejection.  This is accompanied by a transition from a
  narrow absorption line spectrum to emission lines, often with broad
  or asymmetric P Cygni profiles.  These changes imply that the
  pre-1845 luminosity spikes are distinct from the 1850s plateau.  The
  key reason for this change may be that shock interaction with
  circumstellar material (CSM) dominates the plateau.  The spectral
  evolution of $\eta$ Car closely resembles that of the decade-long
  eruption of UGC~2773-OT, which had clear signatures of shock
  interaction.  We propose a 2-stage scenario for $\eta$ Car's
  eruption: (1) a slow outflow in the decades before the eruption,
  probably driven by binary interaction that produced a dense
  equatorial outflow, followed by (2) explosive energy injection that
  drove CSM interaction, powering the plateau and sweeping slower CSM
  into a fast shell that became the Homunculus.  We discuss how this
  sequence could arise from a stellar merger in a triple system,
  leaving behind the eccentric binary seen today. This gives a
  self-consistent scenario that may explain interacting transients
  across a wide range of initial mass.

\end{abstract}

\begin{keywords}
  circumstellar matter --- stars: evolution --- stars: winds, outflows
\end{keywords}

\section{INTRODUCTION}

The underlying physical mechanism for $\eta$ Car's astounding
brightness variation and prodigious mass ejection has been the central
mystery associated with this object since John Herschel first drew
attention to its erratic flashes and relapses in the mid-19th century
\citep{herschel1847}.  Because an extremely luminous and massive star
appears to have survived this event, it has been discussed as a
prototype for a growing and diverse class of non-terminal eruptive
transients seen in external galaxies that have luminosities between
traditional novae and supernovae (SNe), often referred to as giant
eruptions of luminous blue variables (LBVs) or ``SN impostors'' (see
\citealt{smith+11,vdm12}).

Unlike these extragalactic transients, though, $\eta$ Car is nearby
enough that it affords us the opportunity to dissect the properties of
its spatially resolved bipolar ``Homunculus'' nebula \citep{gaviola}
that was ejected in the event \citep{currie96,morse01,sg98,smith17}.
There is a vast literature concerning multiwavelength observational
details of the Homunculus (see a recent review by \citealt{smith12}),
but the main ingredients to note here are its high ejected mass of
about 15 $M_{\odot}$ \citep{smith03,sf07}, its high expansion speeds
that also imply a large kinetic energy \citep{smith06}, and that the
majority of the mass is concentrated in very thin walls of the mostly
hollow bipolar shell \citep{smith06}.  Such extreme mass loss suggests
that brief eruptions may be important in the evolution of massive
stars \citep{so06}.

There are also complex ejecta outside the Homunculus, called the Outer
Ejecta \citep{thackeray50,walborn76}, which have elevated N abundances
\citep{davidson82,davidson86,sm04}.  Some of these Outer Ejecta have
very fast expansion speeds of 3000-5000 km s$^{-1}$ indicating an
origin in the 19th century Great Eruption \citep{smith08}, while the
majority are slower and implicate at least two major mass-loss
eruptions 300-600 years before the 19th century event
\citep{kiminki16}.  Were it not for this last point of recurring major
eruptions, the one-time merger of a binary system (discussed several
times; \citealt{jsg89,iben99,pz16,smith16b}) might seem like a natural
explanation for the energy and mass ejection of the Great Eruption.
The system is still a close and highly eccentric binary system today
\citep{damineli,dcl97}, which requires a triple system initially in
any merger model. The orbital parameters of the surviving binary are
constrained surpisingly well \citep{madura12}, considering that we
have not yet detected the secondary star.  The high eccentricity
dictates that the two stars come very close to one another at
periastron and may even collide or exchange mass during an eruption
\citep{soker01,soker04,ks09,smith11}, adding complexity to any binary
model.  Indeed, brief luminosity spikes in the historical light curve
of $\eta$ Car are seen to coincide with times of periastron passage
\citep{sf11}. Binary interaction is likely to be very important in the
physics of $\eta$~Car's eruption and in other SN impostors, but the
details of a working scenario are still a matter of much debate; our
main interest in this paper is to characterize the observed properties
of the mass loss during the eruption to help guide our understanding
of how so much mass left the system in such a short time.

Two qualitatively different models have emerged for the driving
physics of $\eta$ Car's mass-loss that can be summarized plainly as
either a wind or an explosion, although perhaps neither is quite so
simple.  The eruptive mass loss exhibited by $\eta$ Car occupies a
grey area between winds and explosions --- it is either a heavily
mass-loaded and energy starved wind, or a relatively weak explosion
that only unbinds the outer envelope.  This is between the opposite
extremes of either a line-driven wind or a core-collapse SN explosion.

The more traditional interpretation (traditional in the sense that it
has been around longer and is more developed) involves a strong
radiative luminosity that pushes the star above the classical
Eddington limit and initiates a strong outflow of matter. This is
interpreted in the context of the theory for continuum-driven
super-Eddington winds
\citep{owocki04,owocki17,os16,quataert16,shaviv00,so06,vanmarle08}.
In this picture, the outflow is a result of the increased radiative
luminosity, and the emitting surface is expected to be a relatively
cool pseudo-photosphere in the outflowing wind
\citep{davidson87,hd94,dh97,os16}.  Consequently, the roughly 20-year
duration of the eruption indicates that the star was exceeding its
classical electron-scattering Eddington limit by about a factor of 5
the entire time.

The other type of scenario for the Great Eruption mass loss is
primarily as a hydrodynamic explosion \citep{smith13}.  This picture
is different from the previous one in the sense that in the former, it
is the momentum of escaping photons that accelerates the outflowing
material.  In an explosion model, the radiation we observe is largely
a byproduct of heating by the shock interaction between fast
explosively ejected matter that overtakes slower circumstellar
material (CSM).  This scenario is generally referred to as ``CSM
interaction'', and is similar to the standard model for CSM
interaction in SNe~IIn, but with a non-terminal and lower-energy
explosion.  A key point is that in the CSM interaction model, we avoid
the puzzle of how a star's envelope can persist in a strongly
super-Eddington state for 20 years, because here the emitting material
is not bound.  Instead, the primary source of luminosity during the
plateau of the eruption resides in the shock itself as it plows
through the dense CSM. A simple 1-D model shows that one can account
for the observed decade-long plateau of $\eta$ Car's eruption while
also matching the present-day observed properties of the massive shell
nebula \citep{smith13}.

Both types of models have the shortcoming that they lack a clear
explanation for the ultimate source of the energy.  In the
super-Eddington wind model, the star's radiative luminosity is assumed
to increase temporarily and then decrease as dictated by the observed
light curve, and is the primary agent driving the mass loss.  The
source of this extra luminosity is unknown. In the CSM interaction
scenario, the power source is the relatively sudden deposition of
energy deep inside the star for some unknown reason, and the radiation
is a byproduct.  That energy source may be from binary orbital energy,
as mentioned above, or from nuclear burning instabilities akin to
those that have been suggested in eruptive SN progenitors
\citep{qs12,sq14,sa14,woosley17}.  The wind model may have problems
accounting for several aspects of the observed nebula (see below),
whereas the CSM interaction requires us to invoke some slow
pre-existing CSM for the fast ejecta to collide with.  Both models can
potentially account for the bipolar shape if we invoke either rapid
rotation \citep{do02,og97,st07} or equatorial CSM
\citep{frank95,langer99}.  A key ongoing challenge is to understand
how a merger or some other physical model might provide the required
energy on the appropriate timescale, in a way that yields the observed
results.

Primary sources of empirical information that have guided these two
mass-loss scenarios involve the historical visible-wavelength light
curve \citep{sf11} and the physical parameters of the remnant of the
explosion - the ``Homunculus Nebula'' and its surrounding debris,
which can be studied in exhaustive detail.  The historical record
provides the observed fact that the object's luminosity did exceed the
Eddington limit for the star's presumed mass for more than a decade,
motivating the wind model.  On the other hand, continued study of the
present-day nebula gives several clues that together point strongly
toward a hydrodynamic explosion.  These are: (1) the large mass of
12-20 $M_{\odot}$ in the Homunculus combined with its fast expansion
speeds gives a large kinetic energy of order 10$^{50}$ ergs
\citep{smith03,smith06}, which exceeds the radiative energy budget of
$\sim$10$^{49}$ ergs.  This low ratio of luminious to kinetic energy
is more characteristic of radiation from expanding and cooling SN
envelopes than of stellar winds (although winds with extreme photon
tiring might also achieve this; \citealt{owocki04}). (2) Observations
of material outside the Homunculus indicate very high expansion
speeds reaching 5000 km s$^{-1}$, which is easier to explain with
shock acceleration \citep{smith08}.  This outer fast material also
raises the total kinetic energy budget of the event even more. (3)
Most of the mass in the Homunculus resides in the extremely thin walls
of the bipolar lobes, which points to compression in a radiative shock
\citep{smith06,smith13}.  Other details of the structure in the nebula
also point toward a shock rather than a steady wind (see discussion in
\citealt{smith13}).


\begin{figure*}
\includegraphics[width=6.3in]{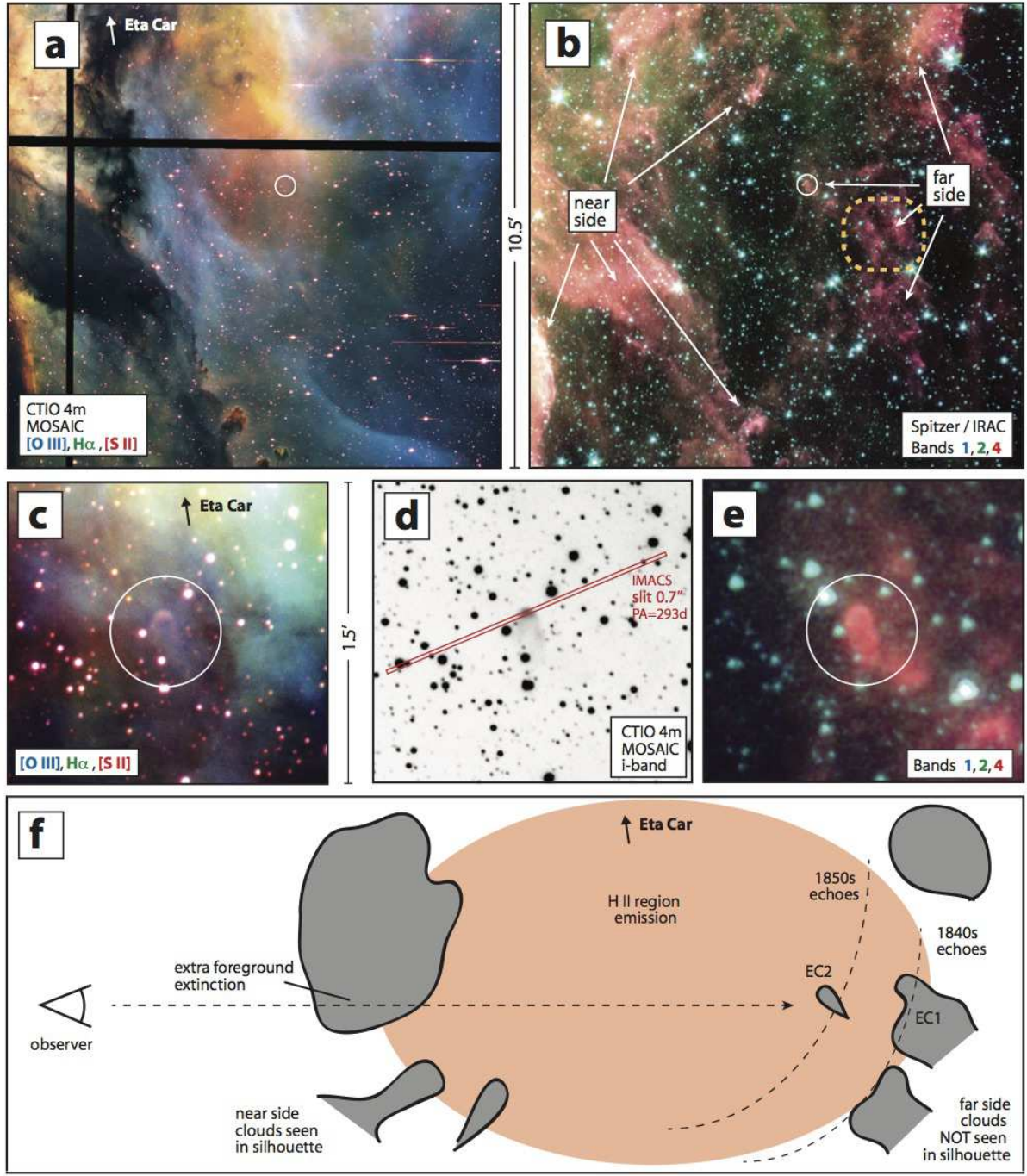}
\caption{The environment around the EC2 light echo.  (a) Large field
  of view 3-color composite image at visible wavelengths, with [O~{\sc
      iii}] $\lambda$5007 in blue, H$\alpha$ in green, and [S~{\sc
      ii}] $\lambda\lambda$6717,6731 in red.  The images were obtained
  in 2003 with the MOSAIC camera on the CTIO 4m telescope
  \citep{smith03a}. An arrow points toward the location of $\eta$ Car
  itself, off the top of the image.  (b) Same field of view as (a),
  but showing IR images obtained with the IRAC camera on {\it
    Spitzer}, in Bands 1 (blue), 2 (green), and 3 (red)
  \citep{smith10spitz}.  (c) Same image and color scheme as (a) but
  zoomed in on a smaller field around EC2.  (d) same field as (c) but
  showing only the $i$-band CTIO4m/MOSAIC image in grayscale, also
  obtained in 2003.  The red box shows the location of our most
  commonly used IMACS slit aperture.  (e) Same field as (c) and (d)
  but in the IR, with the same {\it Spitzer} images and color scheme
  as (b). (f) A sketch showing the global geometry involved.  An
  Earth-based observer is to the left, looking through the cold clouds
  on the near side of the nebula, which appear dark in optical images
  and glow in PAH emission in the IR.  They are seen in silhouette
  against the bright screen of H~{\sc ii} region emission that fills
  the interior of the nebula. Cold clouds on the far side are also
  seen in PAH emission in the IR, but cannot be seen in silhouette at
  visible wavelengths, because they are behind the line emission.
  This is the case for the EC2 cloud, as well as the EC1 group of
  echoes discussed in our previous papers \citep{rest12,prieto14}.
  Dashed curves denote the rough locations of the light echo parabolas
  corresponding to the light curve peaks in the 1830s-1840s, as well
  as the 1850s plateau.}
\label{fig:img}
\end{figure*}

Recent studies have added significantly to the already tremendous
repository of observational information about $\eta$~Car.  Namely, the
discovery\footnote{Historical aside: Light echoes from $\eta$ Car have
  been reported previously.  \citet{walborn+liller} discovered that
  clouds in the Keyhole nebula were reflecting the peculiar spectrum
  of $\eta$ Car. \citet{elliott79} obtained spectra of these features
  as well, but interpreted the relatively broad line wings as evidence
  that the Keyhole was a supernova remnant.  Additional spectra of
  these reflected echoes were also obtained and interpreted as echoes
  with minor spectral variability over time
  \citep{lm84,lm86,boumis98}.  However, these were not strongly
  variable echoes of the Great Eruption, but rather, reflected light
  from the star in its modern post-eruption state.  Interestingly,
  though, \citet{walborn+liller} pointed out that if these nearby
  clouds are scattering light from the star today, then this may
  explain why drawings of the Keyhole by \citet{herschel1847} look
  different from its appearance today \citep{gratton63}.  If so, then
  John Herschel was arguably the first to record light echoes from the
  Great Eruption.} of light echoes from $\eta$~Car's Great Eruption
\citep{rest12} and the evolution of light echo brightness and spectra
over time \citep{prieto14} allow us to probe deeper, providing a
unique and crucial link between the historical brightness record, the
kinematics and structure of the nebula, and potential similarity to
modern extragalactic analogs. \citet{rest12} showed that light echo
spectra near the peak of the eruption showed a characteristic
temperature that was significantly cooler (G-type) than published
expectations for pseudo-photospheres of LBV eruptions
\citep{davidson87} and observed spectra of LBV eruptions \citep{hd94}.
This sparked a debate.  \citet{dh12} argued that if one were to
extrapolate the published pseudo-photosphere models of
\citet{davidson87} in the appropriate way, the wind photosphere might
be consistent with temperatures as cool as observed.  \citet{os16}
noted inconsistencies in the analysis by \cite{davidson87}, but also
showed that by properly accounting for opacities in radiative
equilibrium, wind mass-loss rates of the order of that inferred for
$\eta$ Car's Great Eruption are compatible with temperatures around
5000~K after all.  In any case, spectroscopy of the subsequent fading
of that same echo \citep{prieto14} showed that the temperature became
cooler still, dropping to 4000-4500~K and forming molecular bands
commonly seen in extremely cool carbon stars.  This behavior with time
contradicts simple expectations for a pseudo-photosphere
model, where the apparent temperature should increase as the
photosphere recedes to deeper wind layers \citep{davidson87}.

The development of such cool temperatures and molecular features in
the spectra presented by \citet{prieto14} correspond to one of the
brief luminosity spikes (e.g., 1843, 1838, etc.) observed in the early
stages of the eruption \citep{sf11}.  As described in this paper, the
temporal evolution of echo spectra shows clear disagreement with a
wind pseudo-photosphere interpretation of the eruption, but gives
unambiguous evidence of an explosive component to the mass loss.
There are a number of important implications for the nature of the
Great Eruption and the evolutionary history of the $\eta$ Car system.

\section{OBSERVATIONS}

In this paper, we investigate the spectral and photometric evolution
of an echo from $\eta$ Carinae that is different from the echoes
discussed in our previous papers \citep{rest12,prieto14}.  The new
echo, which we designate EC2 (EC1 was the group of echoes discussed by
\citealt{rest12}), is located at $\alpha$(J2000) = 10:44:28.80,
$\delta$(J2000) $-$60:15:30, and was discovered in the same difference
imaging that we used to discover other echoes; see \citet{rest12} for
details.  EC2 arises on the surface of a cometary shaped dust cloud.
This echo is somewhat brighter than the other echoes we've studied
previously, but is especially distinct in that it fades much more
slowly and shows different spectral characteristics.  EC2 is unique
among the echoes we found in that it was brighter in the first-epoch
2003 image and has faded steadily since then.  All other echoes have
brightened compared to 2003.  As we detail below, EC2 likely
corresponds to the main 1845-1858 plateau in the Great Eruption,
rather than the initial pre-1845 luminosity spikes.  It therefore
provides unique new information about the physics and evolution of the
Great Eruption.

\subsection{Emission-line and IR Imaging}

For context in understanding the location, geometry, environment, and
nearby background emission associated with the light echo studied in
this paper, we include an analysis of multiwavelength images of the
Carina Nebula.  EC2 is located in the southern part of the Carina
Nebula, about 2{\arcmin} away (or about 1.3 pc in projection) from the
EC1 group of light echoes discussed in our previous papers
\citep{rest12,prieto14}, which sample earlier times in the Great
Eruption than the light being reflected now by EC2.  EC1 and EC2
echoes trace roughly the same viewing angle to the star.

Relatively wide-field (10$\farcm$5$\times$10$\farcm$5) color composite
images of this region are shown in Figure~\ref{fig:img}.
Figure~\ref{fig:img}a shows an image in visible-wavelength emission
lines of [O~{\sc iii}] $\lambda$5007 (blue), H$\alpha$ (green), and
[S~{\sc ii}] $\lambda\lambda$6717,6731 (red) that are commonly used to
image H~{\sc ii} regions.  These were obtained in 2003 March with the
MOSIAC2 camera Cerro Tololo Inter-American Observatory (CTIO) 4m
Blanco telescope, which uses a 2$\times$4 array of 2048$\times$4096
pixel CCDs giving a roughly half-degree field of view with small chip
gaps.  A portion of these images is shown in Figure~\ref{fig:img}a.
The reduction and analysis of the images have been described elsewhere
in previous papers that used these same data
\citep{smith03a,smith04,smith05}.  Figure\ref{fig:img}b shows the same
field of view in the infrared (IR) in images obtained with the {\it
  Spizer Space Telescope} in 2005 January using the Infrared Array
Camera (IRAC).  The reduction and analysis of these images were
presented in a previous paper \citep{smith10spitz}.  The color image
shown here combines Band 1 (3.6 $\mu$m) in blue, mostly containing
stellar photospheric light and polycyclic aromatic hydrocarbon (PAH)
emission, Band 2 (4.5 $\mu$m) in green, containing starlight,
Br$\alpha$, and some hot dust emission, and Band 4 (8.0 $\mu$m) in
red, dominated by PAH emission from the surfaces of molecular clouds
illuminated by UV radiation.  PAH emission from clouds appears pink or
magenta in this image, while stars appear blue/green.  The location of
EC2 is circled, and the clouds that give rise to the echoes we have
studied previously are in the dashed yellow box in
Figure~\ref{fig:img}b.

Figures~\ref{fig:img}c, \ref{fig:img}d, and \ref{fig:img}e show a
zoomed-in (1$\farcm$5$\times$1$\farcm$5) region around EC2, where the
color schemes in Panels~\ref{fig:img}c and \ref{fig:img}e are the same
as the larger field images.  Figure~\ref{fig:img}d is a negative
greyscale image of the $i$-band MOSAIC2 image also taken in 2003
March, which is dominated by reflected continuum starlight.  This
$i$-band image has served as our first-epoch template image that we
initially used to make difference images to discover light echoes
around $\eta$ Car \citep{rest12}.  These zoomed images clearly show a
small comet-shaped dust cloud, the near surface of which gives rise to
the light echo discussed in this paper.  The bright end of this cloud
spans about 5$\arcsec$, or roughly 0.2 ly across.  This is important
when considering possible smearing of the echo signal by light travel
time.

The bottom panel in Figure~\ref{fig:img}f shows a sketch of our
interpretation for the viewing geometry for the EC2 light echo.  The
rationale for this geometry is explained later in Section 3.1.

\begin{figure}
\includegraphics[width=3.3in]{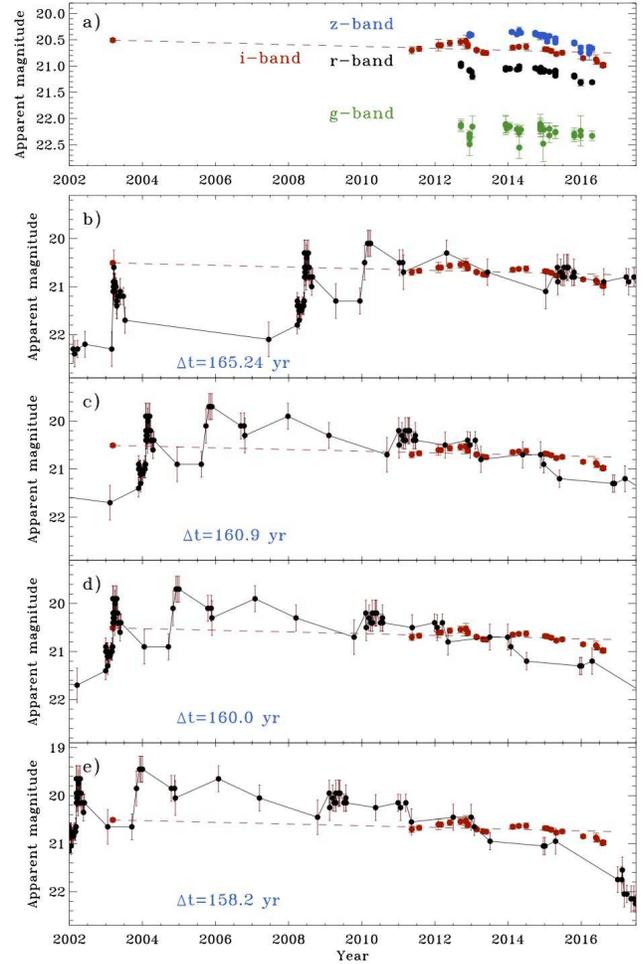}
\caption{Photometry of EC2 compared to the historical light curve of
  $\eta$ Carinae from \citet{sf11}.  Panel (a) shows our $griz$
  photometry of EC2.  The first $i$-band point refers to the apparent
  magnitude of EC2 in our first epoch March 2003 $i$-band reference
  image obtained with the MOSAIC camera on the CTIO 4m telescope. The
  dashed line shows a representative slope of the fading EC2 echo,
  forced to pass through the 2003 point and then fit to the later
  measurements.  The decline rate is 0.062 mag yr$^{-1}$.  Panels
  (bcde) show the same $i$-band light curve and decline rate as panel
  (a), compared to the historical visual light curve from
  \citet{sf11}.  In each successive panel, the historical light curve
  is shifted through EC2's light curve by different amounts
  ($\Delta$$t$ = 165.24, 160.9, 160.0, and 158.2 yr, respectively).}
\label{fig:phot}
\end{figure}

\begin{figure}
\includegraphics[width=3.3in]{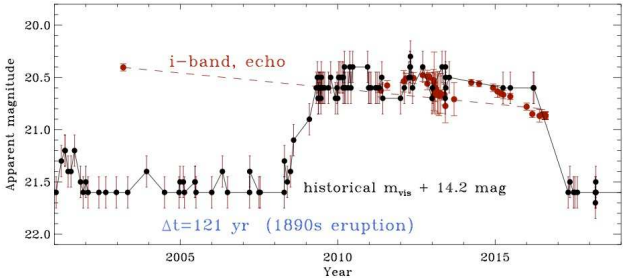}
\caption{Same as Figure~\ref{fig:phot}, but showing the historical
  light curve with a smaller delay time of 121 yr, comparing EC2's
  light curve to that of the Lesser Eruption in the 1890s.  Clearly
  EC2 cannot be reflecting light from the 1890s eruption, because it
  was far too bright in 2003. (In addition, the expansion speed
  observed directly in spectra of the 1890s eruption was far slower
  than in EC2's echo spectra; see text.)  We assume fairly generous
  0.15 mag error bars for the historical light curve in the 1890s (see
  \citealt{sf11}).}
\label{fig:phot1890}
\end{figure}

\begin{table}\begin{center}\begin{minipage}{3.0in}
      \caption{Optical Broadband Imaging}
\scriptsize
\begin{tabular}{@{}lcccc}\hline\hline
Instrument &g & r & i & z \\ \hline
CTIO 4m Mosaic II &... &... &5  &... \\
CTIO 4m DECam     &29  &30  &67 &36  \\
LCO-2 2m FTS      &2   &1   &24 &... \\
LCO-1 Swope       &... &... &19 &... \\
\hline
\end{tabular}\label{tab:imaging}\end{minipage}
\end{center}
\end{table}

\subsection{Broadband Imaging}

The images from which we measured the broadband photometric
lightcurves were obtained with several different telescopes and
instruments: MOSAIC II and DECam \citep{decam} wide-field cameras (4
and 69 epochs respectively) mounted on the Blanco 4 m telescope at
CTIO, the direct CCD camera mounted on the Swope 1 m telescope at Las
Campanas Observatory (LCO-1), and the Spectral CCD camera mounted on the
2~m Faulkes Telescope South (FTS; \citealt{brown13}) at the Las
Cumbres Observatory (LCO-2) Siding Spring site.
These totaled 31, 31, 115, and 36 epochs in $g$, $r$, $i$, $z$ bands,
respectively, with details given in Table~\ref{tab:imaging}.  Standard
image reduction was performed on all the images, including
bias/overscan subtraction and flat-fielding using skyflats and
domeflats.

The photometric data were processed with the \emph{photpipe} pipeline
\citep{rest05a}, which is the same pipeline that has been used to
discover and analyze the light echoes of historical SNe (e.g.,
\citealt{rest05b,rest08}) and other echoes of $\eta$~Car
\citep{rest12,prieto14}.  Images are kernel- and flux-matched,
aligned, and \emph{swarped} \citep{bertin02}, to match a template,
then pairs of images are subtracted to create difference images, and
bright stars are masked. This process produces clean images that
contain, ideally, only the light echo flux. Most of our sampling was
obtained in SDSS $i^{\prime}$ band. The $i^{\prime}$-band lightcurve
was processed using a CTIO Blanco telescope image from 2003 as a
reference template \citep{smith03a}. The SDSS $grz$ imaging campaign
did not begin until 2012. DECam images from 2012 are used as
reference templates for $g^{\prime}$ and $r^{\prime}$, and from 2013
for $z^{\prime}$ band.

Five 3$\times$3 pixel regions are selected along the light echo, and
away from bright stars. We sample multiple regions in order to
decrease noise. The flux from these regions is averaged, and compared
to standard star photometry to produce the lightcurve shown in
Figure~\ref{fig:phot}.  While sampling the flux in 5 locations
increases the number of data points, thus decreasing the noise, it
samples the event at slightly different epochs, due to the slightly
different distance between each dust region observed and the event
source (i.e. light travel time across the reflecting cloud). However,
photometry generated from a single region centered where the
spectroscopic slit is centered (a 3$\times$3 pixel region at
$\alpha$(J2000) = 10:44:28.80, $\delta$(J2000) -60:15:30) produces a
lightcurve with identical time evolution within the errors.  The total
light travel time across the reflecting cloud is small - only 0.2 yr.
The fluxes thus obtained were transformed into the DECam AB magnitude
system using observations of SDSS standards obtained in 2014 January.
We compare this light curve to the historical light curve \citep{sf11}
with various shifts in time corresponding to times during the Great
Eruption (Figure~\ref{fig:phot}) and the Lesser Eruption in the 1890s
(Figure~\ref{fig:phot1890}).

\begin{table}\begin{center}\begin{minipage}{3.0in}
      \caption{Optical Spectroscopy of $\eta$ Car's EC2 light echo}
\scriptsize
\begin{tabular}{@{}lccccc}\hline\hline
UT Date     &Tel./Intr.     &grating &slit     &PA            \\ \hline
2011 Dec 23  &Baade/IMACS f2 &200  &0$\farcs$9 &270$^{\circ}$  \\
2012 Mar 18  &Baade/IMACS f2 &200  &0$\farcs$9 &270$^{\circ}$  \\
2012 Jun 26  &Baade/IMACS f2 &200  &0$\farcs$9 &270$^{\circ}$  \\
2012 Oct 14  &Baade/IMACS f4 &300  &1$\farcs$2 &340$^{\circ}$  \\
2013 Jan 07  &Clay/MagE      &ech. &1$\farcs$0 &240$^{\circ}$  \\
2013 Apr 05  &Baade/IMACS f4 &300  &0$\farcs$7 &270$^{\circ}$  \\
2014 Jan 07  &Clay/MagE      &ech. &1$\farcs$0 &240$^{\circ}$  \\
2014 Feb 05  &Baade/IMACS f4 &1200 &0$\farcs$9 &293$^{\circ}$  \\
2014 Feb 06  &Baade/IMACS f4 &300  &0$\farcs$9 &293$^{\circ}$  \\
2014 May 19  &Baade/IMACS f4 &1200 &0$\farcs$7 &293$^{\circ}$  \\
2014 Nov 03  &Gemini/GMOS    &R400 &1$\farcs$0 &293$^{\circ}$  \\
2015 Jan 20  &Baade/IMACS f4 &1200 &0$\farcs$7 &293$^{\circ}$  \\
2015 Jan 20  &Baade/IMACS f4 &300  &0$\farcs$7 &293$^{\circ}$  \\
2016 Mar 04  &Baade/IMACS f4 &1200 &0$\farcs$7 &293$^{\circ}$  \\
2016 Mar 05  &Baade/IMACS f2 &300  &0$\farcs$7 &293$^{\circ}$  \\
2016 Mar 25  &Clay/MagE      &ech. &1$\farcs$0 &240$^{\circ}$  \\
\hline
\end{tabular}\label{tab:spec}\end{minipage}
\end{center}
\end{table}

\subsection{Optical Spectroscopy}

Following the discovery of light echoes from $\eta$ Carinae
\citep{rest12}, we initiated a followup campaign to study the spectral
evolution of these echoes.  So far in previous papers, we have
discussed the initial spectra and spectral evolution of the EC1 group
of echoes that are thought to arise from pre-1845 peaks in the light
curve \citep{rest12,prieto14}, but we have monitored a number of other
echo systems as well.  EC2 is among the brightest of these targets,
which allowed us to obtain some observations with higher dispersion
than we could use for fainter echoes.

We obtained low- or moderate-resolution spectra of EC2 on a number of
dates from 2011 to the present, as listed in Table~\ref{tab:spec}.
Many of our spectra were obtained using the Inamori-Magellan Areal
Camera and Spectrograph (IMACS; \citealt{dressler11}) mounted on the
6.5m Baade telescope of the Magellan Observatory located at LCO-1. The
chosen slit width depended on seeing conditions and trade-offs between
signal and resolution, but was usually between 0$\farcs$7 and
1$\arcsec$.  With IMACS in f/2 mode, we used the 300 lpm grating to
obtain a single spectrum across the full optical wavelength range of
3900-9500 \AA \ at a low resolution of $R \simeq 200-400$ . With the
f/4 camera, we used either the 300 lpm grating to sample a wider
wavelength range at moderate $R \simeq 500$ resolution, or the 1200
lpm grating to sample a smaller wavelength range with higher
resolution of $R \simeq 6000$.  Usually the 1200 lpm grating was
centered on H$\alpha$, but we also obtained some 1200 lpm spectra of
the Ca~{\sc ii} infrared triplet.  The 2-D spectra were reduced and
extracted using routines in the IMACS package, and also standard
spectral reduction routines in IRAF.\footnote{IRAF is distributed by
  the National Optical Astronomy Observatory, which is operated by the
  Association of Universities for Research in Astronomy, Inc., under
  cooperative agreement with the National Science Foundation.}

We also obtained a relatively high-resolution echellette spectrum of
EC2 using the Magellan Echellette spectrograph (MagE; Marshall et
al. 2008) mounted on the Clay 6.5-m telescope at LCO-1.  These
echellette spectra are listed as ``ech.''  for the grating name in the
third column of Table~\ref{tab:spec}.  The spectra were obtained with
a 1$\arcsec$ slit, which yielded a resolution $R\sim 5000$, and
covered the full optical wavelength range ($\lambda=3200-10000$~\AA),
although with lower signal-to-noise than most of the IMACS 1200 lpm
spectra. The spectra were reduced, combined, and extracted using the
Carnegie pipeline written by D.~Kelson.

We obtained a low-resolution spectrum of EC2 on 2014 Nov 13 using the
Gemini Multi-Objects Spectrograph \citep{hook02} at Gemini South on
Cerro Pachon.  Nod-and-shuffle techniques \citep{gb01} were used with
GMOS to improve sky subtraction.  Standard CCD processing and spectrum
extraction were accomplished with IRAF.  The
spectrum covers the range $4540-9250$~\AA\ with a resolution of
$\sim9$~\AA.  We used an optimized version\footnote{\tt
  https://github.com/cmccully/lacosmicx} of the LA Cosmic algorithm
\citep{vandokkum01} to eliminate cosmic rays.  We extracted the
spectrum using the optimal extraction algorithm of \citet{horne86}.  Low-order
polynomial fits to calibration-lamp spectra were used to establish the
wavelength scale.  Small adjustments derived from night-sky lines in
the object frames were applied.  We employed our own IDL routines to
flux calibrate the data using the well-exposed continua of the
spectrophotometric standards \citep{wade88,matheson00}.

\begin{figure*}
\includegraphics[width=6.8in]{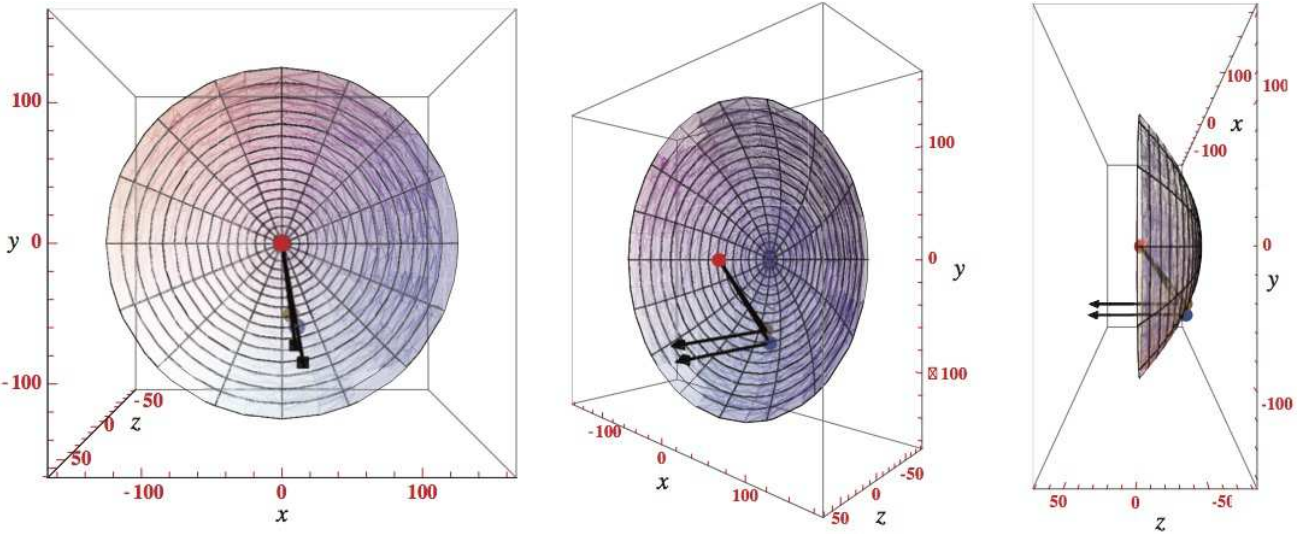}
\caption{Plots of the 3-D light path.  North is toward the positive
  $y$-axis (up), east is toward the negative $x$-axis (left), and the
  positive $z$-axis points toward the observer with the origin at
  $\eta$~Car. The red, brown, and blue circles indicate $\eta$ Car,
  EC2, and the EC1 echoes from our first paper \citep{rest12},
  respectively.  The parabolic relation between the spatial parameters
  of the scattering dust and the time since outburst is described by
  the well-known light echo equation \citep{couderc39}.  Assuming a
  time since outburst of 160 yr for EC2 and 169 yr for the EC1 echo,
  and a distance of 7660 light-years \citep{smith06}, we find that the
  scattering dust is at a position ($x$,$y$,$z$)=(10,-77,-61) ly for
  EC2 and ($x$,$y$,$z$)=(14,-78,-66) ly for EC1.  The black lines show
  the path of the light scattering from the light-echo-producing dust
  concentrations.  The two echoes both view $\eta$~Car from a similar
  direction near the equatorial plane of the Homunculus, but EC2 is
  closer to $\eta$ Car, and therefore sees later times in the
  eruption.}
\label{fig:3d}
\end{figure*}

\section{RESULTS}

\subsection{Environment}

The EC2 light echo, as well as the echoes we have discussed previously
in the literature \citep{rest12,prieto14} are seen in the southern
part of the Carina Nebula among the so-called ``South Pillars''
\citep{smith00} region.  This is a region of active ongoing star
formation in clouds exposed to feedback from the massive O-type stars
that have formed in the region, with this feedback shaping the clouds
into elongated globules and dust pillars.  The structure of these
clouds and dust pillars can be seen in mid-IR PAH emission from the
photodissociation regions on their surfaces in wide-field IR imaging
of the region \citep{smith00,smith10spitz}, as well as the image in
Figure~\ref{fig:img}b.  It is the surfaces of these dense star-forming
clouds in the Carina Nebula that are illuminated by light from the
eruption of $\eta$~Car and scattered toward us \citep{rest12}.  This
is different from the thin dust sheets in the ISM that produce light
echoes observed from a number of SNe \citep{rest05a,rest05b,rest08}.
This region has spatially varying line emission from the ionization
fronts and diffuse gas inside the H~{\sc ii} region, as well as very
patchy and highly variable line-of-sight extinction through the
nebula, which is larger on average than the extinction toward the
central clusters Tr14 and Tr16 \citep{sb07}.

With knowledge of EC2's position on the sky relative to $\eta$ Car,
combined with the constraints on its most likely delay time from
Section 3.2, we can use the understood behavior of a light echo
parabola to constrain its 3D geometry and viewing angle relative to
$\eta$ Car using the same method explained in our previous paper
(\citealt{rest12}; and references therein).  Figure~\ref{fig:3d} shows
resulting plots of the 3D geometry of EC2 and previously studied EC1
echoes \citep{rest12,prieto14} relative to $\eta$ Car, similar to the
plots presented in our earlier papers.  The two echoes both view
$\eta$~Car from a similar direction that is near (probably within
20$^{\circ}$ of) the equatorial plane of the Homunculus, but EC2 is
closer to $\eta$~Car and therefore sees more recently emitted light
(in other words, it lies along a slightly smaller light echo
paraboloid, tracing a later epoch in the eruption).  With the same
definitions for spatial coordinates as in \citet{rest12}, and as
defined here in the caption to Figure~\ref{fig:3d}, we find that the
scattering dust is at a position ($x$,$y$,$z$)=(10,$-$77,$-$61) ly for
EC2 and ($x$,$y$,$z$)=(14,$-$78,$-$66) ly for the EC1 echoes
\citep{rest12}.

%

These coordinates indicate that the scattering dust associated with
EC2 is on the far side of the Carina Nebula, well behind the plane of
the sky running through $\eta$ Car itself.  This has two important
implications.  First, it makes sense in terms of EC2's surroundings as
seen in images. Second, it will introduce more line-of-sight
extinction and contamination from diffuse nebular emission than for
some other parts of the Carina Nebula or $\eta$~Car itself.  These two
considerations are discussed below.

A cartoon depicting the global geometry involved is shown in
Figure~\ref{fig:img}f.  An Earth-based observer is to the left,
looking through the cold clouds on the near side of the nebula, which
appear dark in optical images and glow in PAH emission in the IR.
They are seen in silhouette against the bright screen of H~{\sc ii}
region emission that fills the interior of the nebula. These dark
clouds can be seen as patchy extinction in the optical emission-line
images shown in Figure~\ref{fig:img}a and in PAH emission in the IR
(Figure~\ref{fig:img}b).  Cold clouds on the far side are also seen in
PAH emission in the IR (Figure~\ref{fig:img}b), but they cannot be
seen in silhouette at visible wavelengths, because they are behind the
diffuse visible line emission within the H~{\sc ii} region.  This is
the case for EC2, as well as the EC1 echoes discussed previously.
Dashed curves in Figure~\ref{fig:img}f denote the rough locations of
the light echo parabolas corresponding to the light curve peaks in the
1830s-1840s, as well as the 1850s plateau (not to scale).  This
agreement between the geometry inferred from the light echo parabola
and from imaging of the environment gives an independent indication
that the delay time adopted is roughly correct (i.e. during the Great
Eruption).  If the light echoes were from a later time in $\eta$ Car's
history, such as the 1890s eruption, the younger paraboloid would
place them on the near side of the nebula, and the scattering dust
would be seen in extinction.  There are other reasons why EC2 cannot
be associated with the 1890 eruption as well, as mentioned later
(Section 3.2).

Located on the far side of the nebula, our line of sight to EC2 passes
all the way through the interior of the Carina Nebula, which provides
a long ($\sim$50 pc) path length of diffuse ionized gas that
contaminates the spectrum. This adds to the difficulty of interpreting
light echo spectra.  Fortunately, most of the diffuse emission can be
subtracted (with large residuals for the brightest lines like
H$\alpha$, [N~{\sc ii}], and [O~{\sc iii}]) by carefully sampling the
adjacent emission along the long-slit aperture.  This H~{\sc ii}
region emission is more of a problem than sky lines.

The large path length through the nebula may also add a great deal of
extra line-of-sight extinction.  In Figure~\ref{fig:img}a, one can see
evidence of patchy diffuse extinction within the Carina Nebula.  It is
therefore likely that the required extinction correction for EC2 is
larger than the value of $E(B-V)$=0.47 mag that is usually adopted,
derived from the average for O-type stars in the central Carina Nebula
\citep{walborn95}.  Since the extinction is lowest toward the center
of the nebula, $E(B-V)=0.47$ mag is a minimum value; below we adopt
$E(B-V)=1.0$ mag to deredden all our spectra of EC2.  This value
(within roughly $\pm$0.2 mag, dominated by the noise in the blue
wavelength range) brings the continuum shape in spectra into agreement
with the apparent temperature deduced from spectral features (see
\citealt{rest12}).  The true value of the line-of-sight extinction
could be even higher, but is also mitigated because of scattering by
dust that tends to make the light bluer.  Note also that this dust
within the Carina Nebula has a different reddening law than the
average ISM value, with $R = A_V / E(B-V) = 4.8$ \, \citep{smith02},
rather than the usually assumed value of 3.1.

\begin{figure*}
\includegraphics[width=6.0in]{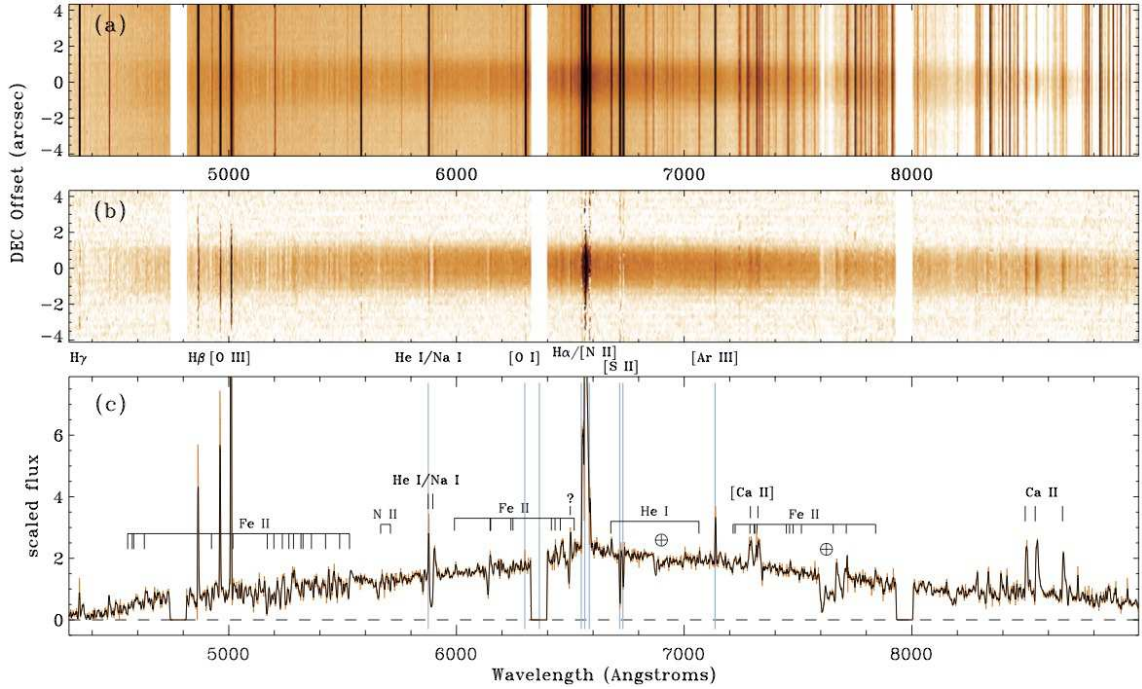}
\caption{An example of our 2D spectra before (a) and after (b)
  background subtraction.  These are from the Magellan/IMACS
  observation on 2015 Jan 20, taken with the low-resolution 300 lpm
  grating. Wavelengths where strong H~{\sc ii} region lines leave
  residual emission are noted below panel (b).  The background
  subtraction removes sky lines quite well, but there is some residual
  emission from the ionization front and photoevaporative flow off the
  surface of the reflecting globule itself.  The bottom panel (c) is
  the 1-D spectrum of EC2 extracted from the background-subtracted 2D
  spectrum in (b), before any flux calibration or sensitivity
  correction.  Several likely line identifications are given.}
\label{fig:extract300}
\end{figure*}

\begin{figure*}
\includegraphics[width=6.0in]{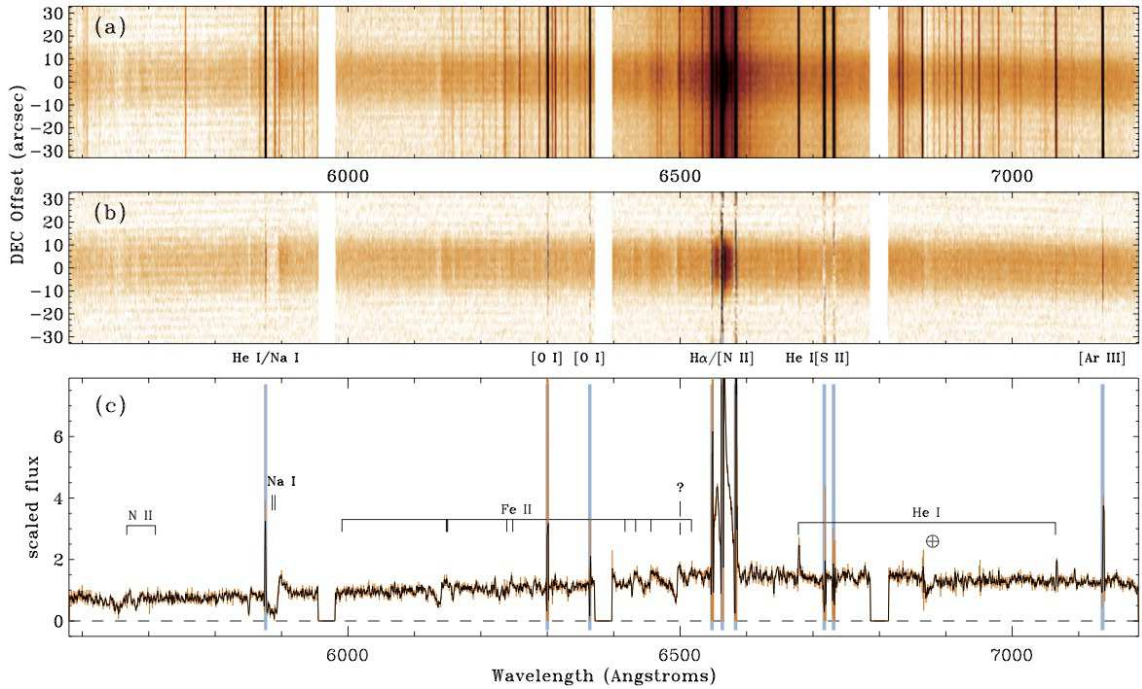}
\caption{Same as Figure~\ref{fig:extract300}, but for the higher
  resolution 1200 lpm grating, obtained with IMACS on the same night.}
\label{fig:extract1200}
\end{figure*}

\begin{figure*}
\includegraphics[width=6.0in]{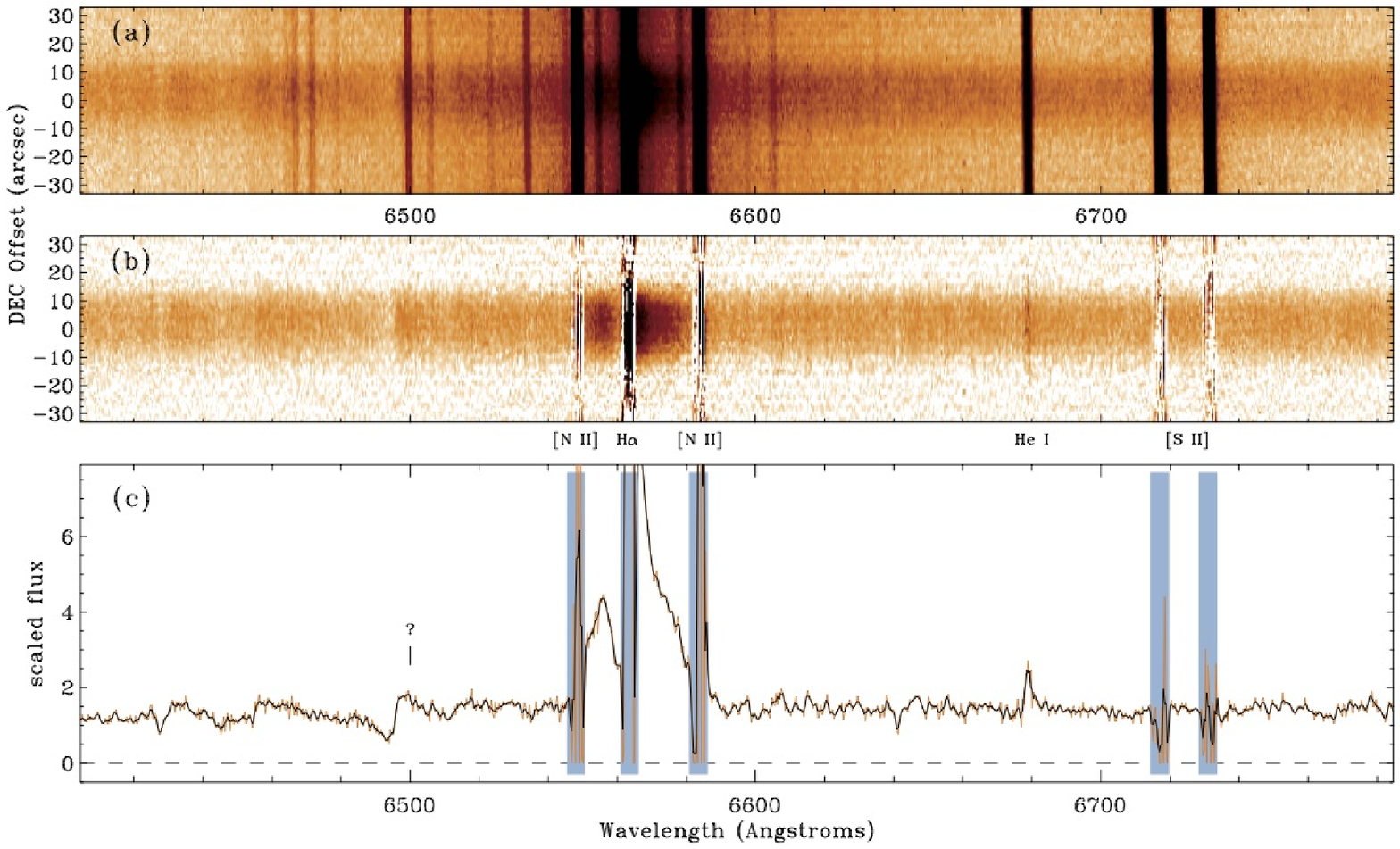}
\caption{Same as Figure~\ref{fig:extract1200}, but zoomed-in on the
  region immediately around H$\alpha$.}
\label{fig:extractHa}
\end{figure*}

\subsection{Light Curve of the EC2 Echo}

As noted earlier, EC2 is so-far unique among the set of light echoes
we have discovered in the Carina Nebula in the sense that it was
brighter in our template 2003 images, and it has stayed bright while
fading only slightly over more than a decade since then.  (In our
initial difference images of 2010$-$2003, it was the only echo
candidate to have a negative subtraction residual because it was
brighter in the reference template.)  It also has a relatively high
surface brightness among echoes discovered so far.  Despite the slow
changes, initial and continued spectroscopy of this feature (see
below) confirm that it is indeed an echo from $\eta$ Car's giant
eruption, since broad emission-line wings are seen, and the spectrum
changes significantly even though the brightness fades slowly.

We can use imaging photometry of EC2 to constrain the most likely time
period that it samples during the Great Eruption.  The historical
light curve \citep{sf11} shows a few brief luminosity spikes in 1843
and before that time, seeming to occur every 5.5 years, and possibly
coinciding with grazing collisions or more violent interaction events
at periastron passage in the highly eccentric binary system
\citep{smith11,sf11}.  Other echoes discussed so far
\citep{rest12,prieto14} brighten and fade on a $\sim$1 yr timescale,
and most likely correspond to some of these $\le$1843 luminosity
spikes.  The complicated effects of time delays introduced by the
geometry and thickness of the reflecting dust layer and how they
influence the light curve shape of an echo will be discussed in a
forthcoming paper (Bianco et al., in prep.).  Two of these brief
events were clearly observed in 1843 and 1838 and were well timed for
extrapolated periastron passages \citep{smith11}.  There is also
potentially one earlier event that was poorly sampled in 1827, and
which would also correspond to a time of periastron.  The available
historical record cannot rule out the possibility that there were many
such events occuring in the decades before the eruption, and that
these contribute to the light echoes we are finding.  One might expect
these events to increase in violence as the instability of the system
grows leading up to the 1840s event.

EC2, however, fades at a much slower rate than any of these brief
luminosity spikes.  It has stayed consistently bright for over a
decade, as shown in Figure~\ref{fig:phot}, ruling out an association
with these brief early interaction events.  Moreover, in the lead-up
to the Great Eruption, $\eta$ Car was slowly brightening in the
intervening quiescent time periods between these brief spikes,
culminating in the 1845 peak of the eruption.  EC2 slowly and steadily
fades, so it cannot be associated with the pre-1845 time period.

Figure~\ref{fig:phot} compares the observed light curve of EC2 to the
historical visual light curve \citep{sf11}, exploring the feasibility
of various potential time delays between the two.  Even though there
is a large gap in our observations between the first epoch in 2003 and
our light echo hunt that began in 2010, it is clear that EC2 has
faded, probably ruling out the time delay shown in panel (c), where
the initial 2003 epoch occurs before 1843.  The options that are
feasible are that our first 2003 epoch corresponds fortuitously with
either the brief 1838 luminosity spike ($\Delta$$t$=165 yr;
Figure~\ref{fig:phot}b), the 1843 spike ($\Delta$$t$=160 yr;
Figure~\ref{fig:phot}d), or later, with time delays of $\le$158.2 yr
as in panel (e).  Since we don't have spectra in 2003, and we don't
have suitable images in the intervening time period, we can't choose
confidently between panels (b), (d), or (e).

In any case, the main result is the same: we can be confident
that the light from EC2 that we have been observing since our campaign
began samples the main 1850s plateau phase of the Great Eruption.  The
fact that this echo is brighter and that its spectra are qualitatively
different from other echoes has important physical implications for
the mechanism of the eruption.

Could the EC2 echo be reflecting light from the so-called ``Lesser
Eruption'' \citep{hds99} in the 1890s?  This is very unlikely for
several reasons.  (1) Although the 1890 eruption is also a
long-duration plateau, it is not long enough.
Figure~\ref{fig:phot1890} shows the observed light curve of EC2
compared to the historical light curve from \citet{sf11} shifted so
that the 1890s eruption overlaps in time.  If we match the 1890s
eruption to photometry of EC2 at the present epoch, we see that EC2
was far too bright in our first epoch in 2003, which would correspond
to echo light from before the Lesser Eruption began.  This rules out
an association with light from the 1890s eruption.  This was unlikely
anyway because: (2) EC2 is the brightest echo we detect, while the
1890s eruption was several magnitudes fainter than the peak of the
Great Eruption, (3) images of the environment suggest that EC2 is on
the far side of the nebula (in very close proximity to echoes that
trace the pre-1845 peaks; \citealt{rest12}), making the path length
and delay time too long, and (4) historical spectra of the 1890s
eruption discussed by \citet{walborn+liller} show a cooler effective
temperature and (more definitively) slower velocities than the slowest
speeds we observe in our spectra of EC2 (discussed below).  Together,
these factors rule out the possibility that EC2 is an echo of the 1890
eruption.  Henceforth, we assume an approximate time delay of 160 yr
for EC2, tracing the light emitted by $\eta$ Carinae during its 1850s
plateau phase.

The light curve is admittedly not a perfect match to the historical
light curve in the 1850s either.  Aside from possible large
uncertainties in the various transformations applied to the historical
accounts \citep{sf11} or to actual observer error, there are three key
reasons why the echo light curve might differ from the historical
account:

1.  Light travel time will smear out a reflected light curve.  As
noted above, however, EC2 has a size of 5{\arcsec}, or only 0.2 ly.
This may smooth-out sharp peaks in the light curve, but will not
drastically change the fading rate.

2.  The historical light curve is visual eye estimates (mostly
blue/yellow wavelengths) by multiple observers, whereas the standard
$i$-band is much redder, so there may be significant color
differences.

3.  There may be real viewing angle differences.  EC2 views $\eta$ Car
from a vantage point close to the equator.  Since the ejection speed
and density varies strongly with latitude, it is plausible that
different latitudes could actually see a different light curve than we
see in the historical record from our vantage point, which traces a
latitude of about 40$^\circ$ \citep{smith06}.  With latitude-dependent
ejecta speeds and densities, dust could form at a range of delay times
from one latitude to the next, and so extinction could vary
substantially with time and viewing angle.  This could cause the light
curve to fade from one direction while remaining bright for a longer
time as seen from another direction, and different directions may have
different reddening, related to point (2) above.  In that case, the
``excess'' luminosity from EC2 seen in $\sim$2015
(Figure~\ref{fig:phot}), as compared to the more rapidly fading
historical curve, may not be a significant discrepancy.

\subsection{EC2's Apparent Color}

While EC2's photometric variability in a given filter is highly
reliable and informative, interpreting the photometric color of a
light echo is somewhat complicated and less reliable as a diagnostic
of the source.  This is because the light emitted by $\eta$ Car may
suffer various amounts of reddening on its way to the reflecting
cloud, it may get bluer due to the wavelength dependence of
scattering, and then it suffers additional reddening again as it
traverses the Carina Nebula and then passes through the ISM between
Carina and Earth.  Moreover, broadband filters are contaminated by
very bright nebular emission lines from the Carina Nebula itself,
making the absolute colors somewhat suspect for such a faint echo.

EC2 has a red color of $g-i \simeq +1.5$($\pm$0.2) mag in 2015
(Figure~\ref{fig:phot}a).  This is similar to the apparent color of
$g-i$=1.4 mag for a different echo in a similar region of the nebula
\citep{prieto14}.  Since that echo at peak had spectral signatures
indicating a temperature and intrinsic color similar to the Sun, it is
likely that most of this color is attributable to extinction by dust
along the light path.  The color of EC2 in broadband filters shows
little change from 2012-2016, consistent with the similar continuum
slope seen in spectra.  In our analysis of spectra below, we correct
all EC2 spectra for $E(B-V)$=1.0 mag.  Since the minumum for the line
of sight to Carina is already $E(B-V)$=0.47 mag, as noted above, this
includes a small amount of additional extinction.  (It is clear from
examining images that the patchy extinction toward the South Pillar
region may be considerable; Figure~\ref{fig:img}a.) This is a
convenient correction, since it causes the continuum shape of the 1843
echo to roughly match 5000-5500~K as indicated by the spectral
diagnostics (see \citealt{rest12}), and this same amount causes the
EC2 continuum shapes in spectra to approximately match a 6000 K
blackbody shape.  Given the similarity between EC2 and spectra
of UGC~2773-OT (see discussion below), this seems reasonable.  Since
we concentrate our analysis mostly on line strength variability and
velocity structure, this choice has little impact on our main results.

\begin{figure*}
\includegraphics[width=6.3in]{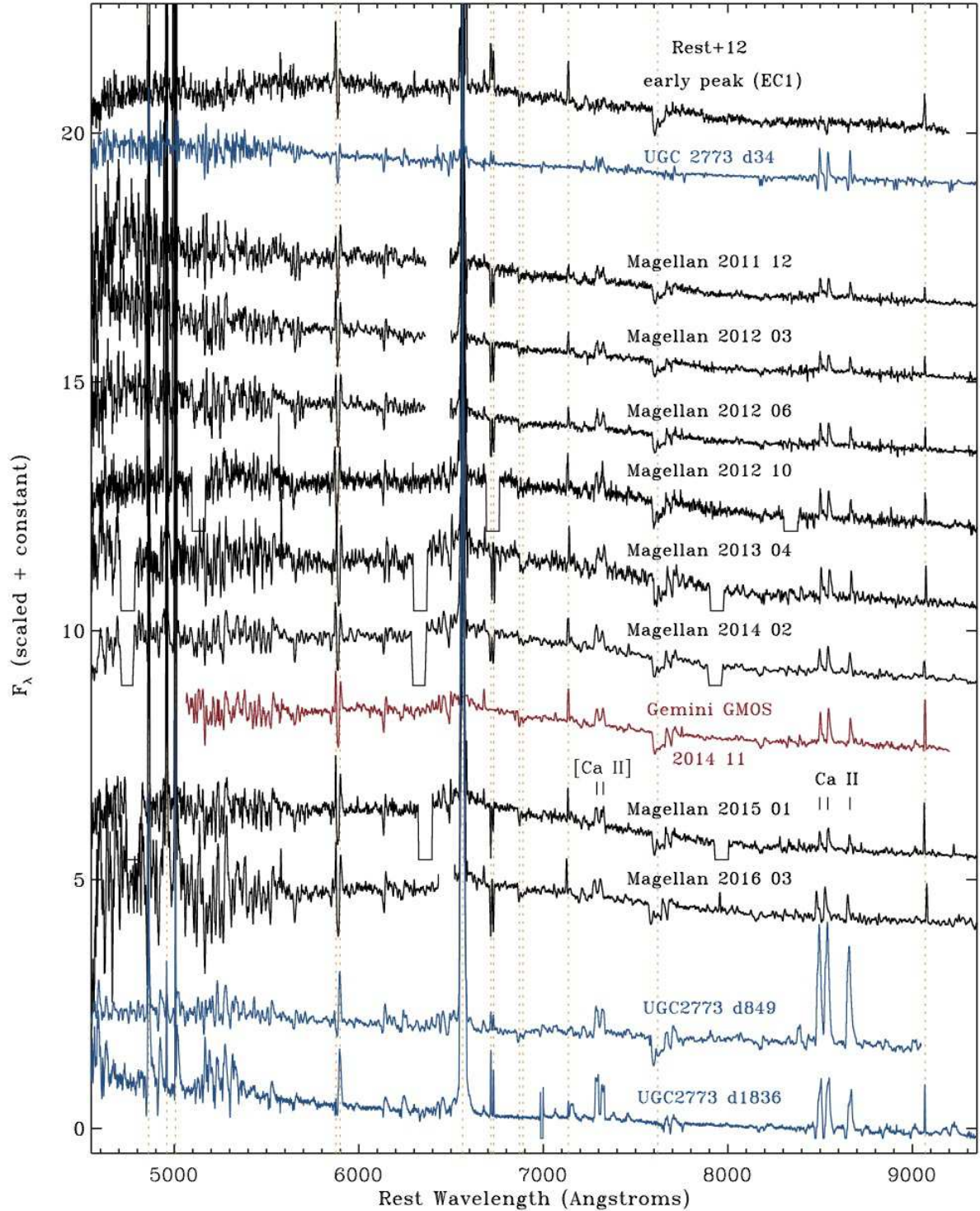}
\caption{Evolution of the low-res spectra of EC2.  Included for
  comparison are the first epoch spectra of the EC1 echo discussed by
  \citet{rest12}, which samples an early peak in 1843 or 1838, and
  both early and late-time spectra of UGC~2773-OT (in blue), from
  \citet{smith16}.  All spectra of $\eta$ Car's echoes have been
  dereddened by $E(B-V)$=1.0 mag; this amount of reddening is probably
  a minimum, but the true line-of-sight reddening is complicated by
  the fact that we are also observing reflected light.  The vertical
  orange dashed lines identify wavelengths of bright emission lines in
  the H~{\sc ii} region and telluric absorption (see previous
  figures).  Several of the Magellan/IMACS spectra have breaks in
  wavelength coverage due to gaps between detector chips.}
\label{fig:spec}
\end{figure*}

\subsection{Spectral Morphology}

\subsubsection{Background Subtraction}

The fact that the reflecting dust resides in clouds or dust pillars
embedded within the Carina Nebula, rather than a cold and thin dust
sheet in the ISM (as for many other SN echoes detected so far),
presents an added difficulty for spectroscopy of $\eta$ Car's echoes.
Although the echoes are brighter than many SN light echoes (EC2 has a
surface brightness of $\sim$20.5 mag arcsec$^{-2}$), they reside in a
region with extremely bright, spatially extended, narrow nebular
emission lines from the H~{\sc ii} region.  The reflecting dust is on
the surface of an opaque cloud, which has its own nebular emission
from the ionization front on its surface.  Therefore, even if the
diffuse background H~{\sc ii} region emission and sky emission is well
subtracted (sampled from adjacent regions and interpolated), there may
still be remaining intrinsic narrow nebular emission from the
reflecting cloud surface itself.  Moreover, the reflecting cloud may
have its own photoionized photoevaporative flow (see, e.g.,
\citealt{smith04finger}), so accurately sampling the adjacent
background emission might lead us to oversubtract any nebular lines
that are bright in the photoevaporative flow.  This becomes a tricky
process of how much to scale the sampled background that is
subtracted, in order to get rid of the narrow nebular emission that
contaminates the echo.  For this reason, our analysis mostly ignores
the very narrow emission or absorption associated with lines that are
bright in the H~{\sc ii} region.

Figures~\ref{fig:extract300}, \ref{fig:extract1200}, and
\ref{fig:extractHa} show examples of 2D spectra before and after
subtraction of the sky and diffuse H~{\sc ii} region emission, as well
as the corresponding 1-D extraction for each.  These correspond to
examples of lower-resolution spectra with broad wavelength coverage,
and higher resolution spectra focussed on the region around H$\alpha$
(Figure \ref{fig:extractHa} is exactly the same as
\ref{fig:extract1200}, but zoomed-in on H$\alpha$ to show differences
between broader reflected line profiles and narrow nebular emission).
The most important H~{\sc ii} region lines are marked in the figures:
H$\alpha$, H$\beta$, [N~{\sc ii}] $\lambda\lambda$6548,6584, [S~{\sc
    ii}] $\lambda\lambda$6717,6731, [Ar~{\sc iii}] $\lambda$7136, and
[O~{\sc i}] $\lambda\lambda$6300,6364. There is also narrow emission
from He~{\sc i} lines at 5876, 6680, and 7065 \AA, although some of
this may also be in the echo (see below).  For each of these, the
dominant residual emission is from nebular emission on the globule
itself, and not from noise in the subtraction of diffuse H~{\sc ii}
region lines.  The sky emission, which is uniform across the slit, is
cleanly subtracted in all our data.

\begin{figure*}
\includegraphics[width=4.4in]{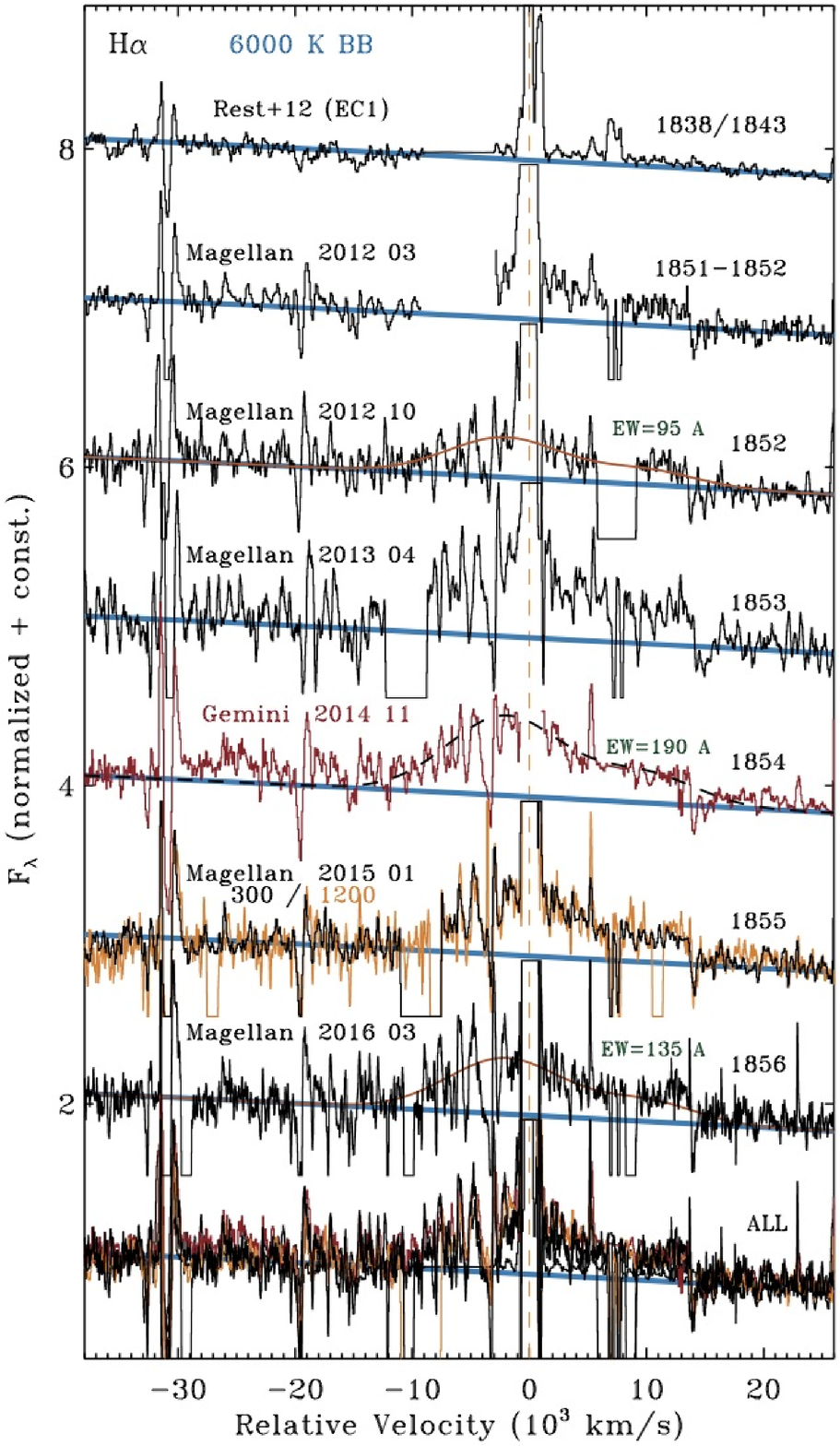}
\caption{Spectra of light echoes concentrating on the region around
  H$\alpha$, showing the broad H$\alpha$ wings plotted as a function
  of velocity relative to the centroid of the narrow H$\alpha$ line
  denoted by a vertical dashed orange line (note that the horizontal
  axis is in units of 1,000 km s$^{-1}$).  Several epochs are shown,
  including a spectrum of the EC1 echo from \citet{rest12}
  corresponding to an early peak in 1838 or 1843.  The other spectra
  are for EC2 obtained on the dates shown, and corresponding roughly
  to epochs during the Great Eruption noted at right.  Each spectrum
  has been dereddened by $E(B-V)$=1.0 mag and is compared to a 6,000 K
  blackbody (blue).  For three epochs, an example of a composite
  Gaussian that approximates the shape of the broad component is shown
  for comparison, and the equivalent width of the broad feature is
  noted near those three spectra.  The bottom shows all epochs of
  spectra overplotted.  As in previous figures, some of the
  Magellan/IMACS spectra have detector chip gaps.}
\label{fig:fastALL}
\end{figure*}

\subsubsection{Low-resolution Spectra}

Figure~\ref{fig:spec} shows a time series of low-resolution spectra of
the EC2 echo, extracted from the 2-D spectra as in
Figure~\ref{fig:extract300}, but also flux calibrated and corrected
for reddening adopting $E(B-V)$=1.0 mag.  These trace the spectrum of
$\eta$ Car in the late 1840s (upper) through the mid 1850s (lower).
For comparison, Figure~\ref{fig:spec} also includes a spectrum of an
early peak in the Great Eruption (1843 or 1838) from the EC1 echo
discussed previously by \citep{rest12}.  Except for the Gemini
spectrum (red), all the echo spectra of $\eta$ Car (black) were
obtained with IMACS.  Also for comparison, Figure~\ref{fig:spec} shows
the early and late phase spectra of the SN impostor UGC~2773-OT (blue)
from \citet{smith16}, which is thought to be a close analog of
$\eta$~Car.

Through the duration of our observations of EC2, the continuum slope
remains roughly constant.  It is, however, somewhat bluer than the
earlier epoch in the eruption.  \citet{rest12} found that the spectral
signatures were best matched by a G spectral type and an effective
temperature around 5000~K.  When we deredden this spectrum by
$E(B-V)$=1.0 mag, the continuum slope is consistent with a blackbody
temperature of 5,000-5,500~K.  The spectra of EC2 with the same
reddening correction have only slightly warmer temperatures around
6,000~K.  Inferring a reliable temperature from the continuum slope is
difficult, however, because of the forest of absorption/emission lines
in the blue, and because the blue end of the spectrum has relatively
poor signal to noise.

EC2 shows interesting differences compared to the EC1 echo that traces
an early 1838/1843 peak in the eruption.  That spectrum was dominated
by a forest of narrow absorption lines in the blue.  Many of these
change from pure absorption into emission, or into P Cygni profiles
(Figure~\ref{fig:spec}).  In the red part of the spectrum, for
example, the Ca~{\sc ii} IR triplet was seen in pure absorption in the
early peak, but shows strong emission with a P Cygni profile in the
EC2 echo.  Similarly, the [Ca~{\sc ii}] $\lambda\lambda$7291,7324
doublet was absent in the earlier peak \citep{rest12}, but is strongly
in emission in all epochs of EC2.  \citet{prieto14} presented a time
series of spectra of the EC1 echo.  As that echo faded from peak over
the subsequent 1-2 yr, it showed a gradual change from absorption to
emission in the [Ca~{\sc ii}] doublet and the Ca~{\sc ii} IR triplet,
with an end state somewhat similar to that seen in EC2. A key
difference, though, is that this occurred as the continuum faded
significantly, unlike EC2.  Moreover, this change from absorption to
emission in the Ca lines was accompanied by the appearance of strong
molecular absorption bands of CN and a further drop in effective
temperature to 4000-4500~K \citep{prieto14}.  These properties are
definitely not seen in EC2 during the time of our observations.
Whereas the changes seen in the EC1 echo may have been due to a shell
ejection with a rapid drop in optical depth and a cooling of the
shell, the EC2 echo appears to trace a significantly different
physical scenario.

All epochs of EC2 spectra show narrow emission of He~{\sc i}
$\lambda$5876, $\lambda$6678, and $\lambda$7065 that remains after
background subtraction.  This is interesting, because He~{\sc i}
emission requires relatively high ionization and should not be seen
from a $\sim$6,000 K atmosphere.  The 2-D spectra in
Figures~\ref{fig:extract300}, \ref{fig:extract1200}, and
\ref{fig:extractHa} show that the background H~{\sc ii} region
emission from He~{\sc i} is cleanly subtracted from the echo spectra.
However, some residual may remain from intrinsic He~{\sc i} emission
that arises on the ionization front or photopevaporative flow
associated with the globule itself, which is exposed to
ionizing UV radiation from the O stars in the Carina Nebula.  Indeed,
some residual emission of [O~{\sc iii}] and [Ar~{\sc iii}] is also
seen from the globule.  It would be tempting to dismiss the residual
He~{\sc i} as arising on the surface of the globule except for three
facts.

First, the He~{\sc i} lines are slightly broader ($\sim$150 km
s$^{-1}$) than the resolution limit of our 1200 lpm spectra with
IMACS, and they show a subtle asymmetric profile shape, with a hint of
a P Cygni profile. This can be seen in the He~{\sc i} $\lambda$6678
emission in Figure~\ref{fig:extractHa}, for example.

Second, the dereddened flux ratio of the He~{\sc i} lines
$\lambda$5876:$\lambda$6678:$\lambda$7065 in EC2 (roughly 2:1:0.5) is
different from the same ratio in the Carina Nebula H~{\sc ii} region
(roughly 3:1:0.75; \citealt{smith04finger}), such that He~{\sc i}
$\lambda$6678 is relatively stronger than the other lines in the echo.
Among the three lines, He~{\sc i} $\lambda$6678 also has the clearest
P~Cyg profile.

Third, similar narrow He~{\sc i} emission was seen in spectra of
UGC~2773-OT in its later phases dominated by CSM interaction
\citep{smith16}. The He~{\sc i} emission was much narrower (about 100
km s$^{-1}$) than H$\alpha$ and other lines (600-1000 km s$^{-1}$),
very much like the case here.  In that extragalactic $\eta$ Car
analog, there is no echo, and the He~{\sc i} emission is not due to
H~{\sc ii} region contamination, because it is seen to change
substantially in strength while the brightness of the transient
remained roughly constant (i.e. the He~{\sc i} emission is absent at
early times).  In UGC~2773-OT, the He~{\sc i} emission is thought to
arise in the pre-shock CSM, photoionized by X-rays from the shock
front \citep{smith16}, and this may be the case for some of the
He~{\sc i} emission in $\eta$~Car's echo as well.

We suspect that the narrow residual He~{\sc i} emission seen in
spectra of EC2 is a mix of intrinsic narrow emission from $\eta$~Car
and emission from the photoionized surface of the globule.  It is
difficult to confidently disentangle these two with available data,
but it will be possible if a spectrum of EC2's position can be
obtained at late times after the echo has faded.  This may take
another decade, however.

In the dereddened low-resolution spectra, the flux ratio of the
[Ca~{\sc ii}] doublet (F1 + F2) to the Ca~{\sc ii} IR triplet (XYZ),
where their flux ratio is denoted as F1+F1 / XYZ in the standard
nomenclature \citep{sl74}, is $\sim$0.46 (varying by $\pm$20\% in
various spectra of EC2).  This implies an electron density of $n_e
\simeq 3 \times 10^9$ cm$^{-3}$ \citep{fp89}.  This density is higher
than the density implied by the ratio of He~{\sc i}
$\lambda$5876/$\lambda$7065.  These lines would be roughly equal in
strength for densities of 10$^{9.5}$ cm$^{-3}$, as seen in some SNe
Ibn events \citep{matheson00}, so the He~{\sc i} lines must come from
more distant and lower density ejecta, if they are intrinsic to $\eta$
Car.

\citet{rest12} noted that the spectrum of UGC~2773-OT from
\cite{smith10} was the closest match to the properties seen in the EC1
spectrum.  This comparison is shown again in the top two spectra in
Figure~\ref{fig:spec}.  That comparison was based on an early spectrum
shortly after discovery of UGC~2773-OT (day 34 in \citealt{smith10}),
but the spectrum changed as UGC~2773-OT evolved over subsequent years.
While maintaining a roughly constant continuum slope and fading very
slowly, UGC~2773-OT's spectrum morphed from a forest of narrow
absorption lines to much stronger emission lines throughout the
spectrum, including many Fe~{\sc ii} lines \citep{smith16}.  Later
epochs of UGC~2773-OT are shown at the bottom of Figure~\ref{fig:spec}
for comparison.  It developed very strong and increasingly broad
emission from H$\alpha$ and Ca~{\sc ii}, and also showed increasing
strength of narrow He~{\sc i} emission as noted above.  Remarkably,
echo spectra of $\eta$~Car show very similar changes from the early
EC1 spectrum discussed by \citet{rest12} to the sequence of EC2
spectra in Figure~\ref{fig:spec}, also occurring while the continuum
luminosity faded only slightly.  Although the Ca~{\sc ii} IR triplet
in EC2 is not as strong as in UGC~2773-OT, the transition from
absorption to emission with P~Cyg profiles is qualitatively similar.
As we discuss below (Section 3.4.4), the similarity to UGC~2773-OT
also extends to the detailed behavior of line profile shapes.  We
discuss the overall similarity between $\eta$~Car and UGC~2773-OT
later in Secion 4.2.



\subsubsection{Broad H$\alpha$ Wings}

Perhaps the most surprising discovery in our study of the EC2 echo
spectrum is the presence of extremely broad emission wings of the
H$\alpha$ line.  To our knowledge, these are the fastest outflow
velocities seen in any eruptive transient, reaching $-$10,000 km
s$^{-1}$ to the blue, and roughly $+$15,000 to $+$20,000 km s$^{-1}$
on the red wing (the red wing is strongly affected by atmospheric
B-band absorption).  In our previous paper \citep{smith+18} we
demonstrated that when corrected for telluric absorption in the
B-band, there is clear excess emission above the continuum in a red
wing that extends to at least +20,000 km s$^{-1}$.

The interpretation and significance of these broad wings is discussed
in more depth in a separate paper \citep{smith+18}.  Briefly, the high
velocities are not an instrumental artifact and they are inconsistent
with electron scattering wings.  Because EC2 views $\eta$~Car from
near the equator, these broad wings are also inconsistent with an
origin in a fast bipolar jet that might arise from accretion onto
companion star, if such a jet is invoked to explain the bipolar
Homunculus \citep{soker01,ks09}.  Velocities of the broad wings are
much faster than any expected escape velocity in the $\eta$~Car
system. Combined with the presence of fast polar ejecta seen in the
Outer Ejecta \citep{smith08}, the broad wings instead suggest the
presence of a wide-angle explosive outflow during the Great Eruption.
The broad wings are relatively faint, and may correspond to a small
fraction of the total outflowing mass accelerated to high speeds.
This is reminiscent of the high speeds seen in SN~2009ip in the
precursor outbursts before its 2012 SN event
\citep{smith10,foley11,pastorello13}.

Here we detail the time dependence of this broad emission.
Figure~\ref{fig:fastALL} shows a sequence of spectra similar to
Figure~\ref{fig:spec}, but zoomed-in on the region around
H$\alpha$. This includes the EC1 echo spectrum of an early peak in the
eruption (probably the 1838 or 1843 peak), discussed already
\citep{rest12,prieto14}.  The broad emission is not present or much
weaker in the EC1 spectrum.  The other spectra are the EC2 echo taken
over several years, all of which show the broad wings at various
strengths.  The broad emission component appears to strengthen and
then fade as time progresses, with a maximum in 2013-2015,
corresponding roughly to the mid 1850s.

Figure~\ref{fig:fastALL} shows a composite Gaussian curve overplotted
on the Gemini spectrum on 2014 Nov.  This has two components: one with
a FWHM of 14,000 km s$^{-1}$ centered at $-$2,000 km s$^{-1}$, and the
other with FWHM of 12,000 km s$^{-1}$ centered at $+$10,000 km
s$^{-1}$ (with about 40\% of the strength of the main component).  The
total emission equivalent width of this broad emission is $-$190 \AA.
The same composite Gaussian is plotted against the 2012 spectrum with
about 50\% of the strength, and over the later 2016 spectrum with 70\%
of the strength (as compared to 2014/2015).  If it did reach its peak
strength in 2014/2015, it will be interesting to see if this broad
component in the EC2 echo continues to fade or moves to slower
velocities as time proceeds.

\begin{figure}
\includegraphics[width=3.0in]{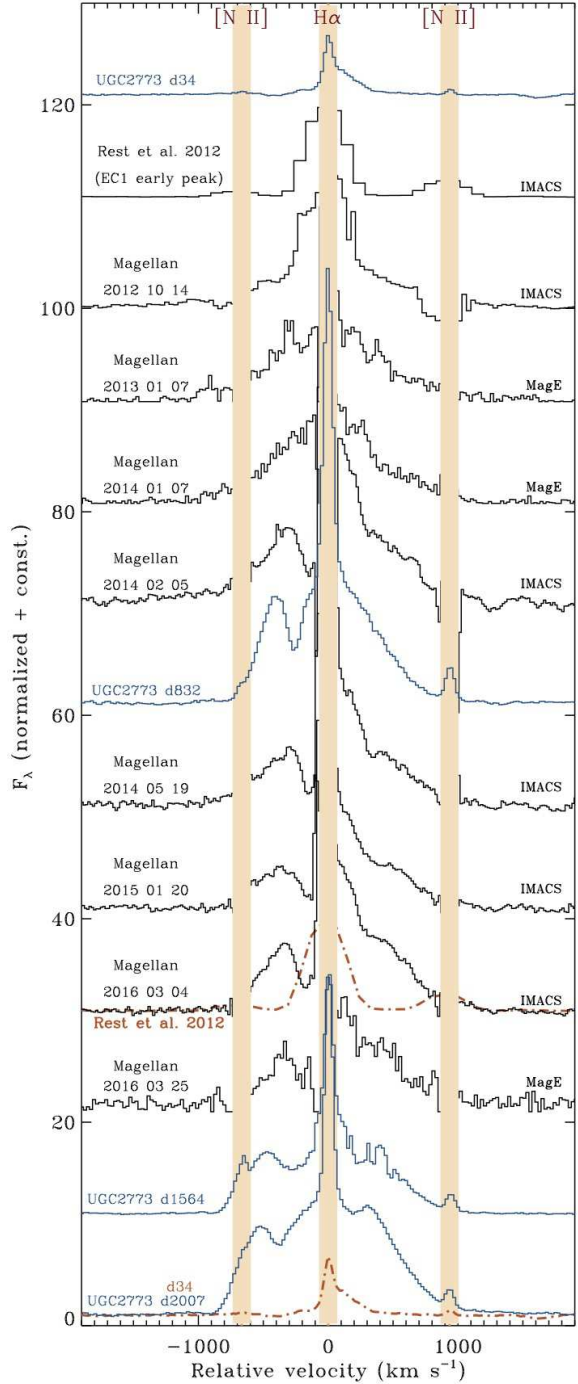}
\caption{Evolution of the narrow and intermediate-width H$\alpha$ line
  profile throughout the eruption, including an early EC1 spectrum
  \citep{rest12} corresponding to a peak in 1843 or 1838. (Note that
  the very broad emission wings extend far outside the velocity range
  plotted here.)  Except for this first spectrum, the rest have higher
  dispersion with the IMACS 1200 lpm grating or with MagE.  We also
  include spectra of the $\eta$ Car analog transient UGC~2773-OT
  (blue) at early and late times \citep{smith16}.  Toward the bottom
  (in orange, dot-dashed) we show the first epoch spectra superposed
  on the late-time spectra, to emphasize differences. Note that echo
  spectra of $\eta$ Car are contaminated by artifacts from the
  imperfect subtraction of bright nebular lines like H$\alpha$ and
  [N~{\sc ii}] $\lambda\lambda$6548,6583, so these regions of the
  spectra are masked in light orange.}
\label{fig:halpha}
\end{figure}

\begin{figure}
\includegraphics[width=3.1in]{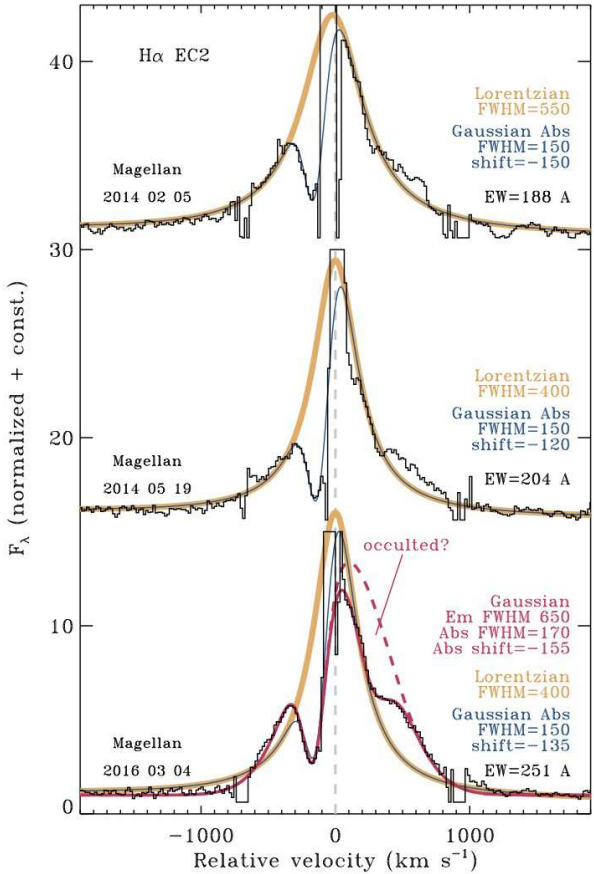}
\caption{A few of the H$\alpha$ profiles of the EC2 echo from
  Figure~\ref{fig:halpha} with Lorentzian and/or Gaussian fits for
  comparison.  Values for the centroid shift and FWHM of these
  components are noted at right.  Values of the total equivalent
  widths (EWs) of the fits are also noted (values in \AA); a caveat is
  that the ``continuum'' level for these fits is actually the wings of
  the broad component, and so the EW values of the fits quoted here
  have had a correction applied (factor of 1.4, 1.4, and 1.35 for the
  top, middle, and bottom, respectively) such that the EW corresponds
  to the underlying continuum and not the level of the broad wings.}
\label{fig:fit}
\end{figure}

\subsubsection{H$\alpha$ Line Profiles (narrow component)}

A majority of the H$\alpha$ line flux is contained in the narrow
component, and this is the emission that most directly traces the
formation of the Homunculus with speeds of several 10$^2$ km s$^{-1}$.
In our spectroscopy of $\eta$ Car's echoes (especially in the
higher-resolution 1200 lpm grating spectra with IMACS and with MagE),
the narrow component is resolved outside the regions of the spectrum
that are heavily contaminated by nebular emission residuals.  It shows
growing strength and changes in line profile shape with time.

Figure~\ref{fig:halpha} shows a time series of the H$\alpha$ line
profile seen in echoes, zooming in on the narrow component.  We have
blocked-out regions of the spectrum within 50 km s$^{-1}$ of the
narrow nebular lines of H$\alpha$ and adjacent [N~{\sc ii}] emission,
because these are heavily contaminated by subtraction residuals (or
over subtraction) of the narrow nebular emission that is much brighter
than the echo.  The most interesting information about the echo light
is therefore the region in between H$\alpha$ and the [N~{\sc ii}]
lines.  Most of the spectra in Figure~\ref{fig:halpha} show the
evolution of the EC2 echo, but for comparison, we also include the EC1
spectrum of an early peak in the light curve (presumably 1838 or 1843)
from \citet{rest12}, and also a few spectra of the extragalactic
$\eta$ Car analog UGC~2773-OT (plotted in blue), from \citet{smith16}.
(UGC~2773-OT is not affected by a light echo, so we do not block the
narrow emission in this spectrum.)

There are a few notable changes to the narrow H$\alpha$ line that
occur with time.  First, the line gets much stronger and broader with
time, moving from the 1840s through the 1850s plateau phase.  Second,
the line profile changes from being fairly narrow and symmetric to
being broad and asymmetric, with a prominent blue bump, a more subtle
red bump, and a central partially resolved component becoming more
clear at late times.  At early phases, the narrow component has a
width of only about 200-250 km s$^{-1}$ and does not show clear P
Cygni absorption.  At later times, the line emission spreads to around
1000 km s$^{-1}$ with clear blueshifted absorption.  It is interesting
that the changes to a broader profile and the appearance of strong
asymmetry in the line profile shape occur over a time period when the
very broad wings grow in strength.  These may be related, as discussed
later.

Examining Figure~\ref{fig:halpha}, it may be clear why we have
included the comparison to the spectral evolution of UGC~2773-OT.  The
changes in the H$\alpha$ line strength and profile shape are almost
identical in these two objects.  The similarity is uncanny; a minor
difference is that the blue bump and blueshifted absorption, as well
as the wings of the line (the wings of the narrow component, at
least), are at somewhat higher speeds in UGC~2773-OT.  This may, of
course, be attributed to a viewing angle effect.  This further
emphasizes the similarity of these two objects \citep{smith16},
perhaps suggesting that we can use the observed properties of
UGC~2773-OT to fill in some of the remaining gaps in $\eta$~Car.

At early phases, when no P Cygni absorption is seen, the line profile
can be approximated within the noise with either a symmetric
Lorentzian or multiple Gaussians.  The emission gets broader with
time, suggesting outflow speeds increasing from 200 km s$^{-1}$ at the
earliest times \citep{rest12,prieto14}, to about 500 km~s$^{-1}$ in
2013/2014.  After 2013/2014, P~Cygni absorption strengthens and the
emission components become asymmetric.  Figure~\ref{fig:fit} shows
some examples of simple fits to the line shape of the narrow H$\alpha$
component in EC2.  As the P~Cygni absorption appears (first seen
clearly in our high-resolution spectrum in 2014 Feb;
Figure~\ref{fig:fit}, top), the line profile can be matched fairly
well by a symmetric Lorentzian with a width of 550 km s$^{-1}$ (shown
in orange) and a blueshifted Gaussian absorption at $-$150 km s$^{-1}$
(the total of Gaussian absorption subtracted from the Lorentzian
emission shown in blue).  Interestingly, this 550 km s$^{-1}$
Lorentzian emission is identical in strength and width to the
H$\alpha$ line in early spectra of SN~2009ip (see \citealt{smith10};
their Figure 10), although that object did not show the narrow $-$150
km s$^{-1}$ absorption (which, again, could potentially be a viewing
angle effect).

Over the subsequent couple of years, however, a symmetric Lorentzian
becomes a poorer description of the emission-line shape.  In the 2014
May spectrum (Figure~\ref{fig:fit}, middle), the observed electron
scattering wings at $-$1000 km s$^{-1}$ become weaker than the
Lorentzian profile, there is a deficit of flux in the narrow component
at $+$100 km s$^{-1}$, and there is a red bump of excess emission at
$+$300 to $+$900 km s$^{-1}$ compared to the Lorentzian.  The narrow
absorption is similar, however, with only a slight change in velocity.

By the third epoch of 2016 March (Figure~\ref{fig:fit}, bottom), a
Lorentzian is a clearly inadequate approximation of the line shape.
The faint blue wing is gone, and the red emission excess is even
stronger.  A better description of the line shape is a Gaussian with
FWHM of 650 km s$^{-1}$, which has its centroid shifted 170 km
s$^{-1}$ to the red, and with a similar $-$155 km s$^{-1}$ P Cygni
absorption component as before.  Even this, however, overestimates the
red side of the center of the line (dashed magenta curve).  Could this
be redshifted gas at low velocities that is occulted by the expanding
photosphere?  We can account for the missing flux by subtracting
another Gaussian with FWHM 250 km s$^{-1}$ centered at $+$260 km
s$^{-1}$ (solid magenta curve). Or, on the other hand, is this
indicative of a very asymmetric or bipolar outflow that is better
matched by simply adding several Gaussian emission components?  The
latter appears to be the case in UGC~2773-OT \citep{smith16}.  Clearly
there is justification for assuming a bipolar outflow in the case of
$\eta$ Car.  The point here is that a symmetric emission component
becomes an increasingly poor description of the line shape; this is a
case where detailed radiative transfer models may lead to a better
understanding of the origin of the line profile \citep{dessart15}.

This change in the emission profile shape from a Lorentzian to a
Gaussian (or asymmetric multi-component Gaussian), is physically
significant, and reminiscent of changes seen commonly in SNe~IIn
\citep{smith+08}.  In the case of SNe~IIn, the interpretation
is that the change corresponds to two different stages in the course
of CSM interaction.  At early times, the shock is buried inside dense
CSM, so X-rays and far-UV radiation from the shock propagate ahead to
make a photoionized precursor in the slow pre-shock CSM.  The
resulting narrow H$\alpha$ line photons must scatter out through high
optical depths in the wind, producing strong electron scattering wings
and a symmetric Lorentzian shape.  At later times, as the shock
marches outward and begins to emerge from the CSM, reaching outer
radii with lower optical depths, line photons from the accelerated
post-shock gas can escape directly, so the Lorentzian wings fade and
intermediate-width Gaussian line shapes (sometimes highly asymmetric)
emerge.  The similar behavior in line profiles from $\eta$ Car's
echoes are probably telling us that a similar CSM interaction scenario
is important \citep{smith13}.  This is discussed below.

The fits to the line shape allow a way to quantify the increasing
H$\alpha$ strength (since measuring it directly from the data requires
one to interpolate over the residuals from imperfect H~{\sc ii} region
subtraction).  The equivalent width (EW) of narrow component fits is
about $-$200 \AA \ (Figure~\ref{fig:fit}), but increasing with time
through epochs corresponding to the mid 1850s (this EW value is
relative to the true underlying continuum, not to the adjacent
apparent continuum level, which is actually the very broad emission in
H$\alpha$; see Figure~\ref{fig:fastALL}).  Interestingly, an EW of
about 200 \AA \ (emission EW is positive here) is very similar to
SN2009ip during its precursor event in 2009, and also similar to the
H$\alpha$ EW in the main 2012b peak powered by CSM interaction.  In
fact, this EW is similar to most SNe~IIn shortly after peak luminosity
(see Figure 7 in \citealt{smp14}).

\begin{figure}
\includegraphics[width=3.1in]{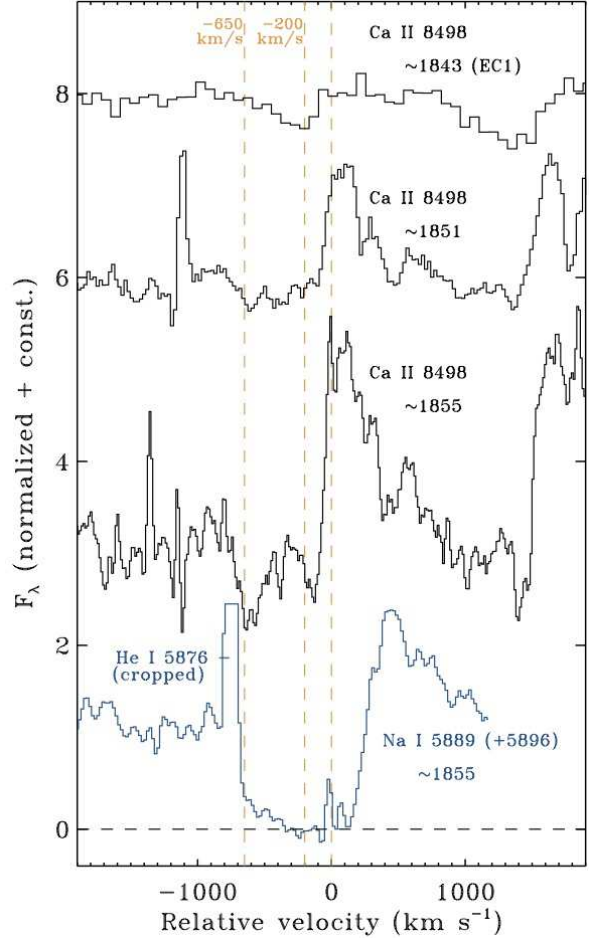}
\caption{Line profiles of Ca~{\sc ii} $\lambda$8498 at three different
  epochs, as well as Na~{\sc i} D at the last epoch.  The top spectrum
  is from the EC1 echo studied by \citet{rest12}, and is thought to
  represent one of the early peaks in the light curve in 1843 (or
  perhaps 1838).  This shows only absorption at a relatively slow
  speed of $-$200 km s$^{-1}$.  Later epochs are spectra of EC2 that
  trace the main plateau in the 1850s; these show higher speed
  absorption out to $-$650 km s$^{-1}$ and also emission from Ca~{\sc
    ii}, both of which were absent at earlier epochs.  The Na~{\sc i}
  line shows strong absorption all the way out to $-$650 km s$^{-1}$.}
\label{fig:Ca2}
\end{figure}

\begin{figure}
\includegraphics[width=3.1in]{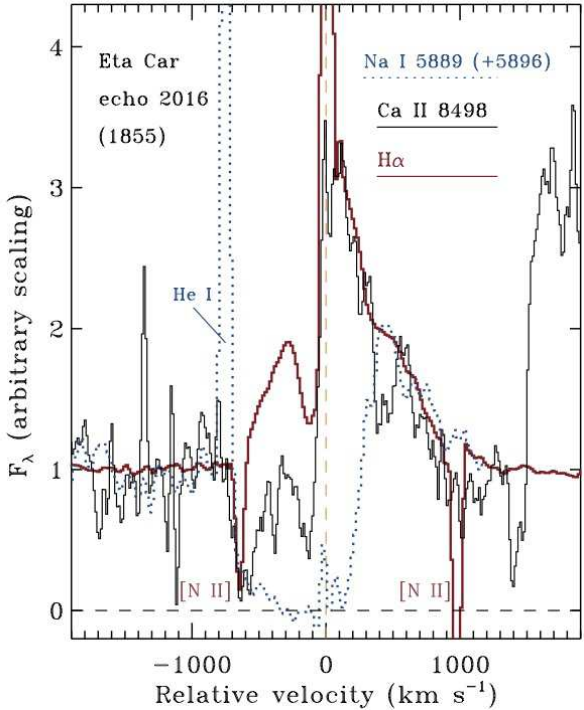}
\caption{Line profiles of H$\alpha$ (red), as compared to Ca~{\sc ii}
  $\lambda$8498 (black) and Na~{\sc i} D (blue dotted).  These were
  obtained in March 2016, which corresponds to roughly 1855 (give or
  take a few years), midway through the main plateau in $\eta$ Car's
  Great Eruption.  The intensity of the lines is scaled arbitrarily
  for display in order to compare their velocity structure, since the
  lines have very different emission strengths above the continuum.
  Features unrelated to the kinematics of the line profiles are the
  strong narrow emission from He~{\sc i} $\lambda$5876, and the
  oversubtraction of nebular [N~{\sc ii}] $\lambda\lambda$6548,6583
  (resembling narrow absorption) at $-$750 and +1000 km/s.  These are
  marked on the figure.  Also note that Na~{\sc i} D is a blend of two
  lines; the velocity is plotted for $\lambda$5889 to correctly
  demonstrate the blue edge of the absorption, so that the companion
  line $\lambda$5896 causes extra absorption at 0 to +300 km s$^{-1}$
  that is not representative of a true P Cygni profile from a single
  line.}
\label{fig:Ca2halpha}
\end{figure}

\begin{figure}
\includegraphics[width=3.0in]{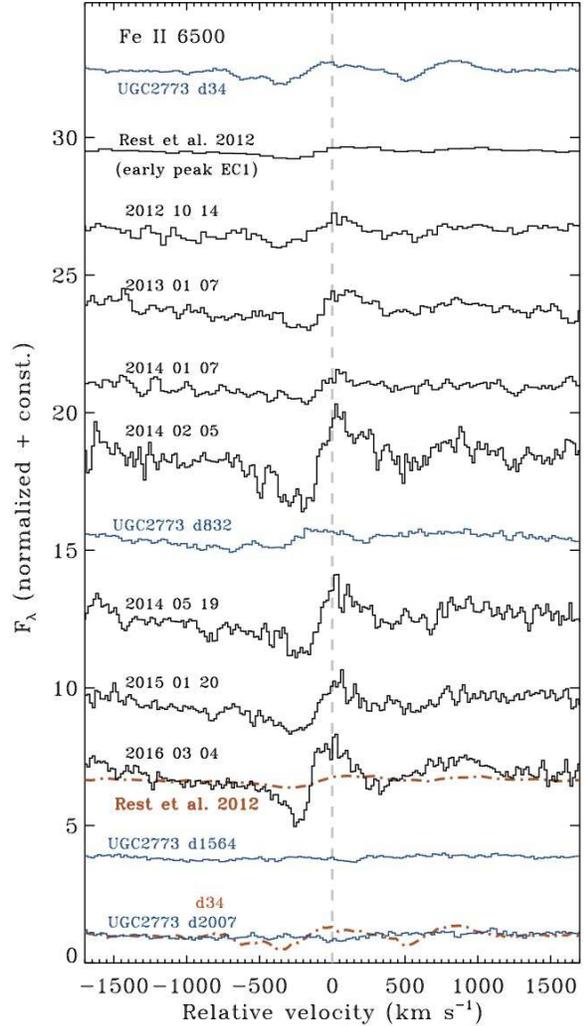}
\caption{Same as Figure~\ref{fig:halpha}, but for the P Cygni line at
  6500~\AA, which is presumably Fe~{\sc ii}.}
\label{fig:fe2}
\end{figure}

\subsubsection{Other Lines}

The Ca~{\sc ii} IR triplet is seen in emission at all epochs in the
EC2 echo at low resolution (Figure~\ref{fig:spec}), and we also
obtained a few epochs of higher resolution spectra with MagE and
IMACS, shown in Figure~\ref{fig:Ca2}.  These lines provide important
tracers of the outflow.  The IR triplet was present in the echo
spectra of an early peak, but was seen in pure absorption blueshifted
by $-$200 km s$^{-1}$ \citep{rest12}, consistent with the narrower
H$\alpha$ emission at this epoch, and similar to the P Cygni
absorption seen at later epochs.  This early EC1 spectrum is shown in
Figure~\ref{fig:Ca2}.  As that echo from an early peak faded over the
next 1-2 years, the IR triplet changed into emission with weak P Cygni
absorption (moving to somewhat slower velocities), and then to pure
emission \citep{prieto14}.  In the EC2 echo, the Ca~{\sc ii} IR
triplet shows a somewhat different behavior, first becoming stronger
with time in pure emission and a broader profile, and then continuing
to strengthen and broaden in its emission, while also developing
strong P Cygni absorption (Figure~\ref{fig:Ca2}).  By the latest
spectrum (corresponding to the mid-1850s), the Ca~{\sc ii} IR triplet
lines show pronounced blueshifted absorption, with a slow component at
around $-$150 to $-$200 km s$^{-1}$ as before, but also showing
absorption out to $-$650 km s$^{-1}$ (Figure~\ref{fig:Ca2}).

Unfortunately, the [Ca~{\sc ii}] $\lambda\lambda$7291,7324 doublet was
not included in the wavelength range of our higher resolution 1200 lpm
spectra with IMACS.  We did trace these [Ca~{\sc ii}] lines with MagE
at moderate resolution at early epochs.  Although noisy, the lines
have a resolved width of about 500 km/s, similar to the Ca~{\sc ii} IR
triplet.  On average, the two lines have roughly equal intensity
within the signal to noise.  The spectra do not have sufficient signal
to noise to determine if the [Ca~{\sc ii}] $\lambda\lambda$7291,7324
doublet shows the interesting slanted asymmetric profile shapes that
are seen in later stages of UGC~2773-OT \citep{smith16}.

The Na~{\sc i} D doublet provides a sensitive tracer of absorption
along the line of sight.  Its evolution is less clear, however,
because our echo spectra generally have lower signal to noise at
shorter wavelengths. At early epochs in the EC1 echo of the 1843 peak
and even in early spectra of EC2, the Na~{\sc i} line is only weakly
in absorption and narrow if present at all in the echo light (Galactic
absorption and residuals from subtraction of sky emission make it
unclear).  At later epochs, however, the Na~{\sc i} feature becomes
much stronger, showing a P Cygni profile with saturated absorption out
to $-$650 km s$^{-1}$, and a moderately strong broad emission
component (Figure~\ref{fig:Ca2}).  This can also be seen clearly in
the 2-D spectrum in Figure~\ref{fig:extract1200}.  (Note that Na~{\sc
  i} D is a doublet; Figure~\ref{fig:Ca2} is plotted as a function of
velocity for the $\lambda$5889 line, and so some of that line's
emission near zero velocity is absorbed by the P Cygni trough of the
$\lambda$5896 line.  The difference in velocity between the two lines
is about 350 km s$^{-1}$.)  In any case, this shows clear evidence for
a dense outflow of $\sim$650 km s$^{-1}$ along the line of sight in
the equator.  This is interesting, since material in the pinched waist
of the equator of the Homunculus is moving much more slowly as seen
today \citep{smith06}. Some features in the Outer Ejecta in the
equator are moving faster along the path seen by EC2. Perhaps the
Na~{\sc i} D absorption is tracing this outer material in the
so-called ``S Condensation'', or perhaps this fast material seen in
the mid-1850s has yet to be decelerated.

Figure~\ref{fig:Ca2halpha} shows a comparison between the line
profiles of H$\alpha$, Ca~{\sc ii} $\lambda$8498, and Na~{\sc i}
D. The three lines have been scaled arbitrarily to compare the line
shape (i.e. they are not normalized the same way).  The emission
components on the red side of the line are basically the same for
H$\alpha$ and Ca~{\sc ii} (and also Na~{\sc i} where it is not
absorbed by the other line in the doublet). The blue side of each line
differs markedly.  Whereas Na~{\sc i} D shows basically saturated
absorption all the way out to $-$650 km s$^{-1}$, H$\alpha$ shows
emission out to the same velocity and only shows a relatively narrow
absorption notch at $-$150 km s$^{-1}$.  Ca~{\sc ii} is in between
with absorption at $-$650 and $-$150 km s$^{-1}$ that is filled-in
with extra emission between those two components.  This comparison
points to simultaneous multiple velocities along the same line of
sight near the equator.  Strong absorption at $-$150 km s$^{-1}$ is
present in all lines, perhaps suggesting that it originates at the
outermost radii.  Altogether, this adds weight to a view where fast
material is expanding and crashing into slower material at larger
radii along the same direction.  We will return to such implications
in the discussion below.

Finally, the echo spectra show a number of weaker lines in the
spectrum that exhibit increasing emission strength and P Cygni
profiles at later epochs (Figure~\ref{fig:spec}).  Many of these are
Fe~{\sc ii} lines typical of warm supergiant LBV winds. One line that
is fairly bright and exhibits this typical behavior is at $\sim$6500
\AA, which can be seen clearly in the 2D spectra in
Figures~\ref{fig:extract300}, \ref{fig:extract1200}, and
\ref{fig:extractHa}.  This line at 6500 \AA \ is also seen in the
present-day spectrum of $\eta$ Car's wind \citep{hillier01}, although
interestingly, it is weak and has no secure identification.
\citet{hillier01} suggest that it may be an Fe~{\sc ii} line, and we
proceed with the same assumption here.  Figure~\ref{fig:fe2} shows the
time evolution of this line, which changes from a very weak line (if
present at all) in the EC1 echo, to having a clear and strengthening
P~Cygni profile.  The absorption trough indicates outflow speeds
around $-$250 km s$^{-1}$, which is in between the slow absorption at
$-$150 km s$^{-1}$ seen in H$\alpha$ and the Ca~{\sc ii} IR triplet,
and the faster absorption at $-$650 km s$^{-1}$ in the Ca~{\sc ii}
triplet and Na~{\sc i}.  This is the same range of blueshifted
velocities where H$\alpha$ and Ca~{\sc ii} are in emission, but
Na~{\sc i} is fully in absorption.  Interestingly, this line appears
to be absent or weak in all epochs of spectra for UGC~2773-OT
(Figure~\ref{fig:fe2}).  While the line is present but weak from
$\eta$~Car in the central star's spectrum today, it is much stronger
in the spectrum of HDE~316285, which is otherwise similar to the
spectrum of $\eta$ Car \citep{hillier01}.  Without a secure ID for the
carrier of this line, it is difficult to draw significance from this
intriguing trend.

\begin{figure*}
\includegraphics[width=5.6in]{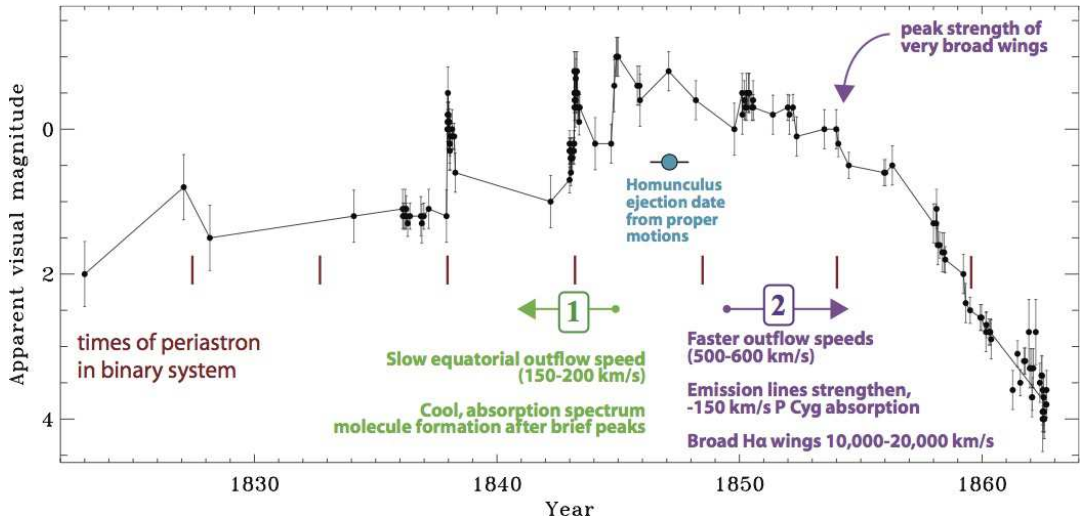}
\caption{The historical light curve of $\eta$ Car's Great Eruption
  \citep{sf11} with annotations summarizing some key points gleaned
  from light echo spectroscopy.  The red hash marks show expected
  times of periastron in the eccentric binary system, extrapolating
  from the orbital cycle observed in modern times.  The recently
  derived apparent ejection date for the Homunculus \citep{smith17}
  derived from proper motions of the nebula (assuming linear motion)
  is noted.  The main developments of the two-stage eruption revealed
  by light echo spectroscopy (somewhat simplified) are noted in green
  (1) and purple (2), corresponding to echoes EC1 and EC2,
  respectively.}
\label{fig:etaLC}
\end{figure*}

\section{DISCUSSION}

\subsection{Overview}

A key result from the observed spectral evolution of the EC2 echo
combined with $\eta$~Carinae's other light echoes is that the basic
character of the spectrum changes considerably during the decade-long
19th century eruption.  These changes are indicative of a line of
sight that views $\eta$ Carinae's Great Eruption from a vantage point
at low latitudes near the equator, and might not be indicative of all
viewing angles.

While there are a number of complicated changes that occur (including
rapid changes during fading from bright peaks), the underlying
evolution is gradual and can be summarized as basically a 2 stage
event: 1) A preperatory wind phase in the lead up to the Great
Eruption during the early 1840s (and perhaps for decades before that),
which may be strongly influenced by binary interaction, and 2)
an explosive event with fast outflow speeds and sustained high
luminosity. These two stages are annotated in Figure~\ref{fig:etaLC}.

{\it Stage 1 (1845 and preceding decade or two):} This initial phase
has slower outflow velocity (150$-$200 km s$^{-1}$) in the equator,
and the overall appearance of the spectrum is dominated by narrow
absorption lines with little or no emission at luminosity peaks,
except perhaps narrow H$\alpha$ wind emission that may be contaminted
by unresolved nebular emission \citep{rest12}.  It has somewhat cooler
apparent temperatures of 5000$-$5500~K near peaks \citep{rest12}, and
cooler temperatures of $\sim$4,000 K and evidence of molecule
formation in the fading after peaks \citep{prieto14}, with increasing
emission-line strength as it fades.  From this viewing direction, the
bulk of outflowing material (traced by the absorption trough) is
moving at about 200 km s$^{-1}$ at times that likely correspond to the
1830s to the early 1840s (although there may be some faster material
at lower density indicated by the absorption line wings extending out
to several hundred km s$^{-1}$).  There may be faster material at
other latitudes as well.

{\it Stage 2 (late 1840s - 1850s plateau):} While the overall
continuum shape is similar to earlier phases (a slight increase in the
apparent temperature to 6000~K), there are distinct changes in line
properties that trace the outflowing material.  The 1850s plateau as
viewed in EC2 spectra developes broader line widths and increasing
emission strength in the narrow components indicating an increase in
the bulk outflow speed to 600 km s$^{-1}$, which is much faster than
the 50$-$200 km s$^{-1}$ speed of the Homunculus at low latitudes near
the equator \citep{smith06}.  This time period also shows much
stronger emission lines in general, a decrease of line blanketing
absorption strength at shorter wavelengths, P Cygni profiles instead
of pure emission in many lines, and signs of higher excitation
(including possible He~{\sc i} emission).  Most remarkably, epochs
corresponding to the mid-1850s show the appearance and strengthening
of very broad emission wings from $-$10,000 to $+$20,000 km s$^{-1}$.
Narrow absorption components at $-$150 km s$^{-1}$ persist from Stage
1, suggesting that this is slower, previously ejected material along
the line of sight at a somewhat larger radius.  {\it A key point is
  that at least three very different expansion speeds are seen
  simultaneously in Stage 2.}

A major implication for the nature of $\eta$ Car's eruption is that
the observed changes show that a steady state wind is clearly {\it
  not} a good approximation.  The bulk outflow speed increases
dramatically over a time period of a few years, and the fastest
material in Stage 2 is two orders of magnitude faster than in Stage 1
(and far exceeds the escape speed).  Line strengths increase while the
apparent continuum temperature and luminosity stay relatively
constant; this probably signifies an increase in density and strong
departures from LTE that may be indicative of shock excitation.

The increasing velocity along the same line of sight also has
important physical implications.  It requires that fast ejecta follows
after much slower material in the same direction, making it inevitable
that fast material will catch slow material and will collide in a
strong shock.  The fact that the outflow speed changes with time
(increasing from slow to fast) is therefore direct evidence supporting
earlier claims, based on multiple lines of circumstantial evidence, of
a strong CSM interaction component that helps power the visible
luminosity of the event \citep{smith13,smith08,smith03}.

Another interesting outcome, noted in Figure~\ref{fig:etaLC}, is that
the ejection date of the Homunculus from its measured kinematic age is
solidly in between Stage 1 and Stage 2.  From a recent study of
available archival {\it HST} imaging over more than a decade, the
proper motion expansion of the Homunculus gives a fairly precise date
of origin for the Homunculus (extrapolating from linear motion
observed today) of 1847.1 $\pm$0.8 yr \citep{smith17}.  Of course, if
there was strong CSM interaction that accelerated slow pre-shock
material and decelerated the fast ejecta, as in a Type~IIn supernova,
the true ejection date might be slightly different (most likely a
short time after this).  Alternatively, if the material was ejected
over a longer period (for example, a more gradually increasing ouflow
speed over many years as opposed to an instantaneous pulse), then this
is a mass-weighted average ejection date.  In any case, it is
remarkable that the Homunculus date of origin lies in between the slow
material and fast material seen in our echo spectra.  This strongly
supports a picture wherein fast ejecta swept up and shocked slower
material, making a radiative shock that cooled rapidly to form the
thin walls of the Homunculus \citep{smith13}.  We consider the
physical interpretation of the 2 Stage event in more detail below.

\subsection{Comparison with UGC~2773-OT}

The relatively nearby LBV-like transient in the dwarf galaxy UGC~2773,
named UGC~2773-OT \citep{smith10,foley11,smith16}, has been compared
with $\eta$ Car's Great Eruption before.  \citet{rest12} showed that
early spectra of UGC~2773-OT at peak luminosity (soon after discovery)
were quite similar to light echo spectra of $\eta$ Car that correspond
to early peaks in the Great Eruption light curve (EC1).
\citet{smith16} showed that after several years had passed,
UGC~2773-OT faded very slowly, sustaining a high luminosity for a
decade, very similar to the slow light curve evolution of $\eta$~Car
during its 1850s plateau.

\citet{smith16} presented a series of spectra of UGC~2773-OT that
document its spectral evolution over several years.  Light echo
spectroscopy of EC2 spanning several years now allows us to extend the
comparison.  The evolution of EC2 spectra shows remarkable similarity
to the spectral evolution of UGC~2773-OT, further supporting the case
that they are close analogs.  As seen in Figures~\ref{fig:spec} and
\ref{fig:halpha} (a more densely sampled series of UGC~2773-OT spectra
can be seen in \citealt{smith16}), the evolution of the overall
low-resolution spectrum and of velocities and excitation is very
similar between the two objects. They have similar continuum
temperatures, both increasing a small amount as they evolve.  They
show many of the same lines, which show mostly narrow absorption at
early times, transitioning into stronger emission at later times.
Both show narrow emission from [Ca~{\sc ii}]
$\lambda\lambda$7291,7324, which is seen in a subset of SN impostors.
Similarities in H$\alpha$ are particularly remarkable: In both $\eta$
Car and UGC~2773-OT, the H$\alpha$ emission line profile starts out as
a weak and narrow P Cygni profile but then gets stronger and broader
(from around 100-200 km s$^{-1}$ initially up to 600-1000 km s$^{-1}$
at later times), with a very similar asymmetric emission line profile
(Figure~\ref{fig:halpha}).

The spectral similarity is interesting because these two also have
similar light curves.  If they are close analogs, perhaps UGC~2773-OT
can help us fill-in gaps in our knowledge due to limitations of light
echo spectra.  The most significant limitations of the echo spectra
are relatively low signal to noise because they are faint, a lack of
access to other wavelengths, and contamination from narrow nebular
line emission from the Carina Nebula.  Data for UGC~2773-OT do not
present the same limitations.

For example, from light echo spectoscopy alone, it is ambiguous if the
residual narrow He~{\sc i} emission (Figure~\ref{fig:extractHa}) comes
from $\eta$ Car itself, or if it is narrow nebular He~{\sc i} emission
arising on the globule's surface that is exposed to radiation from
O-type stars in the Carina Nebula.  The He~{\sc i} $\lambda$6678 line
shows an asymmetric and possibly P Cygni profile, but it is narrow and
weak, so this profile could potentially arise from background
subtraction.  It is therefore extremely interesting that UGC~2773-OT
shows very similar narrow He~{\sc i} emission, which is absent at
first and then grows with time while the eruption has almost constant
luminosity and temperature (it is not H~{\sc ii} region
contamination).  In UGC~2773-OT, the He~{\sc i} emission is also much
narrower than the H$\alpha$ line, qualitatively very similar to $\eta$
Car. In the case of UGC~2773-OT, the narrow He~{\sc i} must be
intrinsic to the object, arising from slow pre-shock gas that is
photoionized by a shock (it must be a shock, since radiative
excitation from a 6000 K photosphere would not produce strong He~{\sc
  i} emission).  This gives a possible indication that the narrow
He~{\sc i} emission in $\eta$ Car arises in a similar fashion.

The very strong contamination from H$\alpha$ in the Carina Nebula
(intrinsically narrow and unresolved in our spectra) makes it
impossible to say anything conclusive about narrow H$\alpha$ emission
in echoes, and hence about narrow H$\alpha$ emission from pre-shock
gas in the eruption.  UGC~2773-OT does not have the same ambiguity,
and narrow emission is present.  Some of this arises from a
surrounding H~{\sc ii} region, but high-resolution echelle spectra
show resolved widths of $\sim$50 km s$^{-1}$, indicating that
expanding CSM is partly responsible for this narrow emission
\citep{smith10}.  Perhaps UGC~2773-OT hosts extended expanding
nebulosity, similar to the Outer Ejecta of $\eta$ Car
\citep{kiminki16,mehner16}.

The light echoes of $\eta$~Car are faint and require a significant
allocation of time on large telescopes in the southern hemisphere; for
practical reasons, this has limited our cadence to roughly 1--2
observations per year. Moreover, different echoes trace different
epochs in the Great Erupton, adding uncertainty to the exact time
evolution.  The time sampling of spectra for UGC~2773-OT is better and
more clearly understood.  UGC~2773-OT exhibits a quite gradual
transition over a few years from Stage 1 to Stage 2.  This is an
important clue; the changes in the spectrum (notably the presence of
broader and stronger emission lines) was not a sudden change on a
dynamical timescale, but rather, the transition happened gradually.
Either the star is changing slowly, or optical depth effects govern
the slow emergence of radiation from faster material deeper in the
expanding envelope.

A limitation of the light echo spectra is that they become very noisy
in the blue part of the spectrum, due to a combination of ISM
reddening and detector/grating efficiency.  This makes it difficult to
study the spectrum at blue wavelengths, while the UV range of the
spectrum is not available to us.  The extreme faintness of the echoes
combined with bright background emission from the surrounding
star-forming region make it difficult to study the echo spectra in the
IR.

For all these similarites, though, UGC~2773-OT is not an identical
twin of $\eta$ Car's eruption.  A few interesting differences between
the spectral evolution of these two objects are:

1) UGC~2773-OT has no brief spikes in its light curve, which in $\eta$
Car have been attributed to periaston interactions as noted above.
Evidently UGC~2773-OT did not experience these sorts of interactions
with a wide companion, but suffered a similar decade-long outburst
anyway.  This provides another indirect suggestion that interactions
with this wide companion were not critical in powering $\eta$~Car's
event.

2) UGC~2773-OT has no 6500~\AA \ (presumably Fe~{\sc ii}) line
exhibiting a P~Cygni profile that develops at late times
(Figure~\ref{fig:fe2}).  The significance of this difference is
unclear, since UGC~2773-OT does show other Fe~{\sc ii} lines with
similar P Cyg profiles.  Since this line is absent at some epochs for
$\eta$ Car, but then appears at epochs when the broad emission is
seen, this line deserves a closer look.

3) The Ca~{\sc ii} IR triplet lines grow in strength at late times
much more than seen in $\eta$~Car's echo spectra.  This is probably an
optical depth effect at late times, and it will be interesting to see
how the echo spectra continue to develop.

4) UGC~2773-OT does not show the absurdly broad emission wings of
H$\alpha$ that are seen in $\eta$ Car.  We do not yet know if this is
a viewing angle effect, a timing issue (the broad lines appear late in
$\eta$ Car's spectra, and begin to fade after 1-2 yr), an optical
depth effect, or a fundamental difference in shock ejection in the two
events.  Examining echoes that view $\eta$ Car from other latitudes
will help clarify any angle dependence of the fast ejecta.

\subsection{Transitioning between two stages of the eruption: Winds
  vs. Explosions}

Continued monitoring of the spectral evolution of $\eta$~Car's light
echoes has demonstrated clearly that observed spectra from later in
the eruption show fundamental differences compared to earlier spectra.
Among the major differences are a faster bulk outflow speed, the
appearance of extremely fast ejecta, stronger emission lines, and
weaker absorption.  This transition occurs well after the eruption was
already underway, and coincides roughly with the ejection date
imprinted on the expanding nebula as measured from its proper motion
expansion (Fig.~\ref{fig:etaLC}).  This points to a dramatic change in
the physical state of the star's envelope after the peak of the
eruption in 1845.  A critical question for interpreting $\eta$ Car's
eruption is what caused this transition from Stage 1 to Stage 2.
Before examining that, we first consider the basic transition of
outflow properties in the two stages as deduced from available light
echo spectra.

The simple fact that a major transition in physical state occured
mid-way through the eruption is an important physical clue.  This
change clearly indicates that a single physical mechanism does not
govern the mass loss throughout the whole eruption.  For example, the
traditional picture of the star increasing its luminosity above the
Eddington limit and driving a strong wind cannot adequately explain
both the precursor luminosity spikes (in 1838 and 1843) and also the
long 1850s plateau, since these two phases clearly have fundamentally
different outflow properties at roughly the same luminosity.
Similarly, the observed transition occurring after a time when the
eruption was already underway indicates that the origin of the
eruption was not as simple as a single, instantaneous deposition of
energy that blasted off the star's envelope on a dynamical timescale
in 1847.  Instead, there was a long preparation phase (years to
decades) leading up to the peak of the eruption, perhaps due to a
building instability inside the envelope, or perhaps due to increasing
intensity of binary interaction as the orbital parameters changed
before a merger (see below).  We must therefore seek a physical
explanation for the eruption that naturally accounts for both of these
observed phases and the transition from one physical regime to the
next in the correct order.

Regardless of the underlying physical trigger of the eruption, the
plain fact that slow outflow velocities observed at earlier epochs
were followed by faster outflow velocities at later epochs along the
same direction (i.e., echoes probing essentially the same line of
sight) necessarily {\it requires that CSM interaction play an
  important role in the event}.  Fast material must overtake the
slower previously ejected material and shock.  In doing so, some
kinetic energy of the fast material is thermalized and converted to
luminosity.

\subsubsection{Stage 2 as an explosion}

The observed spectra in Stage 2 show at least three different outflow
speeds simultaneously (slow $\sim$200 km s$^{-1}$; intermediate
500-1000 km s$^{-1}$; and very fast 10,000-20,000 km s$^{-1}$).  This
fact is not easily explained by a steady wind.  It is, however, a
commonly observed trait of standard SNe~IIn powered by an explosion
crashing into dense CSM.  Thus, the two-stage empirical description of
the eruption from light echoes outlined above is remarkably compatible
with the CSM interaction scenario proposed earlier by
\citet{smith13}. This model envisioned Stage 1 as a relatively slow
(200 km s$^{-1}$) super Eddington wind, which is quite similar to the
value observed in light echoes (note that the absorption speed in
light echoes is seen from the equator; outflow speeds are probably
higher at other latitudes, and some of this could cause the more
extended absorption wings).  This was followed in Stage 2 by an
explosive energy injection of roughly 10$^{50}$ erg, and
\citet{smith13} showed that the ensuing CSM interaction luminosity
could in principle account for the 1850s plateau in the historical
light curve of the Great Eruption.  The changes seen in light echo
spectra therefore provide direct confirmation of a CSM interaction
model like that of \citet{smith13}.

As noted by \citet{smith13}, these numbers are somewhat malleable,
though, and can be adjusted depending on the desired level of
complexity.  The Homunculus nebula of course dictates that there must
be a range of speeds and densities at various latitudes
\citep{smith06}, while the \citet{smith13} model was a simple 1-D
estimate.  Even in 1-D, one can adjust the time dependence of physical
parameters to achieve a similar end result that fits the light curve.
In particular, different choices for the relative amount of mass and
speeds of the outflows in Stage 1 and Stage 2 can lead to similar CSM
interaction luminosities.  Light echoes combined with observations of
the present-day nebulosity help constrain possible values. For
example, light echoes provide strong evidence that the outflow speed
in Stage 1 was indeed roughly 150 km s$^{-1}$.  A signifficant
difference compared to the values adopted by \citet{smith13}, however,
is that the speed of the fast material appears to be even higher than
assumed.  In a CSM interaction scenario, the dominant observed outflow
speed of $\sim$600 km s$^{-1}$ arises from the cold dense shell, where
fast ejecta and shocked CSM pile up in a thin cooled layer.  This also
corresponds well to the final coasting velocity of the Homunculus
nebula \citep{smith06}.  The very high speeds seen in light echoes
give more freedom in this type of model, since we don't know {\it a
  priori} what fraction of the Stage 2 mass loss is contained in the
fastest outflowing material.  If Stage 2 is characterized by outflow
speeds of 10,000 - 20,000 km s$^{-1}$, then it is easy to have a
situation where most of the mass is ejected at slow speeds in Stage 1,
while most of the kinetic energy and momentum is supplied by a fast
wind with much lower mass-loss rate.

The maximum mass in the fastest ejecta can be derived by assuming that
most of the mass is supplied in the slow Stage 1 wind, whereas the
fast ejecta in Stage 2 provide essentially all the kinetic energy that
powered the event.  Under this limiting assumption, the total mass
contained in the fastest ejecta can be expressed as

\begin{displaymath}
M_{fast} \le 0.1 M_{\odot} \times \frac{E_{50}}{V_4^2}
\end{displaymath}

\noindent where $E_{50}$ is the total energy of the event in units of
10$^{50}$ erg, and $V_4$ is the speed of the fast ejecta or wind in
units of 10$^4$ km s$^{-1}$.  This is similar to but slightly smaller
than the mass of the extremely fast Outer Ejecta estimated by
\citet{smith08}.  Thus, the fast material seen in the broad wings in
light echoes must represent a small fraction of the total mass budget
of the Great Eruption, unless the total energy of the event is much
higher than generally believed.  The radiated energy inferred from the
historical light curve (with zero bolometric correction) is only about
2$\times$10$^{49}$ erg \citep{smith+11}, the kinetic energy of the
Homunculus is almost 10$^{50}$ erg \citep{smith03}, and the fast Outer
Ejecta \citep{smith08} are thought to contain a similar amount of
kinetic energy as the Homunculus.  Significantly increasing the total
energy would require either hotter temperatures (and hence, a larger
bolometric correction) during the event, which seems incompatible with
the relatively cool apparent temperatures seen in light echoes
\citep{rest12,prieto14}, or instead, a much larger amount of invisible
mass in the fast ejecta outside the Homunculus.

There must be some additional mass ejected in Stage 2 at lower speeds,
however, in order to account for the final momentum of the Homunculus.
For example, with 10 $M_{\odot}$ ejected in the slow Stage 1 wind at
150 km s$^{-1}$ and only 0.1 M$_{\odot}$ ejected in Stage 2 at 10$^4$
km s$^{-1}$, the final coasting speed of the Homunculus would only be
$\sim$250 km s$^{-1}$.  In any case, the significant changes seen in
light echoes are highly constraining for any model of the event.
While Stage 1 can be explained quite well with existing models of a
quasi-steady, continuum-driven super-Eddington wind
\citep{owocki04,vanmarle08,vanmarle09,os16}, the transition to Stage 2
requires time-dependent energy input well beyond this.

\subsubsection{Stage 2 as a time-dependent wind}

Rather than a slow wind followed by a single hydrodynamic explosion,
as in a Type IIn supernova, the observed properties of the Great
Eruption might be accounted for with a more complicated,
time-dependent wind with slow outflow transitioning to a faster wind.
This has been predicted in models that include super-Eddington energy
deposition below the surface of a massive star \citep{quataert16}.  In
this sort of model, the observed differences between an explosion and
a wind become less obvious.  Much of the radiated energy arises from
internal shocks in the wind, and the photosphere can reside in the
compressed post-shock zone itself \citep{quataert16,owocki17}, which
is qualitatively similar to the prediction of the explosion plus CSM
interaction model.  If one envisions a fast wind rather than a
hydrodynamic explosion, then light echoes require that the fast wind
must be able to achieve extremely high speeds of 10$^{4}$ km s$^{-1}$
and a mass-loss rate (spread over the 3-4 year duration of the broad
wings seen in light echo spectra) of $\dot{M} \simeq 3 \times 10^{-2}$
M$_{\odot}$ yr$^{-1}$.  The corresponding mechanical luminosity of
such a wind is at least 2$\times$10$^8$ L$_{\odot}$, or about
$\Gamma$=40.

\subsubsection{Grey area}

The central question of whether Stage 2 is better described as a wind
or an explosion ventures into muddy waters.  Limiting cases of these
two are well defined.  An explosion will result if energy deposition
occurs faster than the dynamical timescale with a total energy well
exceeding the gravitational binding energy of layers above.  A strong
wind will result when energy is carried efficiently through the star's
envelope by convection, and photon diffusion at the surface can power
a steady radiatively driven wind.  

Observational estimates, however, clearly place the Great Eruption of
$\eta$~Car precariously between these two extremes. This is in an
interesting regime similar to that discussed by \citet{rm17}, where
the energy that eventually emerges from the star's surface as kinetic
energy or radiation must be transported through the envelope by
acoustic waves that steepen to shocks, and may dissipate their energy
in the outer envelope \citep[see also][]{piro11}.  Light echoes paint
a picture where $\eta$~Car was relatively stable at first, but then
underwent a transition past some critical point where the outflow
changed dramatically.  One can imagine a physical scenario where the
rate of energy deposition grows with time to exceed a critical limit,
or where there is a sudden change in the deposition rate or depth in
the envelope.

Regardless of specific mechanism, it is tempting to ascribe the two
stages to (1) an early phase that is trans-Eddington, where radiative
damping or weak shock dissipation is sufficient to inhibit strong
shock formation, depositing energy in the outer envelope at a rate
that can can be carried away by radiation, thus driving a strong
super-Eddington wind, and (2) later phases that exceed the critical
wave luminosity, where radiative damping and shock dissipation are no
longer able to suppress strong shock formation, and shocks grow in
strength, removing mass from the surface of the star hydrodynamically.
Thus, $\eta$~Car is probably an object where {\it we have directly
  witnessed the transition from a quasi-steady wind to explosive mass
  loss}.  This motivates continued theoretical investigation of
time-dependent energy deposition in massive star envelopes, in order
to ultimately reconcile the observed physical parameters with the
central engine that caused the outburst.

The energy deposition required in this picture could in principle
arise from one of multiple possible physical causes, including:
inspiral of a companion during a stellar merger, runaway instability
in shell burning, wave driving, or the pulsational pair instability
\citep{qs12,sq14,sa14,smith+11,woosley17}.  A number of observational
facts (but perhaps most importantly the axisymmetry of the Homunculus
nebula and the decade-long duration of the ramping-up of luminosity in
Stage 1) suggest that a binary merger event is a plausible explanation
for the Great Eruption, although not all the others are necessarily
implausible.  A general model for such a merger with CSM interaction
is discussed in Section 4.6, followed in Section 4.7 by a discussion
of details pertaining specifically to $\eta$ Car.

\subsection{Origin of the Fast outflow?}

The biggest surprise in our study of $\eta$~Car's light echoes has
been the discovery of extremely fast ejecta indicated by the broad
H$\alpha$ line wings extending from $-$10,000 to $+$20,000 km
s$^{-1}$.  This is discussed more in a companion paper
\citep{smith+18}, so the reader is referred to that paper for
additional observational details.  So far, no model proposed for
$\eta$~Car or stellar mergers in general predicts such extreme outflow
velocities that produce a SN-like blast wave.

A speed of 20,000 km s$^{-1}$ is much faster than any escape speed
envisioned in the $\eta$ Car system; it is even 10 times faster than
the wind of a WR star.  Super-Eddington winds that drive
strong mass loss are generally expected to have relatively slow
outflow speeds comparable to the escape speed at large radii where the
wind originates \citep{owocki04,quataert16,vanmarle08,vanmarle09}.
Binary mass loss from L2 predicts relatively slow outflow speeds no
more than several hundred km s$^{-1}$, even at very high luminosities
\citep{pejcha16a}.  Bipolar jets driven by accretion onto a companion
\citep{ks09} would be expected to be no more than a few times the
surface escape speed of the accreting star, and one would not expect
to see a fast outflow from bipolar jets in the equator (especially if
those same bipolar jets are invoked to shape the Homunculus).
Clearly, the fast speeds place important fundamental constraints on
the nature of the event.

If the outflow traces a steady flow, the extremely high speed would
imply an outflow from a massive compact object, such as a jet from an
accreting neutron star or black hole. These are the only objects with
such high escape speeds.  The presence of such a companion in the
1850s might be reconciled with a lack of any such companion seen in
data at the present epoch if the compact object is the thing that
merged with a companion star in the Great Eruption, making the
present-day primary star a Thorne-$\dot{\rm Z}$ytkow object
(T$\dot{\rm Z}$O).  This would open a direction of inquiry far beyond
the scope of this paper, and would be a departure from current ideas
about $\eta$~Car --- but it is interesting to note that a T$\dot{\rm
  Z}$O might be consistent with reports of unusual lines seen in one
particular equatorial region in the nebula that shows very strong
emission from species such as Sr, Y, and Zr, plus Sc, Ti, V, Cr, Mn,
Fe, Co, Ni, etc \citep{hartman04,bautista06,bautista09}.  This is the
so-called ``Strontium Filament'' in the equatorial ejecta. A bipolar
jet from a compact object is not, however, a very satisfying
explanation for the origin of the fast material because of geometrical
reasons.  Namely, such a jet is expected to be highly collimated and
bipolar, as in the case of SS~433 \citep{paragia99}.  Yet, the light
echo in which the very fast material is seen views $\eta$ Car {\it
  from near the equator} of the Homunculus.  Such a jet could
therefore have had little impact on shaping the bipolar Homunculus
nebula (which has an orthogonal orientation), and we would need to
invoke some other explanation for the very fast ejecta with speeds of
$\sim$5,000 km s$^{-1}$ in the polar regions of the Outer Ejecta seen
at the present epoch \citep{smith08}.

Instead of relatively steady mechanisms including jets, the extremely
fast ejecta seen in light echoes of $\eta$~Car more naturally point to
a wide-angle explosive outflow driven by strong shock acceleration,
which is not necessarily expected to be close to any escape speed in a
system (providing that it exceeds the escape speed) because it is
determined mainly by the energy in the shock and the density gradient
where the shock acceleration occurs.  What could be the origin of such
a shock?  Energy deposition deep in the stellar envelope, by whatever
mechanism, will be transported outward by waves and will steepen to a
shock if the energy deposition rate exceeds the steady stellar
luminosity \citep{rm17}.  When such strong shocks exit the star, they
will accelerate a small amount of mass to very high speeds.  So then
the question is shifted to what the source of this energy deposition
would be.

One mechanism that can most likely be ruled out here is energy
deposition by wave driving from core convection in late evolutionary
phases \citep{qs12,sq14,quataert16,fuller17}.  The reason it doesn't
work in the particular case of $\eta$~Car is because of timescales.
While this is an efficient way to suddenly dump energy into the
stellar envelope, it is only expected to be significant in the latest
Ne and O core burning phases \citep{qs12}, which last just a couple
years. It has been 170 years since the Homunculus was ejected, and
$\eta$ Car has apparently not yet undergone core collapse.

A mechanism that is harder to rule out, and which may indeed be a
plausible explanation, is sudden energy deposition via the pulsational
pair instability (PPI).  This is explored more in Section 4.5.

Another possibility is that the Great Eruption of $\eta$ Car was a
stellar merger event, which is an attractive hypothesis for several
reasons, as noted earlier and discussed further in Section 4.6.  If
the lead-up to the eruption corresponds to the inspiral and L2 mass
loss phase, and the decades-long eruption is the common envelope
ejection phase with CSM interaction, then what specifically launches a
small fraction of the total mass to extremely high speeds while most
of the mass is ejected at only 600 km s$^{-1}$?  How does a stellar
merger eject material at speeds much faster than the escape velocity
of either star?  This is a central question for models of massive star
mergers that remain unanswered.

One speculative possibility is that the energy deposition arises from
unsteady nuclear burning as fresh fuel is mixed to deeper shell
burning layers, leading to an outburst \citep{sa14}.  Explosive common
envelope ejection was discussed in a very different scenario (merger
of a post-He core burning star with a low-mass companion) by
\citet{ppod10}, but perhaps something similar might happen when the
cores of the two stars merge, mixing unburned fuel into the core.
Perhaps such a mechanism could lead to explosive energy deposition
that travels outward through the star and steepens to a shock,
mimicking a scenario like a PPI eruption.  While it is still uncertain
how this would work, observations seem to require that some violent
process like this must be an ingredient of any merger model for
$\eta$~Car in order to explain the extremely fast ejecta seen in light
echoes.

An even more speculative origin for the fast ejecta involves an
external mechanism.  Namely, any binary merger model for the Great
Eruption of $\eta$ Car must involve a hierarchical triple system,
since a wide and eccentric binary remains today.  The current
companion on its eccentric orbit would have plunged into the bloated
star or common envelope during the event \citep{smith11}, and perhaps
that violent collision led to the ejection of a small amount of
material to high speeds in certain directions.  Whether this could
provide enough energy to power the fast ejecta seen in light echoes is
uncertain.  This is discussed more in our companion paper on the fast
ejecta \citep{smith+18}.

A related point has to do with the influence of the shock breaking out
of the star.  Broad emission lines in light echoes reveal that a small
amount of material appears to have been accelerated to very high
speeds by a shock.  If a strong shock with $\sim$10$^{50}$ ergs breaks
out of the surface of the star or the outer boundary of the common
envelope, it coud be accompanied by a UV flash from the shock breakout
itself.  It would be interesting to search for observational evidence
of this in future data, either by high-ionization nebular lines in
echo spectra, or in blue light curves of echoes --- especially in
echoes that view the eruption from polar directions that do not need
to peer through dense equatorial CSM.  One could also search for
signals of a UV flash from shock breakout in extragralactic SN
impostors.

Note, however, that the broad H$\alpha$ wings that are seen in echo
spectra probably do not arise from direct emission by the fast ejecta
immediately after ejection, because in that case the high velocities
would be seen for only an extremely brief window of time.  It is more
likely that the broad emission wings arise as the fastest freely
expanding ejecta approach the reverse shock in CSM interaction,
consistent with their more persistent appearance in Stage 2 of the
eruption.  This fast material will either be excited radiatively by
inward propagating X-rays from the shock front, or collisionally as it
crosses the reverse shock.  This is commonly observed in late phases
of CSM interaction in SNe IIn, and is the explanation for the very
broad H$\alpha$ wings still remaining in the spectrum of SN~1987A,
more than a decade after explosion \citep{smith05rs87a,heng06}.

\begin{figure*}
\includegraphics[width=3.3in]{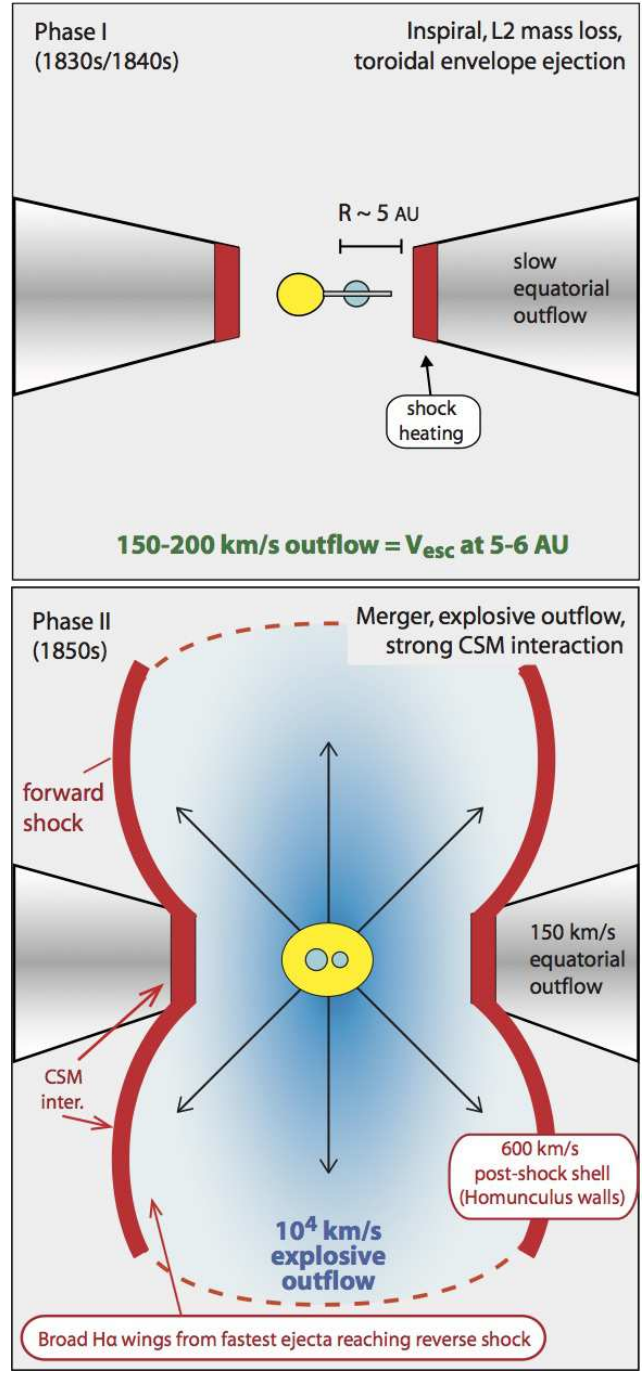}
\caption{A sketch of the possible geometry in a hypothetical stellar
  merger model for $\eta$ Car's eruption, showing the two phases
  discussed in Sections 4.3 and 4.5.  {\it Top:} Phase I corresponds
  to the decades leading up to the Great Eruption, which in a merger
  model is the inspiral phase when the orbit decays and there is
  prodigious mass loss from the system through L2.  This is adapted
  from the scenario for lower-mass mergers like V1309~Sco discussed by
  \citet{pejcha16a,pejcha16b,pejcha17}.  In the case of $\eta$ Car,
  light echoes from this period indicate a relatively slow outflow of
  150-200 km s$^{-1}$.  The luminosity in this phase is a combination
  of shock heating as the L2 outflow collides with itself in a ``death
  spiral'' \citep{pejcha17}, as well as stellar photospheric
  luminosity.  {\it Bottom:} Phase II corresponds to the 1850s plateau
  phase of the Great Eruption, when higher velocities are seen.  The
  very broad wings in H$\alpha$ correspond to a 10$^4$ km s$^{-1}$
  explosive ejection or very fast wind (light blue), which is excited
  as the fast ejecta approach the reverse shock.  The
  intermediate-width $\sim$600 km s$^{-1}$ lines (broader than the 150
  km s$^{-1}$ seen in Phase I) correspond to the thin swept up
  post-shock shell (red).  The post shock gas expanding at 600 km
  s$^{-1}$ will cool and will eventually form the dense walls of the
  Homunculus.  This is adapted from the scenario discussed previously
  by \citet{smith13}, where strong CSM interaction dominates the
  luminosity in this phase.  During Phase II, the slow 150 km s$^{-1}$
  outflow can still be seen in absorption from favorable directions,
  although it is eventually swept up and becomes part of the walls and
  pinched waist of the Homunculus.}
\label{fig:merger}
\end{figure*}

\subsection{Pulsational Pair Instability Eruption}

As noted in the previous section, one potential explanation for the
energy deposition required to power the fast ejecta (and for the
global energetics of the event) is a pulsational pair instability
(PPI) eruption.  Recently this has been discussed in detail --
including applying it to the specific case of $\eta$ Carinae -- by
\citet{woosley17}.  Very massive stars that approach the ends of their
lives with He core masses above about 30 $M_{\odot}$ will encounter
the pair instability during the latest nuclear burning phases
\citep{fh64,barkat67,rs67}.  The core then implodes and triggers
explosive burning, which may lead to a diverse range of outcomes.
When the explosive burning exceeds the core binding energy, a terminal
pair instability supernova (PISN) destroys the star
\citep{bond84,hw02}.

In the lower range of final He core masses (about 30-60 $M_{\odot}$),
the energy depositon from explosive burning may not be energetic
enough to completely unbind the star.  Instead, a repeating cycle
occurs where explosive burning expands the core, which cools, and
eventually contracts again to reignite explosive burning.  As a
result, a series of repeated pulsations occur that cause severe
eruptive mass loss before the star finally collapses to a black hole
\citep{hw02,woosley17}.  The mass ejected, energy, and recurrence
timescales for these eruptions are diverse.  A comprehensive overview
of PPI eruptions has recently been discussed by \citet{woosley17}.  We
will not repeat that discussion here, except to note some pros and
cons of PPI eruptions as an explanation for $\eta$~Car, specifically
informed by clues in our new light echo spectroscopy.

{\it Pro:} A key argument in favor of the PPI eruption mechanism to
explain $\eta$ Car's Great Eruption is that this is a well-established
fundamental mechanism that is predicted to occur in very massive stars
-- even single massive stars.  Indeed, $\eta$ Car is a very massive
star whose likely initial mass, based on its current luminosity, is in
the right ballpark to experience PPI eruptions \citep{woosley17}.
Moreover, the total mass lost \citep{smith03}, total kinetic plus
radiated energy of the Great Eruption \citep{smith03,smith06,smith08},
shock driven mass loss and the presence of CSM interaction
\citep{smith13}, and repeated eruptions over centuries
\citep{kiminki16}, fall nicely within the range of expectations for
PPI eruptions \citep{woosley17}.  The usual assumption that the PPI
and PISN are resticted to low metallicity (because line-driven winds
reduce the star's mass too much at $Z_{\odot}$) may not be such a
concern, since mass-loss rates adopted in most stellar evolution
models are too high anyway \citep{smith14}, and since other effects
such as rotational mixing or binary interaction can influence the
mapping of initial mass to final He core mass \citep{cw12}.
Particularly well-matched to evidence from light echo spectrosopy and
the nebulosity around $\eta$ Car is that the PPI eruptions may produce
strong shocks with fast explosive outflows, giving a natural
explanation for the broad wings we observe in light echo spectra (this
work) and fast Outer Ejecta \citep{smith08}, but there may also be a
wide range of outflow speeds.  Moreover, the recurring nature of the
PPI eruptions allows faster material to overtake previous eruptions
and power a transient with strong CSM interaction \citep{woosley07},
as inferred for $\eta$ Car \citep{smith13}, and seems consistent with
$\eta$ Car's eruptive history \citep{kiminki16}.

{\it Con:} One potential objection arises because the PPI is generally
expected, over most of the applicable initial mass range, to occur
only within a few years before the star's final collapse to a black
hole \citep{woosley07}.  Yet, it has been $\sim$170 yr since the Great
Eruption and $\eta$~Car doesn't seem to have collapsed to a black hole
yet; moreover, $\eta$ Car appears to have suffered previous major
eruptions 300-600 yr before the Great Eruption \citep{kiminki16}, so
its delay between pulsations would need to be quite long.  As noted by
\citet{woosley17}, however, the expected outcomes of the PPI are
diverse, and the delay between pulsations can in some cases be much
longer -- up to centuries or even millennia.  These longer delays
occur at the highest part of the initial mass range just below the
threshold for true PISNe, where the PPI flashes are not energetic
enough to fully unbind the massive core -- but they {\it almost} do
it, leading to very long recovery times as the expanded and cooled
core undergoes Kelvin-Helmholtz contraction. Here we run into a
potential snag, though, because applying these long PPI delays to
$\eta$ Car may be problematic in two ways.  (1) The longest delays
occur for a fairly narrow range of mass at the highest masses (He core
masses of roughly 60-64 $M_{\odot}$ for low metallicity models;
\citealt{woosley17}).  This would make $\eta$ Car's eruption an
extremely rare event, which is in itself perhaps not a debilitating
objection.  However, the requirement that this long delay occurs for
the highest part of the mass range exacerbates the difficulty
mentioned above with reaching this end at roughly $Z_{\odot}$.
Getting massive stars at $Z_{\odot}$ into the lower end of the PPI
range despite mass loss is difficult enough, but getting them to reach
their end with the most massive He cores seems unlikely.  (2) By
necessity, the PPI flashes that are {\it almost} energetic enough to
unbind the star, thereby achieving a long delay before the next pulse,
also produce explosive events with high energy exceeding 10$^{51}$ erg
\citep{woosley17}.  The kinetic plus radiative energy budget of $\eta$
Car's 19th century Great Eruption can accomodate about 10$^{50}$ erg
\citep{smith03}, but a total energy exceeding 10$^{51}$ erg seems
difficult to reconcile with observational estimates.

A different counterargument to the PPI for $\eta$ Carinae has to do
with the properties of its nebula and companion star.  First, the
Homunculus nebula is highly bipolar in shape, with a tightly pinched
waist.  The PPI doesn't give a clear explanation for this geometry.
Instead, this suggests either a strong influence of shaping by
interaction with a close companion star, or perhaps very rapid
rotation (rapid rotation late in life despite already having suffered
extreme mass loss may also require interaction with a companion).
Moreover, previous major mass-loss events had a different geometry
\citep{kiminki16}, which seems hard to reconcile with a single rapid
rotator.  Second, the PPI model gives no explanation for why $\eta$
Car's current companion star is so unusual, with a highly eccentric
wide orbit and having an extremely strong and fast wind.  As discussed
in the final section of this paper, these properties would seem to
require violent binary interaction of some sort.  (Although it adds
complexity, there is no clear reason to rule out a PPI event occuring
in a binary system.) However, it is worthwhile to ask if a binary or
multiple system interaction could account for the Great Eruption on
its own, even without a PPI event.

\subsection{A generic model for eruptive transients: Binary merger
  with CSM interaction}

The $\eta$ Car system, its surrounding nebula, its historical record,
and light echoes represent one of the most observationally rich
objects in massive star research.  Any model for $\eta$~Car must face
a daunting gauntlet of observational constraints.  This may act to
repel conservative theorists, or to prematurely quash potentially
interesting models.  To any proposed simple theory applied to
$\eta$~Car, one can usually respond with ``Yes, but what
about... [insert obscure observational detail here]?''  For now, we
momentarily stow such objections in order to discuss a general
scenario, and we will return to specific complexities of $\eta$~Car
later in Section 4.7.  The two stages of the event described above (in
Section 4.3) arose from an empirical description of the spectral
evolution and historical light curve, but we can also attempt to
ascribe a physical cause to these observed changes. Here we discuss a
promising physical model to account for the properties of $\eta$ Car's
Great Eruption: a massive star merger event.

A merger of two massive stars provides an attractive model for the
trigger and the energy supply of the Great Eruption, and has been
discussed before (see the Introduction).  However, the merger model
has never been reconciled with detailed observational constraints for
$\eta$~Car, and there have been significant unanswered problems with a
simple merger as proposed in previous models.  Here we describe how a
merger model may indeed be reconciled with many of the observed
properties of the Great Eruption. The model described below differs
from previously proposed merger models for $\eta$~Car in that it
adopts the hypothesis that CSM interaction is a main engine for
producing the observed radiative luminosity.  This provides a
self-consistent explanation for the two observed phases of the event
and their properties.  In a simplified scenario involving a binary
merger, these two phases may be understood basically as:

\textit{\textbf{Phase 1 (1840s and before):}} This is the inspiral
phase with mass transfer and/or common envelope, when mass and angular
momentum are shed from the L2 point, allowing the orbit to degrade and
for the binary to eventually merge.  This creates a relatively slow
outflowing disk or torus (Fig.~\ref{fig:merger}; top).  In this phase,
the mass shed in an equatorial spiral from L2 quickly catches and
shock heats previous L2 mass loss as the spiral winds up.  The shock
heating of this torus may make a considerable contribution to the
total luminosity, and gradually rises as the stars move closer
together. A such, the ``photosphere'' in this phase may be a composite
of the two stellar photospheres plus optically thick shock-heated
material in the circumbinary disk.  This inspiral phase and the
associated luminosity has been discussed in detail for lower-mass
stellar mergers \citep{pejcha14,pejcha16a,pejcha16b}.

\textit{\textbf{Phase 2 (1850s plateau):}} Relatively sudden energy
deposition from a final merger event (i.e. the merging of the two
stellar cores inside the common envelope) steepens to a strong shock
in the bloated envelope, and triggers an explosive outflow or very
strong fast wind.  As noted in previous sections, a large deposition
of energy deep inside the star's extended envelope is likely to
steepen to a strong shock \citep{rm17}, and we suppose that this is
the driving mechanism of the fastest ejecta in $\eta$~Car.  The fast
ejecta from this explosive outflow or fast wind then overtake and
shock the slower outflow from Phase 1 (Fig.~\ref{fig:merger}; bottom).
This sudden ejection will naturally lead to a sustained phase of high
luminosity that could last for years, because the fast ejecta take
time to expand through the previous mass loss as in a scaled-down
version of a Type~IIn supernova \citep{smith13}.  This is the plateau
phase of the Great Eruption, when $\eta$~Car was thought to have
exceeded its own classical Eddington limit for about a decade.  A key
point, however, is that in the model proposed here, the observed
``super-Eddington'' light comes largely from CSM interaction
\citep{smith13}, and not from stellar luminosity diffusing out through
a bound stellar atmosphere as in super-Eddington wind models
\citep{owocki04,owocki17,os16,quataert16}.\footnote{Note, however, that the
  super-Eddington model discussed by \citet{quataert16} is somewhere
  in between, where much of the emergent radiation has been processed
  by internal shocks as the fast super-Eddington wind overtakes slower
  outflowing material upstream, similar to that described here. An
  interesting direction for future work on this model would be to
  determine if a super-Eddington wind with such internal shocks could
  achieve the extremely fast speeds observed in our light echo
  spectra.}  A few key attributes make a simple model like this
plausible:

First, it is self consistent, in the sense that the initial inspiral
during Phase I must shed mass and angular momentum in order for the
orbit to degrade and for a merger to proceed, but in doing so, it also
naturally provides the CSM required for the interaction that will
occur in Phase II.  We do not need to invoke a different mechanism or
a previous outburst to provide the CSM into which the shock
expands. Lost from the L2 point, this Phase I mass loss should be
relatively slow and concentrated in the equator.  CSM interaction
where a fast outflow overtakes a slow and dense equatorially
concetrated outflow (i.e. a torus) may naturally lead to a bipolar
shape in the resulting nebula \citep{frank95,langer99}.

Second, CSM interaction is an extremely efficient engine for
converting outflow kinetic energy into radiation.  Whereas a
recombination plateau may typically tap only $\sim$1\% or less of the
available kinetic energy, CSM interaction is much more efficient,
typically converting $\sim$50\% or more of the available kinetic
energy into light.  The conversion is most efficient when a fast and
relatively low-mass explosion collides with slow and massive CSM.
Thus, for a given explosion or eruption energy, transients with CSM
interaction will be much more luminous than those where the radiation
is powered by a recombining H envelope.  This is a key way to get
super-luminous SNe II from normal core-collapse energy \citep{sm07},
and it is what converts kinetic energy to radiation in PPI eruptions
\citep{woosley17}.  Similarly, it will allow events with modest
kinetic energy to achieve substantial outburst luminosity.  In any
magnitude-limited sample of transients found in nearby galaxies, then,
the ones powered by CSM interaction are likely to make a dominant
contribution simply because they tend to be brighter than those
without interaction for the same explosion energy.

Third, this two-phase merger scenario is qualitatively consistent with
the time sequence of outflow velocities seen in light echo spectra of
$\eta$~Car, which initially show a slow velocity of 150-200 km
s$^{-1}$.  As time proceeds, extremely fast material appears, while
simultaneously the ``narrow'' emission component broadens from 200 to
about 600 km/s, as if it is being accelerated by a shock.  Some slower
velocities (150 km/s) are still seen in absorption along this line of
sight, even at later times when the broad emission appears, requiring
the simultaneous presence of slow moving CSM outside much faster
ejecta.  It would be difficult to avoid strong CSM interaction, based
on this observational evidence alone.  As noted by \citet{smith13},
the radiative shock that forms will lead to a very thin post-shock
layer as in models of SNe~IIn \citep{vanmarle10,chugai04}, which
explains many attributes of the structure of the Homunculus around
$\eta$ Car that are harder to explain with wind mass loss.


While the $\eta$~Car system has additional complexities that will be
discussed below, this basic sort of merger plus CSM interaction model
may be widely applicable to other non-terminal transients.
\citet{smith13} proposed that CSM interaction may be responsible for a
wide range of non-terminal transients besides $\eta$ Car, including
other LBV giant eruptions, SN impostors, SN~2008S-like events, pre-SN
eruptions, infrared transients, so-called luminous red novae, or other
events. All these objects have quite similar spectra at peak
resembling SNe~IIn, although they exhibit a wide range of outflow
speed and luminosity \citep{adams15,kochanek11,prieto09,smith+11}. A
similar model with CSM interaction was applied recently to luminous
red novae like V1309~Sco as well \citep{mp17}.

In this physical scenario, $\eta$~Car's Great Eruption can be seen as
one of the most energetic examples of stellar merger events that span
a wide range of initial masses for massive stars, extending to similar
transients that have been associated with mergers from low-mass stars.
These include clear mergers like the spectacular example of V1309~Sco
\citep{tylenda11,pejcha14}, and may include other suspected mergers
like V838 Mon \citep{bond03}. Dusty transients like SN~2008S, NGC
300-OT, and their kin have been discussed variously as terminal events
like electron capture SNe, eruptive transients, or possibly mergers as
well
\citep{prieto08,prieto+08,berger09,bond09,botticella09,thompson09,smith09,smith+11}.
\citet{kochanek14} estimated rates and argued that merger events from
low-mass binaries are common, with the rate falling toward higher
initial mass, and they noted that the peak luminosity is a steep
function of the initial mass.  \citet{smith16b} discussed ways that
$\eta$ Car and the recent transient in NGC~4490 seem to be an
extension of stellar merger transients to higher initial mass, and
noted that the duration of the bright transient may also depend on
initial mass.  \citet{smith16b} noted that these objects can show
quite similar spectra at various times in their evolution.
\citet{blagorodnova17} suggested that the recent transient in M101 may
be a massive star merger that fits with this scenario as well.

Although the common envelope phase and stellar mergers are still far
from being well understood, theoretical models of low-mass events are
able to reproduce some aspects of the observed transients, in most
cases with the mass ejection and luminosity powered by recombination
\citep{ivanova13,nandez14}.  For the case of $\eta$ Car, the high
speed outflow suggests strong shock excitation, the decade-long
plateau seems too long to be powered by recombination, and the thin
walls of the resulting nebula appear to have been compressed in a
radiative shock \citep{smith13}, so a merger model with CSM
interaction would seem more appropriate for $\eta$~Car, closer to the
models for V1309 Sco discussed by Pejcha and collaborators (see
above).  In fact, one might argue that a merger model where CSM
interaction does not constitute a significant fraction of the
luminosity could be problematic; the inspiral must eject mass and
angular momentum in order for a merger to occur, so an ejected
envelope will likely collide with it unless the final ejecta are very
slow.  Even in some low-mass meger events like V1309 Sco, some
evidence of relatively high speed outflow is seen \citep{mason10},
making CSM interaction seem likely (although the speeds are not as
extreme as seen in echoes of $\eta$ Car).


In the specific case of $\eta$~Car, a merger scenario as described
above is quantitatively plausible in terms of the energy budget of the
event and the resulting nebula.  \citet{smith13} already showed that
the luminous plateau phase of the Great Eruption (1845-1860) can be
powered by CSM interaction with a $\sim$10$^{50}$ erg explosion
running into previous mass loss.  This energy can easily be supplied
by a massive-star merger event.  The gravitational potential energy of
two massive stars (say $\sim$60 $M_{\odot}$ each) that are about to
merge (separated by roughly 100 $R_{\odot}$, which is roughly the
present-day radius of the primary star; \citealt{hillier01}) would be
about 1.4$\times$10$^{50}$ erg.  This is independent of the source of
energy for that explosive outflow (i.e. a PPI event works as well, as
noted above).  However, a merger scenario provides a natural
explanation for why the $\eta$~Car primary seen today, which would be
a merger product, appears to still be a very rapid rotator, rotating
at a substantial fraction of its expected critial rotation rate
\citep{do02,smith02,smith06,st07,smith03a}.  An explanation for the
high current rotation and for the bipolar nebula is harder to find in
any single-star model (including the PPI model), because the primary
has expanded significantly from its ZAMS radius and it has already
shed huge amounts of mass and angular momentum.

One difference indicated by our light echo spectroscopy, as compared
to the simple CSM interaction model in \citet{smith13}, is that the
post-1843 mass-loss rate may be lower and its bulk outflow speed is
higher.  Radiative transfer models of the light echo spectra are
needed for quantitative estimates of how much mass is contained in the
outflow responsible for the broad H$\alpha$ wings we observe, but with
more kinetic energy per unit mass in the fast outflow, this may
suggest that more of the mass in the Homunculus was provided by the
slow L2 mass loss in Phase 1, while most of its kinetic energy came
from the explosive event in Phase 2.  It will be interesting to see if
future numerical simulations of an explosive outflow expanding into a
compact torus similar to the one formed by L2 mass loss in lower-mass
models \citep{pejcha16a,pejcha16b} can explain observed structures in
the nebula of $\eta$ Car, such as the apparent ``holes'' in the polar
caps \citep{smith06,smith03}.  \citet{gonzalez18} recently simulated
the shaping of the Homunculus via an explosion with shock interaction
following the scenario proposed by \citet{smith13}, finding that it
could indeed explain structural features of the Homunculus.  With the
adopted parameters, the Homunculus simulated by \citet{gonzalez18} was
expanding too quickly, and so that author favored a super-Eddington
wind scenario.  The adapted pre-explosion mass loss and the explosion
parameters are not tightly constrained, however, and a different ratio
of mass-loss rates in Stage 1 and Stage 2 can change the resulting
final speed of the Homunculus.  The fast ejecta speeds reported here,
indeed, alter the expected explosion parameters one might adopt and
point directly to an explosive outflow.

Overall, this simplified two-phase merger scenario therefore gives a
somewhat satisfactory (although incomplete) explanation for the
evolution of observed light echo spectra of the Great Eruption, basic
energetics of the event, and basic structural properties and
kinematics of the Homunculus seen today.  It does not, however,
resolve questions of greater complexities in the presently observed
system, including the unusual surviving companion star, a history of
previous outbursts, and some of the more detailed structures in the
nebula.  These are discussed below.

\subsection{Can we reconcile this generic binary merger picture with
  the complexity of Eta Car and its wide companion?}

In a standard picture of a stellar merger event, guided by spectacular
recent examples such as V1309~Sco, one envisions a relatively long
(many orbits) inspiral phase with L2 mass loss shedding both mass and
angular mementum, followed by a relatively brief plunge-in (common
envelope) phase that may give rise to a sudden but one-off transient
event.  The end product should be a (potentially dust-obscured) rapid
rotator surrounded by a toroidal nebula.  As such, this type of
scenario provides a plausible origin for the unusual properties of
LBVs and B[e] supergiants \citep{justham14,ppod10}, both of which have
blue-straggler environments that are more isolated from O-type stars
than expected in single-star evolution \citep{st15}.  Mergers may also
help account for LBVs as immediate SN progenitors \citep{justham14}.

In some basic respects, $\eta$~Car seems to fit this general two-phase
merger picture, as noted above, and there are attractive aspects of a
merger model for $\eta$ Car as noted previously by several authors
\citep{jsg89,iben99,ppod10,pz16,justham14,smith16}.  It is the most
luminous star in its birth cluster Trumper 16 \citep{smith06b},
consistent with being a blue straggler (even though the lifetimes are
all similarly short for such massive stars), and there are clues that
it is currently a rapid rotator \citep{smith02,smith03wind,do02}.

However, $\eta$~Car also has a lot of observational ``baggage''.
There are several apparent contradictions between simplified merger
models and the vast amount of observational data for $\eta$ Car, which
requires that the situation is more complicated if indeed a stellar
merger is the correct physical description.  To reconcile a merger
model with the specific observed case of $\eta$ Car, the most critical
problems posed by observations are that: (1) the system is still a
binary (and the surviving companion star is weird), and (2) the Great
Eruption was not a one-off event, because the star suffered major
mass-loss eruptions at least twice before.  There are a number of
other complexities as well.  These challenges are discussed below.  We
warn the reader that the following sections are speculative; it is not
our goal here to provide the final word, but to contemplate how many
pieces of the complex puzzle fit together.

\begin{figure}\begin{center}
\includegraphics[width=2.3in]{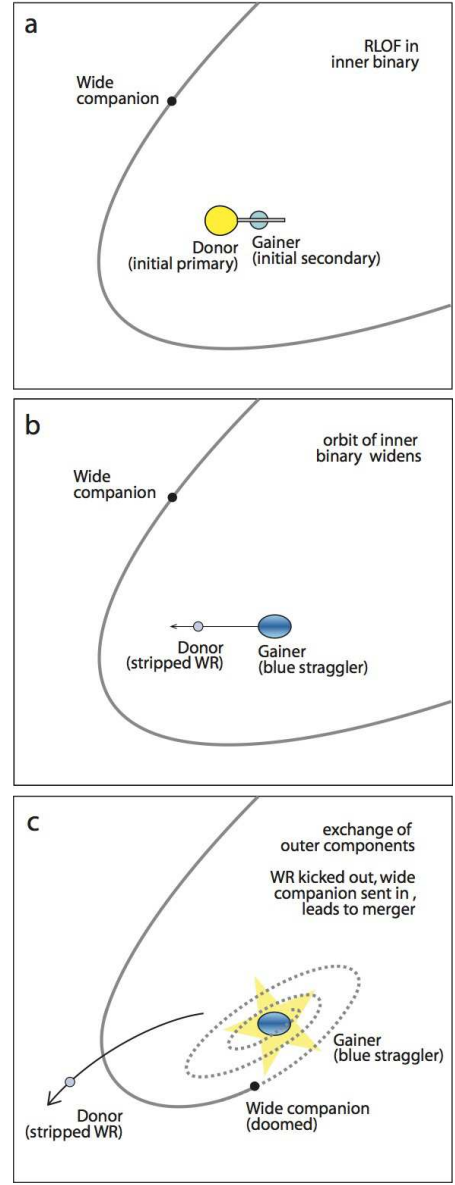}
\end{center}
\caption{A sketch of the proposed orbital interaction in a
  hierarchical triple.  {\it Panel A:} As the primary star nears the
  end of the main sequence, its expansion initiates mass transfer.
  {\it Panel B:} After RLOF ends, the primary is a stripped-envelope
  WR star, and the mass gainer is spun up, overluminous, and enriched
  (a blue straggler).  The orbit of the inner binary widens at the end
  of RLOF, and apastron may widen further due to mass loss or Kozai
  cycles.  {\it Panel C:} Eventually the orbit widens enough that the
  WR interacts with the wide tertiary companion and they exchange
  places (TEDI; \citealt{pk12}), sending the WR star out on an
  eccentric orbit, and sending the tertiary inward.  Various processes
  influence the orbit until a series of colisions and a final merger
  occur.} \label{fig:triple}
\end{figure}

\subsubsection{But what about the weird surviving companion?}

In its presently observed state $\sim$170 yr after the Great Eruption,
$\eta$ Car is known to be a binary system with a roughly 5.5 yr period
and a highly eccentric ($e \simeq 0.9$) orbit
\citep{damineli,dcl97,madura12}.  Thus, if the Great Eruption was
powered by a stellar merger event, then its persisting binarity
requires that the eruption must have occurred in a hierarchical triple
system.  

Explaining $\eta$ Car's eruption as binary merger in a triple system
was proposed about two decades ago \citep{iben99}, although it is now
amusing to note that Iben described such a model as ``preposterous''.
In that model, the wide tertiary (now binary) companion was
essentially an innocent bystander that is a relatively inert main
sequence O-type star.  However the surviving companion is probably not
a normal main sequence O-type star (see below).

More recently, \citet{pz16} proposed a conceptually similar model of a
binary merger in a triple system, although this time with the
surviving companion playing a more active role by helping to initiate
the merger of the inner binary via the Kozai-Lidov effect
\citep{kozai62,lidov62}.  This model, however, has discrepancies with
observational parameters.  For instance, because of the precarious
nature of the original triple system, a merger or collision is
triggered by the Kozai-Lidov mechanism after only 0.1-1 Myr
\citep{pz16}, which is in tension with the 3-4 Myr age of the Tr16
cluster, and also at odds with the fact that the strong enrichment of
nitrogen in $\eta$ Car's ejecta requires that it is a $>$3 Myr old
(post-main-sequence) object \citep{davidson82,davidson86,sm04}.
This model doesn't explain why the orbit of the wide tertiary that
survives today as the eccentric binary companion would be aligned with
the equatorial plane of the Homunculus (presumably the plane of the
binary merger), since the Kozai-Lidov torque is most effective when
the two orbits are misaligned (indeed, \citealt{pz16} adopted a
relative inclination between the inner and outer orbital planes of
about 90$^{\circ}$).
Moreover, in this model \citet{pz16} explained the ejection of the
Homunculus via an enhanced wind caused by tidal heating several
decades before the 1840s with the final merger occuring in 1838; this
is ruled out by the 1847.1 ($\pm$0.8 yr) ejection date from
proper motions \citep{smith17}, and the radiatively driven wind of 500
km s$^{-1}$ in their model provides no explanation for the fast Outer
Ejecta \citep{smith08}, the very fast ejecta seen in light echoes, the
extremely thin walls of the bipolar Homunculus, multiple major
eruptions, or the very luminous 1850s plateau in the light curve.  In
this model as well, the surviving companion should be a normal main
sequence O-type star, which it is probably not (see below).
Nevertheless, it is interesting to pursue the model of a merger in a
triple system, to find a satisfying scenario that is in agreement with
available observational constraints.  \citet{smith16b} showed that, at
least observationally, the light curve and spectra of $\eta$ Car's
eruption have similarities with other transients that have been
proposed as mergers.

Any model that aims to explain the Great Eruption as a merger event
needs to account not only for the highly eccentric orbit of
$\eta$~Car's surviving companion seen today, which is in the same
plane as the equator of the Homunculus, but also its unusual nature
and likely evolutionary state. Previous models assumed that the
companion was a main sequence star, but this seems inconsistent with
its observed wind properties.

The inferred physical properties of the surviving companion star in
the $\eta$~Car system indicate that it was probably a more active
player than previously suggested.  In particular, estimates of the
companion's wind indicate extreme physical parameters.  Typical values
for the mass-loss rate and wind speed of $\eta$ Car's companion
derived by comparing models of the colliding winds with observed X-ray
emission are $\dot{M}$=(1-2)$\times$10$^{-5}$ $M_{\odot}$ yr$^{-1}$
and $v_{\infty}$$\approx$3000 km s$^{-1}$, respectively
\citep{parkin11,parkin09,corcoran10,corcoran05,pc02,okazaki08,russell16,hamaguchi16}.
This is a much denser and faster wind than any normal main-sequence
O-type star, especially when considering constraints on the
companion's luminosity and ionizing flux that would point to an
initially $\sim$30 $M_{\odot}$ star or less
\citep{mehner10,teodoro08,verner05}.  A typical 30 $M_{\odot}$ main
sequence star, by contrast, would have $\dot{M}$ \, = \, 10$^{-7}$
$M_{\odot}$ yr$^{-1}$ and $v_{\infty}$=1000 km s$^{-1}$
\citep{smith14}.  Instead of a main-sequence companion, then, the
extreme wind properties of $\eta$ Car's companion are much more
consistent with it being a fairly typical hydrogen-poor early-type
Wolf-Rayet (WR) star.

The surprising strength and speed of the companion's wind, as well as
its possible similarity to those of WR stars, has been noted several
times before, but to our knowledge {\it the rather profound
  evolutionary implications have so far not been discussed in the
  literature}.  Namely, finding that $\eta$ Car's surviving companion
is likely a WR star forces us to rethink the interaction history of
this system.  If the primary star were an extremely massive single
star of $M_{\rm ZAMS} \simeq 200 M_{\odot}$ or so, or even if it is a
merger product from two 60-80 $M_{\odot}$ stars, then its short
main-sequence lifetime should prohibit its lower-mass companion
(currently the secondary) from being a WR star already as a
consequence of its own unaltered evolution with mass loss.  More
massive stars have shorter lifetimes, and there is not enough time for
this to have happened if the companion has evolved effectively as a
single star.  Indeed, a 30 $M_{\odot}$ star that has made it to the WR
phase already would be older than the 3 Myr stellar population in
the Carina Nebula \citep{smith06b}, so one would need to invoke the
unlikely scenario that a wandering star ejected from another cluster
was captured to make the $\eta$ Car multiple system.  A companion
star of $M_{\rm ZAMS} \simeq 30 M_{\odot}$ should still be midway
through its main-sequence H-burning phase, and it certainly would not
yet have shed its H envelope --- especially when its much more massive
companion is still an LBV that has retained its H envelope.

At this point we must resign ourselves to the fact that something
fairly complicated has happened to $\eta$~Car.  We have what appears
to be an initially $\sim$30 $M_{\odot}$ star that has reached its He
core burning phase as a compact fast-winded WR star, while orbiting
around one of the most luminous blue supergiant stars in the Milky Way
that still retains its H envelope and has presumably just finished its
core H-burning evolution.  How can this be?  Without the assistance of
time travel, {\it we conclude that the surviving wide secondary must
  have participated in close binary evolution in order to alter the
  masses and lifetimes of the system components that have been
  inferred}.  This requires previous close interaction and mass
exchange, followed by dramatic orbital evolution.

Here is a scenario that appears plausible, given observational
constraints, although it is admittedly speculative and probably not a
unique explanation.

\begin{itemize}

 \item Suppose that the $\eta$ Carinae system was originally a
   high-mass hierarchical triple system (Figure~\ref{fig:triple}a).
   Toward the end of its main sequence, the initially most massive
   star in the inner close binary initiates stable mass transfer,
   being stripped of its H envelope in the process, and donating that
   envelope to its mass-gainer companion.

 \item This mass transfer produces a stripped-envelope WR star
   (initially the primary, now the wide secondary), and makes the mass
   gainer into a rapidly rotating, overluminous, N-enriched, blue
   supergiant or LBV that is now the most massive star in the system
   (Figure~\ref{fig:triple}b).
   

 \item Up until this point in the evolution of the system, the wide
   tertiary has been inert.  However, after mass transfer and reversal
   of the mass ratio, the inner binary system's orbit widens
   \citep{paczynski71}.  With the center of mass closer to the more
   massive LBV-like mass gainer, it is the stripped WR star who moves
   outward (Figure~\ref{fig:triple}b).

 \item The speculative part about this scenario is that this widening
   of the orbit may then trigger a violent 3-body interaction, wherein
   the stripped WR star and the outer tertiary companion (still a main
   sequence O-type star) exchange places.  This interaction kicks the
   WR star out on a wide and highly eccentric orbit (as observed
   today), while kicking the previously inert tertiary star inward to
   interact and eventually merge with the mass gainer
   (Figure~\ref{fig:triple}c).

 \item The orbital evolution of the inner binary may then be pushed to
   a merger by Kozai cycles, by grazing collisions with the bloated
   mass-gainer, and/or by interaction with a disk that still surrounds
   the mass gainer.

 \item This final merger, potentially preceded by a few violent
   grazing collisions, eventually culminates in the Great Eruption.

\end{itemize} 

Is this type of model really so far-fetched?  For earthling
astronomers accustomed to orbiting a single star, stellar binary and
triple systems often seem exotic.  However, observed statistics
indicate that for massive stars, binary interaction is the norm, and
hierarchical triple systems are common rather than a rare exception
\citep{moe17,chini12,kiminki12,kk12,sana12,eggleton07,kf07}.  Mass
loss and mass transfer in the inner binary in triple systems can have
dramatic effects in tandem with Kozai cycles and tides
\citep{naoz16,st13,mp14,kem98}, and companions can exchange places
\citep{kea94,keo94,pk12}.  \citet{pk12} have discussed in detail how
mass loss and interaction from the inner binary in a hierachical
triple system can lead to chaotic orbital evolution, including
collisions, exchanging places, and the formation of highly eccentric
systems, in an instability they refer to as the triple evolution
dynamical instability (TEDI).  The present configuration of the $\eta$
Car system seems to naturally fit expectations from TEDI.  Triple
systems have been invoked to help explain the observed fraction of
close binaries \citep{tokovinin06,moe17}, and mergers in triples akin
to the scenario above may be necessary to explain the origin of blue
stragglers \citep{pf09,iben99b,pk12}.  This may be particularly
relevant to LBVs, since the observed environments of LBVs suggest that
they are indeed massive blue stragglers \citep{st15}.  Interestingly,
a triple-system encounter that led to an exchange of the original wide
tertiary and an inner companion was already suggested for $\eta$ Car
\citep{lp98}, although not in the context of triggering a merger.

While the scenario outlined above is admittedly somewhat ad hoc and in
need of further quantitative exploration to assess its probability,
the basic picture is self consistent and provides a single plausible
explanation for a large number of pecularities of the $\eta$ Car
system: The exchange of partners explains the highly eccentric and
aligned orbit of the secondary, it accounts for the apparent age
discrepancy where a lower-luminosity WR star is at a more advanced
evolutionary stage than the more luminous H-rich primary, and the
exchange of partners is the event that sent one of the stars inward to
trigger the merger that powered the Great Eruption.  It may also shed
some light on the previous eruptions (discussed in the next
subsection).  We have a plausible explanation for why this event
occured at the 3-4 Myr age of the Tr16 cluster and not sooner, which
is that the process began as a result of mass transfer when the
initial primary finished its main sequence evolution; the widening
orbital evolution following mass transfer is what initiated the 3-body
interaction in a hierachical system that had been relatively stable
during the main sequence.  Without having mass transfer and an
exchange of partners that kicked out the stripped envelope star
(originally the primary), it is almost unfathomable that we could have
a WR star that has a significantly lower presumed initial mass while
being in a more advanced stage of evolution than its more luminous
primary star, and also in a highly eccentric orbit.  Capturing an
unrelated star from the host cluster might explain the high
eccentricity, but would still conflict with the apparently advanced
evolutionary state of a star whose H-burning lifetime is longer than
the Tr16 cluster.  We encourage quantitative studies of the 3-body
parameter space that might lead to the current observed properties of
the $\eta$ Car system in the qualitative scenario that we described,
but such an exploration is well beyond the scope of our current paper.


\subsubsection{But what about the previous eruptions?} 

Another key challenge for any merger model of the Great Eruption is
that a binary system merging into one star should be a singular event,
whereas observational evidence indicates that $\eta$~Car has erupted
repeatly.  Proper motions of the Outer Ejecta around $\eta$~Carinae
reveal that it suffered at least two major eruptive mass-loss episodes
prior to the 19th century Great Eruption; these occurred aproximately
300 and 600 yr before \citep{kiminki16}. These precursor eruptions
also had a different geometry than the 19th century eruption, with the
first being almost entirely one-sided (to the northeast and
blueshifted) and the second sending material in a few different
directions, but with neither sharing the axisymmetry or orientation of
the bipolar Homunculus \citep{kiminki16}.

In the scenario discussed above, where a triple-system exchange leads
to a stellar merger, one might envision the multiple eruptions as a
consequence of several ``near misses'' or grazing collisions before
the final stellar merger event.  This is an expected outcome of the
TEDI instability described above.  After the exchange of partners that
kicks out the WR star and kicks in the original tertiary, the inner
binary will likely be eccentric and its orbital evolution can be
influenced by Kozai cycles as noted above.  This may drive the binary
to closer periastron distances such that they eventually have a near
miss or their outer envelopes collide, leading to significantly
asymmetric mass ejection from the primary star's loosely bound
envelope.  Additionally, the companion may interact with a remnant
disk or bloated envelope around the rapidly rotating mass gainer
\citep{munoz15}.  The tidal or dynamical friction of these grazing
collisions may degrade the orbit enough that that the binary finally
merges together after a few violent encounters \citep{pk12}.
(Collisions resulting from the TEDI instability were originally
discussed in the context of enlarged red giant envelopes in lower mass
stars, but the bloated envelope of a massive LBV also has a very large
radius.)  Since the first pass is likely to be the most eccentric,
this might be an interesting explanation for why the first of the
three historical eruptions sent ejecta in largely one direction
\citep{kiminki16}.  The final event (the actual merger of the two
stars) produced strong mass loss that was, by contrast, highly
axisymmetric (the Homunculus).

If a significant fraction of the primary star's outer envelope is
ejected in these repeating eruptions, there may be a physical
mechanism for the delay between eruptions.  Namely, the star's
envelope may recover and re-establish equilibrium on a thermal
timescale (centuries for the mass of the outer envelope).  When the
envelope reaches a radius comparable to the periastron separation,
this may trigger the next grazing collision.  The outer companion star
seen today shares the same orbital plane as the equator of the
Homunculus, whereas the previous eruptions did not share the same
geometry.  One might also then envision this series of grazing
collisions or failed mergers as a process by which the merging stars
exchanged angular momentum, so that the inner orbits became aligned
with the primary star's rotation and the orientation of the outer
binary. This is admittedly speculative and unrefined, but at least it
gives some plaubile explanation for the differing geometry in
subsequent historical eruption events.  This is lacking from any other
model for $\eta$ Car's eruption discussed so far, and is a fruitful
topic for numerical simulations.

\subsubsection{But what about the structure of the equatorial skirt
  and the Strontium Filament?}

The structure of $\eta$ Car's highly non-spherical ejecta provide
important clues about its recent violent mass-loss history, as noted
previously (see, e.g., \citealt{smith12,smith13}).  The detailed
structure of the material in the equatorial plane has particular
relevance to any merger model, since the inspiral phase leading to a
merger should shed a large amount of mass through L2
\citep{pejcha16a}.

The ragged spray of debris referred to as the Equatorial Skirt is a
particularly recognizable feature of the Homunculus \citep{morse98},
and the so-called ``Strontium filament'' is one particular location in
the equatorial skirt with unusually strong emission lines from
[Sr~{\sc ii}] and a number of other low-ionization metals
\citep{hartman04}.  Most of the equatorial skirt has the same age as
the rest of the Homunculus \citep{morse01}, but some features appear
to have been ejected decades before or afterward \citep{smith17}.  One
expects these features to have some satisfactory explanation in a
model of the Great Eruption.

At the pinched waist of the Homunculus Nebula, multiwavelength
observations (especially in the infrared) have revealed the existence
of a complicated toroidal structure. This torus has been discussed by
numerous authors, the history of which is summarized by
\citet{smith12}.  More recently, high-resolution observations of
$^{12}$CO 2--1 with the Atacama Large Millimeter Array (ALMA) have
revealed new clues to the structure of the equatorial ejecta
\citep{smithALMA}.  The CO emission shows a toroidal structure with a
size similar to previous IR data, but the density structure of the
CO torus departs strongly from azimuthal symmetry.  In particular, the
torus is a series of clumps, with higher density on the far southest
side and an opening on the northwest side.  It basically shows a ``C''
shape, with the gap in the ``C'' pointing toward us (i.e. the near
side of the equator to the northwest direction).

The connection to any model for the Great Eruption is evident when we
consider the relative orientation of the torus compared to the
currently observed eccentric binary system.  Namely, the direction of
apastron in the eccentric binary seen today is also to the northwest
\citep{madura12}, toward the middle of the gap in the torus
\citep{smithALMA}.  This suggests that the wide eccentric companion
seen today had roughly the same orbital orientation during the lead up
to the Great Eruption (i.e. the inspiral or Phase 1 as discussed
above), which it maintained as the equatorial ejecta expanded past it.
This is consistent with the notion that the orbital period did not
change by more than a few percent before and after the Great Eruption
\citep{smith11}.  In other words, the wide secondary star we study
today was not a major source of energy to power the Great Eruption. It
may be that periastron encounters were able to enhance the mass loss
toward the far side of the nebula, whereas equatorial material ejected
toward us (i.e. toward apastron) by the central binary may have been
diverted or disrupted by the eccentric companion after being shed by
the inner binary.  This is an area where hydrodynamic simulations
could be very useful to understand the interaction.  It is interesting
to note, however, that the polar lobes of the Homunculus do not depart
so strongly from azimuthal symmetry.  In any case, the resulting
density structure of the torus helps explain why the currently
observed equatorial skirt appears as a ragged spray of streamers in
optical images.  Namely, the equatorial skirt is illuminated by
scattered light that escapes preferentially through holes and gaps
between clumps in the inner torus.  Because of the azimuthal
asymmetry, more light escapes in the direction of apastron, which is
why the equatorial skirt is seen mostly on our side of the Homunculus.
The fast post-eruption wind from the secondary can also escape more
easily through the gaps in the torus, influencing structures on the
near side of the equatorial ejecta.  See \citet{smithALMA} for a more
detailed discussion.

The peculiar nature of the Sr filament may have less to do with the
Great Eruption.  As discussed recently by \citet{smithALMA}, the Sr
filament is downstream from the dense inner clumps known as the
``Weigelt knots'', and its pecular low ionization may arise because it
is shadowed by them.  The Weigelt knots appear to have been ejected
later, probably in the 1890 eruption or afterward
\citep{dorland04,smith04acs}, but they are clearly in the direction of
apastron in the equatorial plane.  The asymmetry of the mass ejection
during 1890 is beyond the scope of our paper.

\subsubsection{But what about the S Condensation and NN jet?}

Further outside the Homunculus, emission line images reveal a
spectacular array of clumpy nebular structures known as the Outer
Ejecta \citep{walborn76}.  Most of these are from older eruptions
centuries before the Great Eruption \citep{kiminki16}.  Prominent
among the Outer Ejecta are the so-called ``NN Jet'' and ``S
Condensation'', which resemble collimated outflows in the equatorial
plane \citep{walborn76,walborn95,kiminki16,mehner16}.  The S
Condensation and NN Jet have proper motions that indicate they were
either ejected in the Great Eruption or in the decades leading up to
it, but not in the eruptive events 300 and 600 years before
\citep{kiminki16}.  While detailed study of the kinematics has
revealed that they are ballistic ejections rather than true
hydrodynamic jets
\citep{meaburn93,glover97,morse01,kiminki16,mehner16}, one may still
wonder how $\eta$ Car managed to simultaneously send two large bullets
out in opposing directions in its equatorial plane in an event that
was otherwise highly axisymmetric.

Here, again, it is likely that the wide eccentric secondary plays an
important role in this non-axisymmetric structure.  As noted by
\citet{smith11}, the apparent magnitude in the historical light curve
and apparent color in the years leading up to the Great Eruption peak
dictate that the emitting photosphere was larger than the periastron
separation of the current binary, and that as such, some sort of
violent interaction like a collision must have occurred at times of
periastron.  This remains true whether this photosphere was a true
hydrodynamic surface of the star, or (more likely) the photosphere of
a common envelope or the emitting radius of the shock heated
circumbinary disk during the inspiral phase of a merger.  These
periastron collisions have been interpreted as causing the brief
luminosity spikes seen in the historical light curve in 1838 and 1843
\citep{smith11}.  The wide companion must have plunged into one side
of the dense envelope around the star and popped out the other side,
possibly multiple times.  This is discussed in more detail in our
companion paper on the fast ejecta seen in $\eta$ Car's light echoes
\citep{smith+18}, where we point out that with the known orbital
geometry from \citet{madura12}, the S Condensation roughly matches the
direction of ingress and the NN Jet roughly matches the trajectory of
egress in this collision. The wide companion may have left a tunnel in
its wake as it drilled its way through the extended toroidal envelope,
through which fast ejecta from the Great Eruption may have been able
to squirt.  We noted earlier that the known echo geometry of EC2
indicates that it views $\eta$ Car roughly in the equatorial plane,
and its position angle is within 10-20 deg of the S Condensation's
trajectory (also in the equatorial plane).  In other words, if the S
Condensation is a bullet of fast ejecta, then the EC2 echo seems to be
looking nearly down the barrel of the gun.  This is probably related
to the very broad wings of H$\alpha$ seen in the EC2 echo reported
here.

\subsubsection{But what about the brief 1838 and 1843 luminosity
  spikes?}

Following the scenario discussed in the previous two subsections, the
star that we now see as the wide secondary was orbiting around a close
binary in the process of merging, as it ejected substantial amounts of
material in the equator.  In addition to profoundly influencing the
structure of the outflowing ejecta seen today around $\eta$~Car, the
close periaston encounters may have played an important secondary role
in the light curve.  These periaston passes were not enough to power
the total kinetic and radiated energy of the whole 10$^{50}$ erg
event.  However, shock heating from the companion ripping through the
L2 mass loss disk or plunging into the bloated common envelope of the
merger may have powered the brief luminosity spikes in the light curve
\citep{smith11,sf11}.  In this view, the main 1850s plateau and the
brief 1838 and 1843 precursor spikes have two different specific
physical causes, even though they both stem from the same merger
event.  This may help explain why similar brief luminosity spikes are
absent in UGC~2773-OT, even though the light curve and spectral
evolution of its plateau phase are almost identical to $\eta$ Car's
(i.e. the outer tertiary may have been on a wider orbit that didn't
cause periastron collisions in the case of UGC~2773-OT, or it may have
simply been a merger not in a triple system).  Even if we are
restricted to merger events in triple systems, the influence of
periastron encounters by the outer companion could cause significant
diversity from one merger event to the next, depending on the
configuration of the outer orbit and the mass of that companion.  This
may be related to the tremendous diversity in extragalactic SN
impostors \citep{vdm12,smith+11,pastorello10}.

\subsubsection{But what about the age of the Homunculus?}

In the two-stage merger scenario proposed above, the ejected mass that
constitutes the Homunculus Nebula actually leaves the star over a time
period of several decades or more.  The single, well-constrained
apparent ejection date for the Homunculus of 1847.1 $\pm$0.8 yr from
proper motions \citep{smith17,morse01} arises because a strong shock
from an explosive event sweeps up this previously ejected material
into a thin cooled shell with a single dynamical age.  The shock
cooling, which provides the luminosity of the plateau, also allows
this swept up material to collapse to a very thin layer as seen today
in the walls of the Homunculus \citep{smith06,smith13}.  This shock
essentially erases the previous mass-loss history and creates the
illusion, when we measure its proper motion expansion, that the whole
nebula was ejected instantaneously \citep{smith17}. Doppler velocities
seen in light echoes, on the other hand, show that strong mass loss
was occurring over several years, but that the fastest material
appeared after 1847.  We have associated the fast, explosive outflow
with a shock that results from energy deposition due to the final
stellar merger event, whereas the slower 150-200 km s$^{-1}$ outflow
seen prior to 1847 was due primarily to mass loss associated with the
inspiral phase of the merger.

\subsubsection{But what about...?}

Admittedly, there are remaining open questions associated with $\eta$
Car and its nebula that are not close to being settled by the
speculation in this final part of the paper. The general scenario
outlined above certainly is a challenge for quantitative models to
match, but light echoes combined with the properties of $\eta$ Car's
ejecta narrow the range of possible configurations and free parameters
considerably.  While it may be fun to entertain even more complicated
scenarios (multiple mergers, precessing jets, exotic compact object
interactions, a T$\dot{\rm Z}$O, etc.), the triple-interaction
scenario described above with an exchange of partners leading to a
merger seems like the minimum level of complexity needed to
simultaneously account for a WR-like companion in a wide eccentric
orbit around a much more massive and more luminous rapidly rotating
primary that still retains its hydrogen envelope, which can, moreover,
achieve disk plus bipolar geometry, heavily nuclear-processed ejecta,
and multiple eruptions with one of those being a $\ge$10$^{50}$ erg
explosive ejection event that accounts for the spectroscopic evolution
seen in light echoes.

\section{Summary and Future Work}

In this paper we have analyzed photometry and spectroscopy of a light
echo EC2 that we argue reflects light from the decade-long plateau of
the Great Eruption of $\eta$ Carinae.  This echo views the star from a
vantage point that is near the equatorial plane of the Homunculus
Nebula.  Combined with another echo (EC1) that we reported previously
\citep{rest12,prieto14}, which reflects light from an earlier epoch
along roughly the same direction, this provides the first long-term
spectroscopic time series of the main part of $\eta$ Car's 19th
century eruption.  Briefly, some of the main results are as follows:

1. Echo spectroscopy gives a time series of outflow velocity and line
strength that indicate a two-phase eruption, with a slow 150-200 km
s$^{-1}$ outflow at early times, followed by a faster (600 km
s$^{-1}$) bulk outflow at later times.  The division between these two
phases coincides with the measured dynamical age of the Homunculus.

2.  Spectra also reveal, for the first time, expansion speeds as high
as 10,000-20,000 km s$^{-1}$ in later stages of the eruption.  We
suspect that these are associated with the fastest nebular material
seen today in the Outer Ejecta \citep{smith08}.  These are the fastest
speeds so far reported in any non-terminal SN impostor or LBV-like
eruption.  They clearly indicate an explosive component to the
eruption.

3.  The relatively slow outflow followed in time by a fast outflow
make it highly likely that the plateau of $\eta$ Car's eruption was
indeed powered largely by shock luminosity in CSM interaction,
qualitatively similar to a scaled down version of a Type~IIn supernova
\citep{smith13}.

4.  We interpret the two-phases of the observations (slow vs.\ fast
outflow) in a physical picture that corresponds to a two-stage stellar
merger event, with (1) inspiral and L2 mass loss \citep{pejcha16a},
and (2) explosive outflow and CSM interaction \citep{smith13}.  We
attribute the persistent broad H$\alpha$ line wings in Stage 2 to the
fastest explosive ejecta crossing the reverse shock, as commonly seen
in later phases of SNe~IIn.

5.  In order to reconcile this simple binary merger picture with the
eccentric colliding wind binary system and $\eta$ Car's complex nebula
seen today, we propose a qualitative model involving an exchange of
partners in a hierachical triple system that led to the merger,
kicking out the original primary as a stripped-envelope WR star on a
wide eccentric orbit.  This invokes the TEDI instability discussed for
lower-mass triple systems.  While mergers and triple systems have been
discussed before in the context of $\eta$ Car, this new scenario
differs in several key ways that are more easily reconciled with
observational constraints, as we describe in the text (Section 4.6).
As such, $\eta$ Car would add a relatively extreme example of blue
stragglers formed in triple systems.

We plan to continue monitoring the evolution of the EC2 echo, since it
is expected to fade significantly over the next several years.  If it
does, this will confirm that it was an echo from the plateau of the
Great Eruption, and it may provide important clues about dust
formation in the Homunculus.  For instance, if the fading after 1858
was caused by dust formation, we may see the red sides of strong
emission lines fade as newly formed dust blocks receding parts of the
ejecta as in some SNe and novae.  We may also witness signatures of
the formation of molecules that are pathways to the dust formation
\citep{prieto14}.  After EC2 fades, we can obtain a spectrum of the
reflecting surface of its cloud to serve as a template to subtract
from our earlier spectra, to clearly delineate narrow emission
features that come from the surface of the cloud and to determine
which, if any, narrow emission features (such as He~{\sc i} lines)
were in fact part of the echo light.  Other echoes along similar lines
of sight tracing the pre-1845 peaks \citep{rest12,prieto14} should
brighten again in next 5-10 years and eventually show a spectrum
similar to EC2.  This will definitively confirm that EC2 was light
from the 1850s plateau.

So far we have only analyzed a time series of light echo spectra as
viewed from a single vantage point near the equator.  Given the
present day Homunculus, the outflow and CSM interaction must have been
highly latitude dependent, so we might expect strong differences in
both the light curve and spectra of the Great Eruption as viewed from
the poles or intermediate latitudes.  The evolution of velocities and
excitation with time may be very different, not to mention the
possible formation of dust at different times along various lines of
sight.

If any of the ideas outlined in this paper are on the right track,
then $\eta$ Carinae is a gold mine for understanding the recovery of a
massive-star merger product in the centuries after the event.  This is
something we cannot address in the near term with extragalactic SN
impostors, whereas the tight observational constraints on the physical
parameters of $\eta$ Car's eruption and its spatially resolved nebula
provide fertile ground for theoretical work.  We now have a detailed
record of observations during the aftermath over a timeline of 170
years since the Great Eruption, spectroscopy and photometry of the
event itself, constraints on previous eruptions that occurred 300-600
years before the putative merger, plus the remnant binary star system
and nebula.  If $\eta$ Car's eruption really was a merger, this opens
a new field of inquiry for investigating mergers in the most massive
stars, the formation of massive blue stragglers, non-terminal
explosions, and the progenitors of some extreme types of SNe.

\section*{Acknowledgements}

\scriptsize 

We thank an anonymous referee for a careful reading of the manuscript.
We benefitted from numerous sisyphean conversations about the grey
area between explosions and winds with Stan Owocki, Dave Arnett, Eliot
Quataert, and Chris Matzner, and informative conversations about
binary L2 mass loss and mergers with Ondrej Pejcha, Stephen Justham,
Kaitlin Kratter, and Philipp Podsiadlowski.  We have also discussed
binary evolution, on occasion, with Selma de Mink.  N.S.'s research on
Eta Carinae's light echoes and eruptive transients was supported by
National Science Foundation (NSF) grants AST-1312221 and AST-1515559.
Partial support for this work was provided by NASA grants AR-12618,
AR-14586, and GO-13390 from the Space Telescope Science Institute,
which is operated by the Association of Universities for Research in
Astronomy, Inc. under NASA contract NAS 5-26555.  Support for JLP is
provided in part by FONDECYT through the grant 1151445 and by the
Ministry of Economy, Development, and Tourism’s Millennium Science
Initiative through grant IC120009, awarded to The Millennium Institute
of Astrophysics, MAS. DJJ gratefully acknowledges support from the NSF
through award AST-1440254.

This paper includes data gathered with the 6.5m Magellan Telescopes
located at Las Campanas Observatory, Chile.  This project used data
obtained with the Dark Energy Camera (DECam), which was constructed by
the Dark Energy Survey (DES) collaborating institutions. Funding for
DES, including DECam, has been provided by the U.S. DoE, NSF, MECD
(Spain), STFC (UK), HEFCE (England), NCSA, KICP, FINEP, FAPERJ, CNPq
(Brazil), the GRF-sponsored cluster of excellence “Origin and
Structure of the Universe” and the DES collaborating institutions.
This work makes use of observations from the LCO network.  Based, in
part, on observations obtained at the Gemini Observatory, which is
operated by the Association of Universities for Research in Astronomy,
Inc., under a cooperative agreement with the NSF on behalf of the
Gemini partnership: the National Science Foundation (United States),
the National Research Council (Canada), CONICYT (Chile), Ministerio de
Ciencia, Tecnolog\'{i}a e Innovaci\'{o}n Productiva (Argentina), and
Minist\'{e}rio da Ci\^{e}ncia, Tecnologia e Inova\c{c}\~{a}o (Brazil)
(Program GS-2014B-Q-24).

\end{document}